\documentclass[preprint]{aastex61}
\usepackage{graphicx}

\def\H2{H$_{2}$}

\def\roH2{$\rho_{\textrm{H}_2}$}
     
\def\MH2{M$_{\textrm{H}_2}$}

\submitjournal{AJ}

\shorttitle{Disks in cluster S0 galaxies} \shortauthors{Sil'chenko et al.}

\begin{document}

\title{The structure of large-scale stellar disks in cluster
lenticular galaxies.\footnotemark[1]\thanks{ Based on observations made with
the Las Cumbres Observatory telescope network.}}

\correspondingauthor{Olga Sil'chenko}
\email{olga@sai.msu.su,olgasil.astro@gmail.com}

\author[0000-0003-4946-794X]{Olga K. Sil'chenko} 
\affil{Sternberg Astronomical Institute, M.V. Lomonosov Moscow State University, Universitetsky pr., 13, Moscow, 119991 Russia}
\affil{Isaac Newton Institute, Chile, Moscow Branch}
\email{olga@sai.msu.su}

\author{Alexei Yu. Kniazev}
\affil{South African Astronomical Observatory, PO Box 9, 7935 Observatory, Cape Town, South Africa}
\affil{Southern African Large Telescope Foundation, PO Box 9, 7935 Observatory, Cape Town, South Africa}
\affil{Sternberg Astronomical Institute, M.V. Lomonosov Moscow State University, Universitetsky pr., 13, Moscow, 119991 Russia}
\email{akniazev@saao.ac.za}

\author{Ekaterina M. Chudakova} 
\affil{Sternberg Astronomical Institute, M.V. Lomonosov Moscow State University, Universitetsky pr., 13, Moscow, 119991 Russia}
\email{artenik@gmail.com} 

\begin{abstract}
By obtaining imaging data in two photometric bands for 60 lenticular galaxies -- members of
8 southern clusters -- with the Las Cumbres Observatory one-meter telescope network,
we have analyzed the structure of their large-scale stellar disks. The parameters of radial surface-brightness 
profiles have been determined (including also disk thickness), and all the galaxies have been classified into pure exponential 
(Type I) disk surface-brightness profiles, truncated (Type II) and antitruncated (Type III) piecewise exponential disk
surface-brightness profiles. We confirm the previous results of some other authors that the proportion of 
surface-brightness profile types is very different in environments of different density: in the clusters the Type-II profiles 
are almost absent while according to the literature data, in the field they constitute about one quarter of all lenticular
galaxies. The Type-III profiles are equally presented in the clusters and in the field, while following similar scaling 
relations; but by undertaking an additional structural analysis including the disk thickness determination we note
that some Type-III disks may be a combination of a rather thick exponential pseudobulge and an outer Type-I disk.
Marginally we detect a shift of the scaling relation toward higher central surface brightnesses for the outer segments 
of Type-III disks and smaller thickness of the Type-I disks in the clusters. Both effects may be explained by enhanced radial 
stellar migration during disk galaxy infall into a cluster that in particular represents an additional channel for Type-I disk 
shaping in dense environments. 
\end{abstract}


\section{Introduction}  

The origin of S0 galaxies -- which is the second, after spirals, most frequent morphological 
type in the nearby Universe constituting about 15\%\ of all galaxies  \citep{apm,efigi,s4gmorph}  -- 
remains unclear and controversial. Their global structure -- a presence of two main large-scale components, a bulge 
and a disk, with various contribution of each into the total luminosity, -- resembles the structure 
of early-type spirals very much; however, their disks lack spiral arms, and low-level star formation in 
S0s, if present, is usually organized in ring-like structures. The resemblance of the lenticular and 
spiral scaling relations \citep[see e.g.][]{lauri10,carmen15} and the absence of intense star formation 
in the formers provoke numerous scenaria of S0 (trans-)formation from spirals by removing the gas from 
the disks and by quenching star formation in the disks. The dominance of S0 population in clusters 
at $z=0$ \citep{dressler,wings} implies dynamical mechanisms related to dense environments and massive 
host dark-matter halos as probable ways to transform a spiral galaxy into a lenticular one. However,
despite the dominance of S0s in clusters, the majority of them inhabit loose groups and even very
rarified fields being completely isolated \citep*{isosalt}; their origin cannot be related to
ram-pressure gas removal from spiral disks or tidal disk transformation in dense environments. 
There are some evidences that S0s in clusters and in the field may have different channels
to form \citep[e.g.][]{wilman_erwin}. If it is true, and the dynamical mechanisms shaping large-scale
stellar disks of S0s in clusters and S0s in the field are different, one can expect the
quiescent stellar disks of nearby S0s to have different structure characteristics in environments
of different density. When referring to the `structure characteristics', we mean both radial
structure and vertical structure of the disks.

As for the radial structure of galactic stellar disks, it is now well known that the shape of
stellar surface-brightness (density) profiles in disk galaxies is piecewise exponential. They
can be fitted by a single-scale exponential law up to the border of a stellar disk \citep{freeman},
or by an exponential law with truncation at some radius as the Type II disks in \citet{freeman} or
disks with breaks noted by \citet{vdkruit_searle}, or by
two exponential profiles suitable within different radius ranges, with the outer exponential law
having a larger scalelength -- so called antitruncated disks \citep*{n5533,n80,erwin05}. 
Presently, after \citet{pohlen_trujillo}, these three types of surface-brightness profiles are 
numbered as follows: pure exponential disks are Type I, truncated disks are Type II, and antitruncated 
disks are Type III. It is not clear yet, if the shape of a stellar surface-brightness (density) profile is
an initial condition of the disk formation, or there are some ways of dynamical transition between
the types. Recently \citet*{erwin12} and \citet{pranger} have reported a discovery of environment effect 
on the profile-shape statistics for disk galaxies: in clusters there is a deficit
of Type II profiles and an excess of Type I profiles. Consequently, \citet{clarke_db} have
proposed a dynamical mechanism to transform truncated disks into pure exponential ones
when entering into cluster environment: a combination of gas stripping by ram pressure and of enhanced
stellar radial migration in the disks provides necessary changes in the stellar disk structure. Earlier, 
there were also works with some dynamical simulations transforming pure exponential disks into antitruncated ones 
\citep[e.g.][]{younger,borlaff14}. Cosmological simulations reveal advanced dynamical evolution just of the Type-III disks:
they suffer strong radial migration as well as concentration of newly accreted stars in the outermost parts
\citep{ruizlara}. However, here we must note that not all observational studies of the disk-type proportion
dependence on the environment density find any difference between clusters and field for S0s: for example,
in the STAGES survey results \citep{stages} the absence of the Type-II disks in S0 galaxies has been claimed
both for the cluster and for the field. So probably the question remains open.

Another important property of the galactic stellar disks is their thickness. It is crucial to
have statistics on the thickness of S0 disks because it would allow to restrict strongly a
choice of dynamical mechanisms shaping the large-scale components of lenticular galaxies.
For example, dry minor mergers which have been proposed by \citet{younger} to form an antitruncated
surface-density profile, would thicken stellar disks strongly by increasing their stellar velocity
dispersion \citep*{walker96}. Observational data on the galactic disk thicknesses were rather sparse; 
and up to now individual estimates of stellar disk thickness were made directly only for galaxies seen 
edge-on \citep*[e.g.][]{mosenkov10}. Though some interesting statistics has been derived from these
decompositions -- for example, \citet{mosenkov15} have reported very weak dependence of the
disk thickness on the morphological type, large scatter of the disk thicknesses in intermediate-type
spirals, thicker disks in barred galaxies etc. -- however, on our opinion, it is rather difficult
to discuss radial and especially azimuthal structure of the galaxies seen strictly edge-on.
\citet{thickmeth} have proposed recently a quite novel method allowing to estimate
the thickness of an exponential (or piecewise exponential) stellar disk seen under arbitrary
inclination, if only it is not strictly edge-on or strictly face-on (with our method we explore the disk
inclinations between $10{\degr}$ and $75{\degr}$). We have already begun to study
galactic disk thicknesses and to compare the statistics of disk thickness among the samples with
various types of radial surface-brightness profiles for early-type disk galaxies in the field
\citep*{thickmeth,lcogt1}. In the present paper we continue to apply our method of measuring thicknesses 
of individual galactic disks to a sample of S0 galaxies which are members of several southern clusters 
of galaxies. In Section 2 we describe the sample, in Section 3 we give details of our approach to
the stellar disk structure characterization, in Section 4 we present our quantitative results, in
Section 5 we discuss the consequences of our findings, and in Section 6 we conclude.

\section{The sample}

For our photometric study we have selected 60 S0 galaxies in several southern clusters of galaxies,
spanning a range of masses (X-ray luminosities) but all being not too far from us. The
southern sky offers a rich choice of nearby clusters of galaxies, unlike the
northern sky where only the Virgo cluster is closer to us than 70 Mpc. The list of clusters 
considered here is presented in the Table~\ref{clusters}.

\begin{table*}
\caption{Clusters which host the studied S0s}
\label{clusters}
\begin{flushleft}
\begin{tabular}{lcclr}
\hline\noalign{\smallskip}
The cluster name & $D^1$, Mpc & Scale$^1$, kpc per $^{\prime \prime}$ & $L_X$$^2$, in $10^{44}$ erg/s 
& Number of S0s studied here\\
\hline\noalign{\smallskip}
Abell 194 & 71 & 0.331 & 0.070 & 8 \\
NGC 1550 group & 49.6 & 0.238 & 0.153 & 4 \\
Fornax & 18.2 & 0.088  & 0.012 &  3 \\
Centaurus & 37.5 & 0.212 & 0.721 & 11 \\
Hydra & 54 & 0.272 & 0.297 & 13 \\
Antlia & 35 & 0.205 & 0.034 & 8 \\
Abell 3565 & 55 & 0.262 & 0.008 & 4 \\
Abell S0805 & 61 & 0.292 & 0.029 & 9 \\
\hline
\multicolumn{5}{l}{$^1$\rule{0pt}{11pt}\footnotesize
NASA/IPAC Extragalactic Database}\\
\multicolumn{5}{l}{$^2$\rule{0pt}{11pt}\footnotesize
\citet*{xrayclust}}\\
\end{tabular}
\end{flushleft}
\end{table*}

\begin{figure}
\centering
\includegraphics[width=0.45\textwidth]{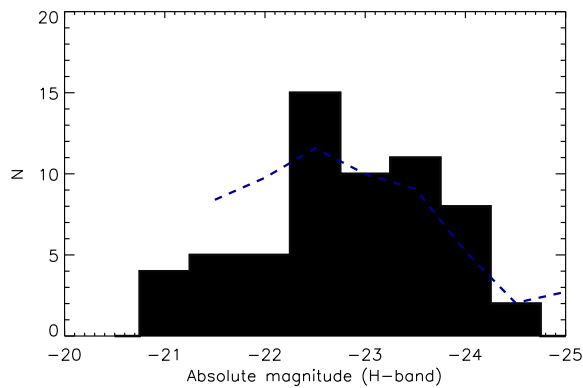}
\caption{The distribution of the galaxy absolute magnitudes in the $H$-band for our sample;
the $M_H$ are taken from NED. The dashed line overposed represents the luminosity function of
the volume-limited sample of early-type galaxies from \citet{atlas3d_1}.}
\label{mh_distrib}
\end{figure}

It is in the clusters listed in the Table~\ref{clusters} that we have selected galaxies classified as S0 or S0/a 
in the NASA/IPAC Extragalactic Database (NED). Some of the galaxies which we have taken to study are assigned the
classification type of E or E$+$ in the NED; but our visual inspection of their images has revealed clear 
disk signatures -- such as bars or blue rings, and we have recognized them as S0s. After deriving the surface
brightness profiles (by the method described in the next Section) we have additionally checked the presence
of large-scale stellar disks in the galaxies selected for the analysis -- since the only distinctive attribute
(definition) of the S0 morphological type is the presence of a large-scale disk without spiral arms. We have fitted
the outer parts of the surface-brightness profiles by an exponential law and have assured that there exists at least
one segment of the profile which lacks any systematic deviations from the exponential law within two
exponential scalelengths. Just this criterion -- the exponential law validity within two-scalelength radial range --
was proposed by \citet{freeman} in his classical work as a feature of exponential stellar disks.
The range of luminosities of the S0s selected for our study is from $M_H=-21$ to $M_H=-24.5$ -- in any case,
our galaxies are not dwarfs. Moreover, if one compare the distribution of the NIR absolute magnitudes
of our galaxies (Fig.~\ref{mh_distrib}) with the luminosity function of the volume-limited sample
of nearby early-type galaxies from \citet{atlas3d_1}, one can see that we have a well-presented
sample at $L\ge L_*$ and an underpresented sample at lower luminosities. However, we have covered all
the luminosity range of non-dwarf S0 galaxies. The galaxies are selected to be not strictly edge-on;
curiously, in some clusters (e.g. in Abell 194 or in Abell S0805) there is unexpectedly large fraction of edge-on S0s, 
and our choice has not been easy. In general, the sample is not full, but rather representative; for example,
in the Antlia cluster, among 25 S0s brighter than $m_b=15.7$ \citep{antlia_cat} we have taken 8 galaxies
for our photometric study.

\section{Observations and data analysis}

\begin{figure*}[t]
\centering
\begin{tabular}{c c}
 \includegraphics[width=8cm]{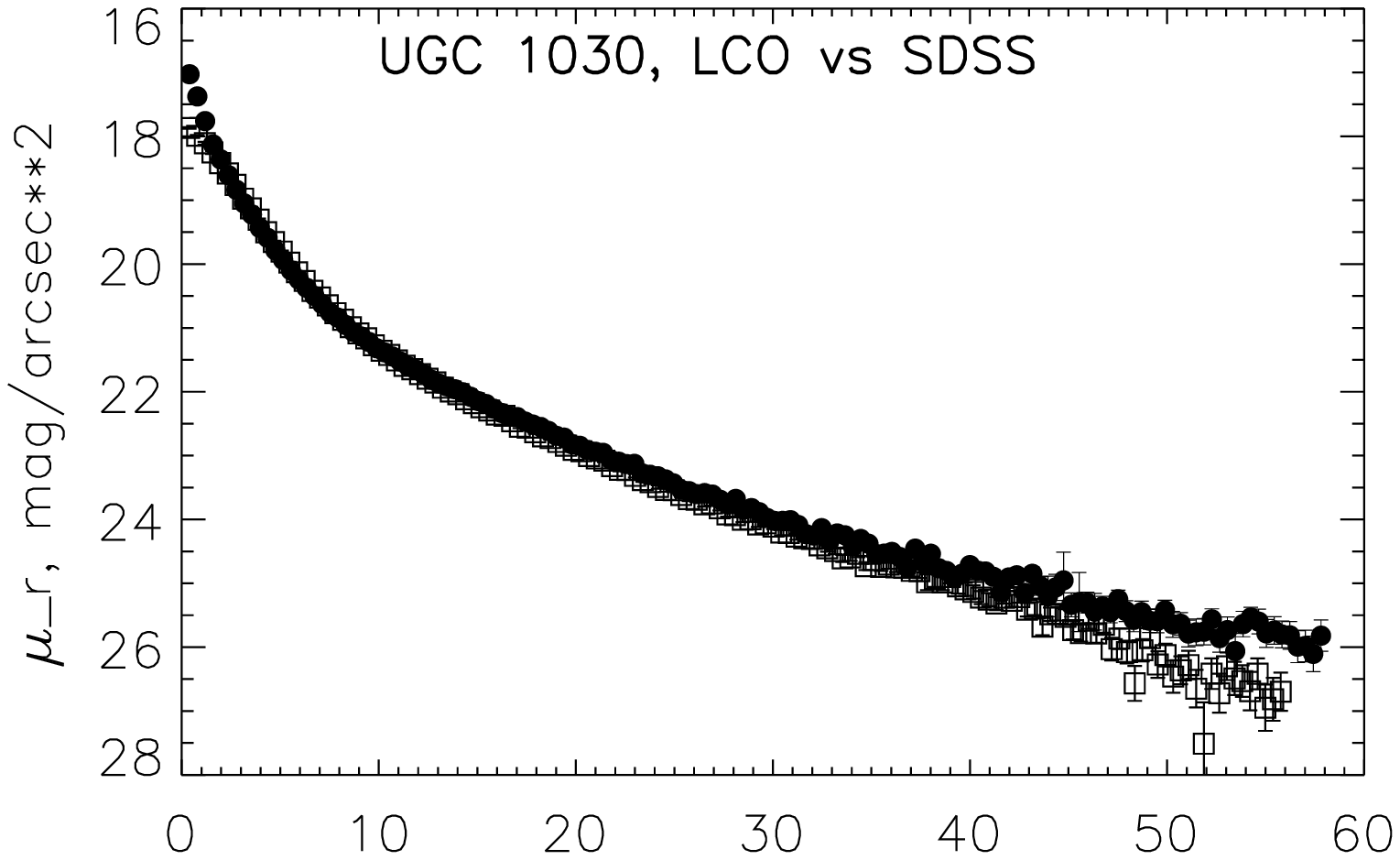} &
 \includegraphics[width=8cm]{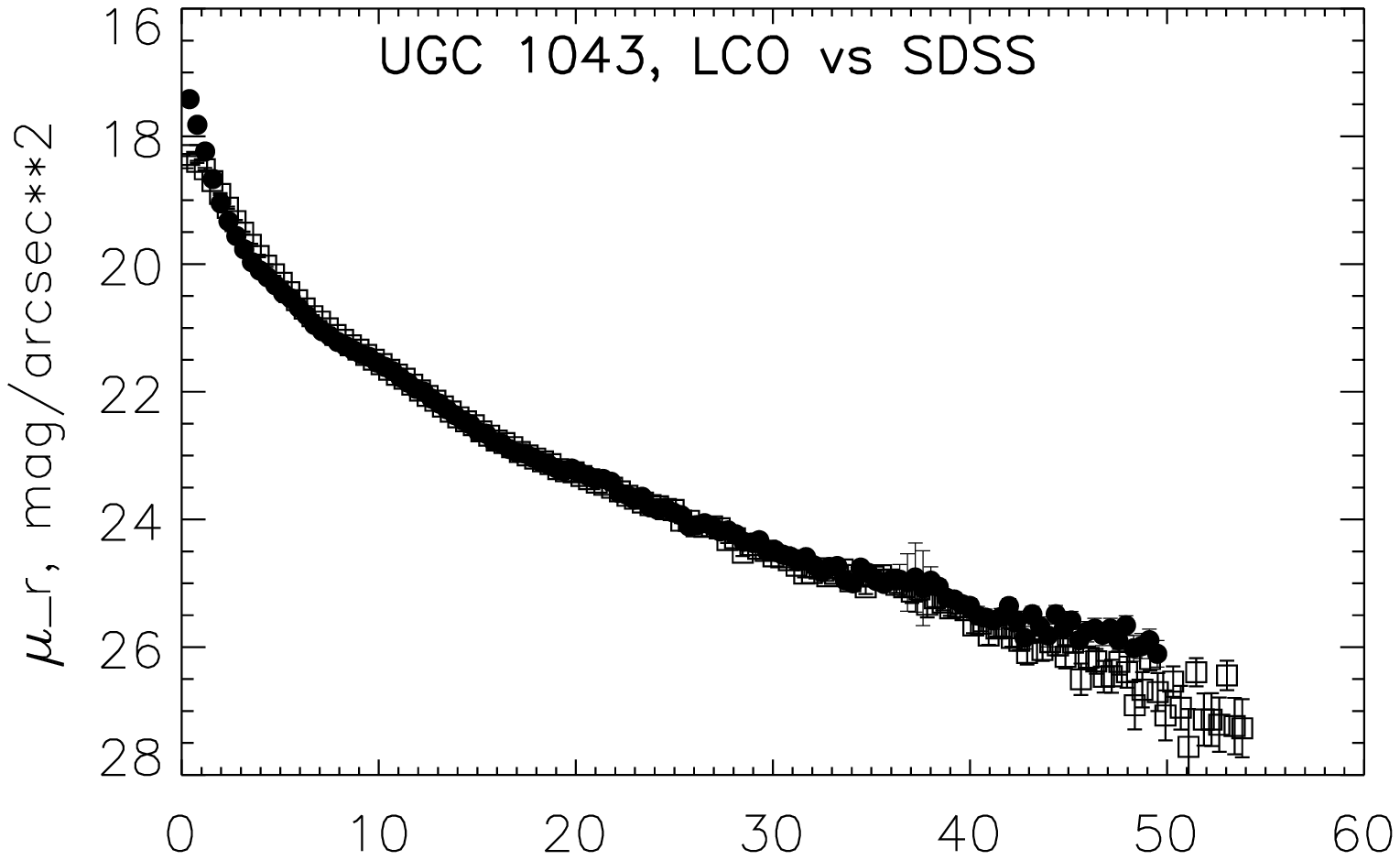} \\
\end{tabular}
\caption{The comparison of the azimuthally averaged surface brightness profiles calculated from the
SDSS/DR9 $r$-images (black points) and from the LCO $r$-images (open squares) for the Abell~194 members UGC~1030
and UGC~1043.}
\label{sdsscomp}
\end{figure*}

The observations have been fulfiled by the Las Cumbres Observatory (LCO) robotic telescope network
\citep{lcogt} between May 2015 and February 2016.
Currently, the LCO consists of two 2-m optical telescopes, of ten 1-m telescopes, 
of eight 40-cm telescopes, and one 83-cm telescope located at six observatories, 
three in the Northern and three in the Southern hemisphere. 
Such distribution of telescopes makes it possible to carry out photometric and spectral observations 
of objects regardless of their declinations, create continuous time series of observations, 
operatively acquire spectra of recently discovered supernovae, and much more \citep{lcogt}.
All our observations were done with LCO meter-class telescopes with standard Sinistro cameras for the
acquisition of direct frames. This camera consists of a 4000$\times$4000 CCD. With the
physical pixel size of this CCD, 15$\mu$m, and standard
1$\times$1 binning, the angular size of each pixel is 0.389 arcsec,
and each frame covers an area 26.5$\times$26.5 arcmin in size.
Each Sinistro camera can obtain images in 21 different bandpasses,
of which we used the $g$ and $r$ filters of the Sloan survey photometric system. 
Our observations have been made mostly in Cerro Tololo with standard exposures
of $900\mbox{s}\times 2$ in $g$ and $600\mbox{s}\times 2$ in $r$.
The seeing quality estimates made over the primarily reduced images ranged from 1.2 arcsec to 2.5 arcsec:
due to the robotic regime of the observations, the focussing and guiding were not always good.
During the observations, photometric standard stars were not exposed, so we calibrated our
images by using the HyperLEDA\footnote{\textrm{http://leda.univ-lyon1.fr}} aperture photometry collections: 
the Johnson-Cousins $BVR$ aperture data for every galaxy,
mostly based on the compilations of the photometric survey of the southern sky \citep{esolv},  
were transformed into the $gr$-system with the interrelations found by \citet*{sdsscal}.
The cluster Abell 194 is in the zone covered by the SDSS survey while the aperture photometric data are absent
in the HyperLEDA for the galaxies -- members of this cluster; so for the members of Abell~194 we have
used the $gr$-photometry of nearby stars from the SDSS/DR9 public data archive as standards. 
Also we have used $gr$-photometry of stars from the Pan-STARRS1 
public data archive as standards for the galaxies in the NGC 1550 group and for the field of NGC 3307 (the Hydra
cluster) since they lack aperture photometry in the HyperLEDA database too.
Figure~\ref{sdsscomp} shows comparison of the azimuthally averaged surface-brightness profiles
in the $r$-band calculated for UGC~1030 and UGC~1043 by exploring our data and by exploring the calibrated SDSS/DR9 frames.
One can see that in general the agreement is good. In the very center the SDSS profiles rise sharper than ours
because of the better spatial resolution. At the edges of the disks the SDSS profiles go slightly above
the LCO profiles, while the latters continue the exponential shape of the disks till fainter surface
brightnesses. The cause of this discrepance is not clear; perhaps we deal with a different level of
scattered light in the SDSS and LCO images. 

The primary reduction provided by the LCO pipeline included bias subtraction and flatfielding
of individual frames. We then extracted smaller pieces of images including the galaxies selected, co-added
two $g$-band images and two $r$-band images proceeding cosmic hit cleaning, and subtracted the sky background
from the combined $g$- and $r$-images. The sky background levels were estimated over large, by $51 \times 51$
pixels, empty square areas, beyond the galaxies, in several, from 4 to 8, directions from the galaxy centers.
These estimates were averaged before subtraction, or, in the cases of noticeable sky background gradients, 
interpolated linearly onto the galaxy position.

\begin{figure*}[p]
\centering
\begin{tabular}{c c}
 \includegraphics[width=8cm]{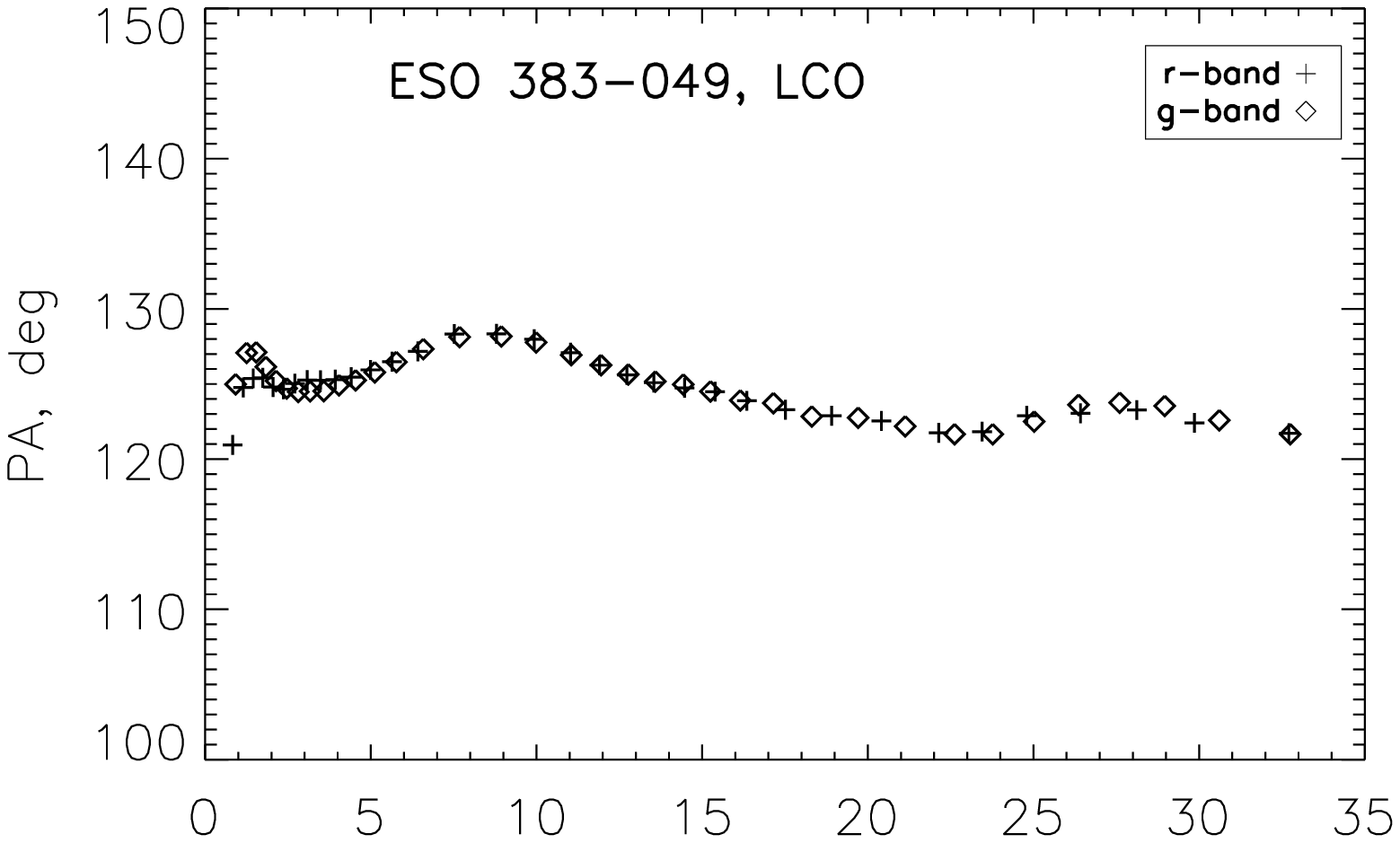} &
 \includegraphics[width=8cm]{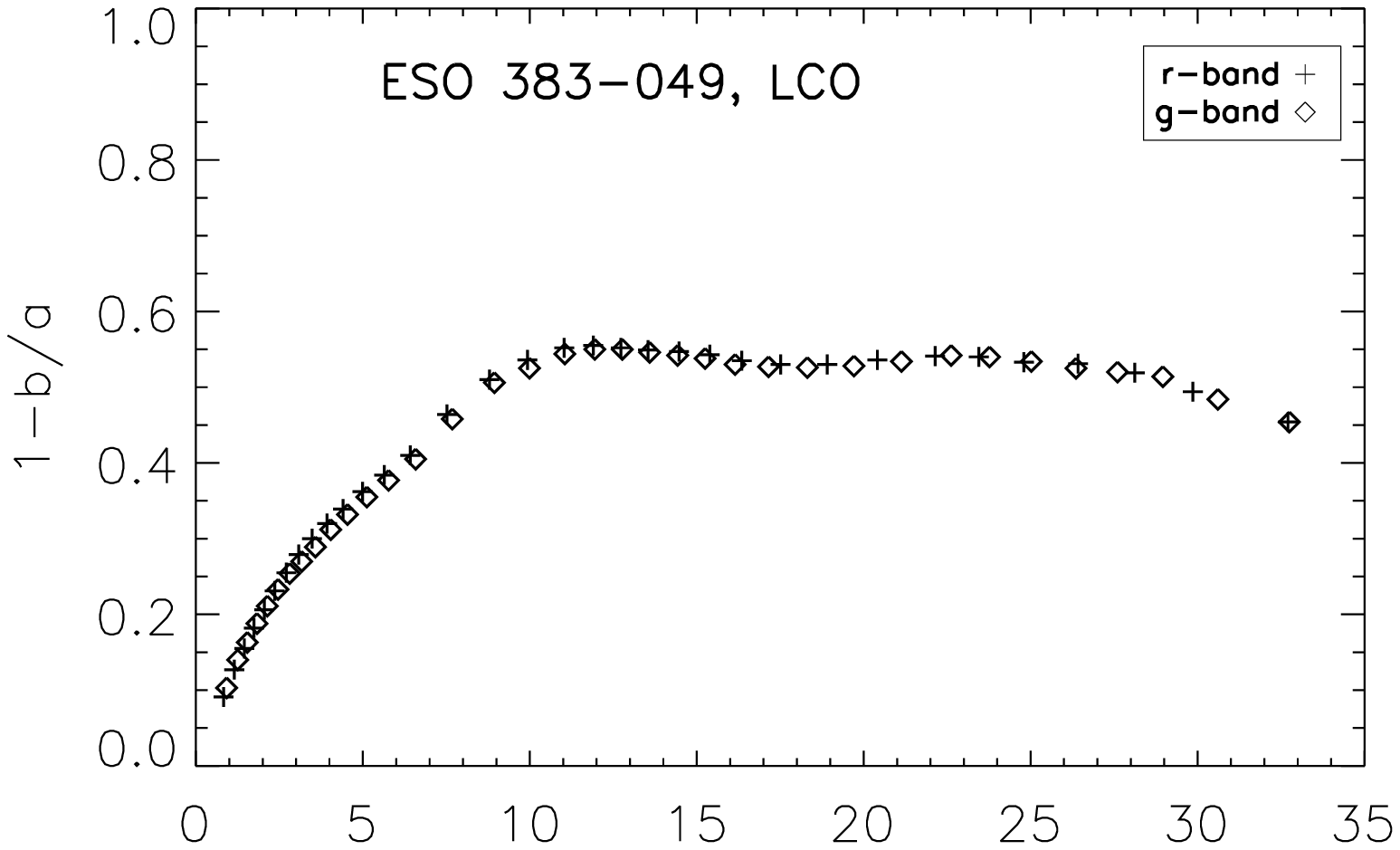} \\
 \end{tabular}
 \includegraphics[width=12cm]{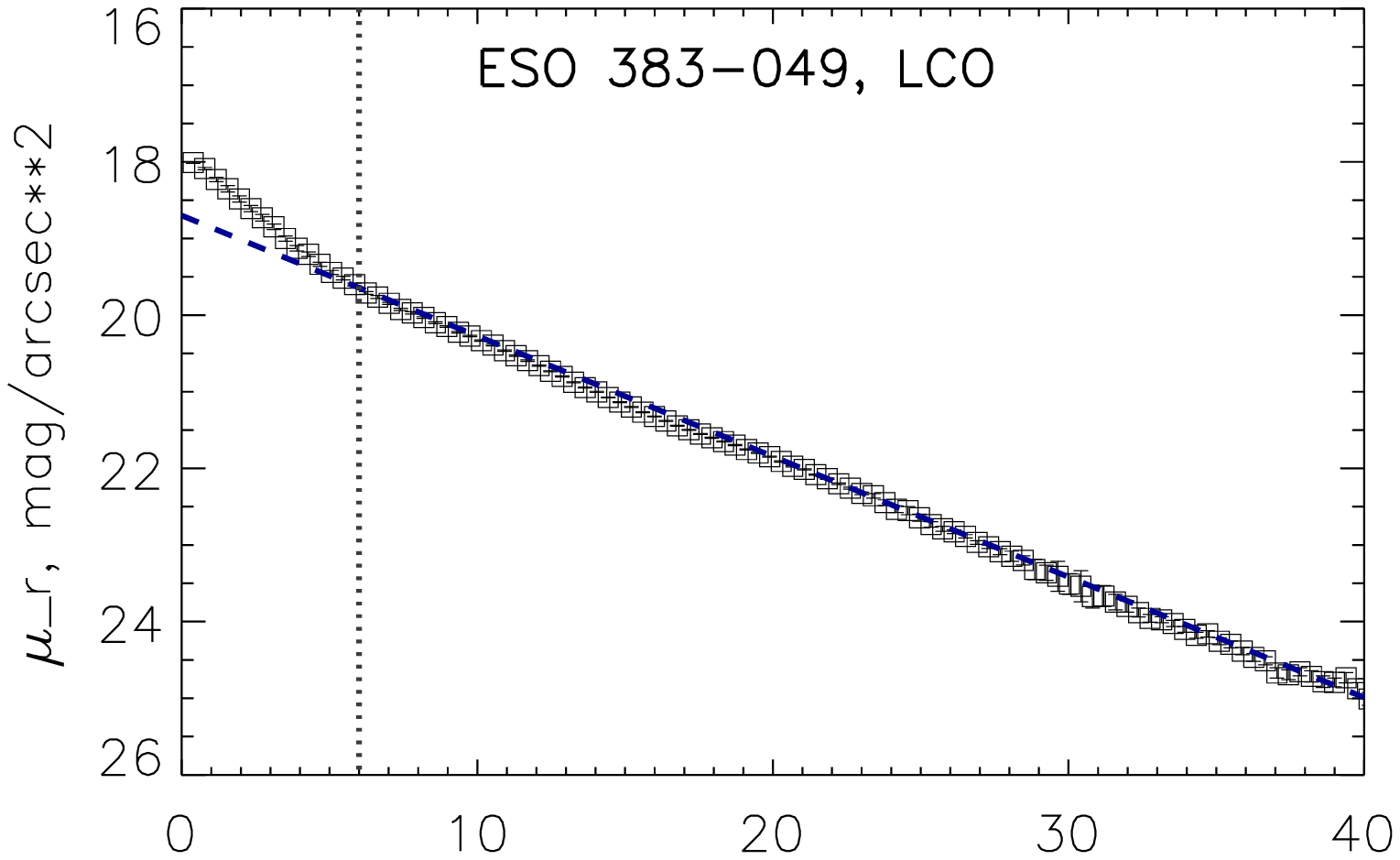} \\
\caption{An example of a single-scale exponential surface-brightness profile (Type I):
ESO 383-049. The isophote ellipticity stops to rise at $\sim 10^{\prime \prime}$, and from
this radius to the optical border of the galaxy at $35^{\prime \prime}$ the 
surface-brightness profile is well fitted by an exponential law with unique scalelength. We show
the boundary between the bulge and the disk by a vertical dotted line.} 
\label{type1}
\end{figure*}

\begin{figure*}[p]
\centering
\begin{tabular}{c c}
 \includegraphics[width=8cm]{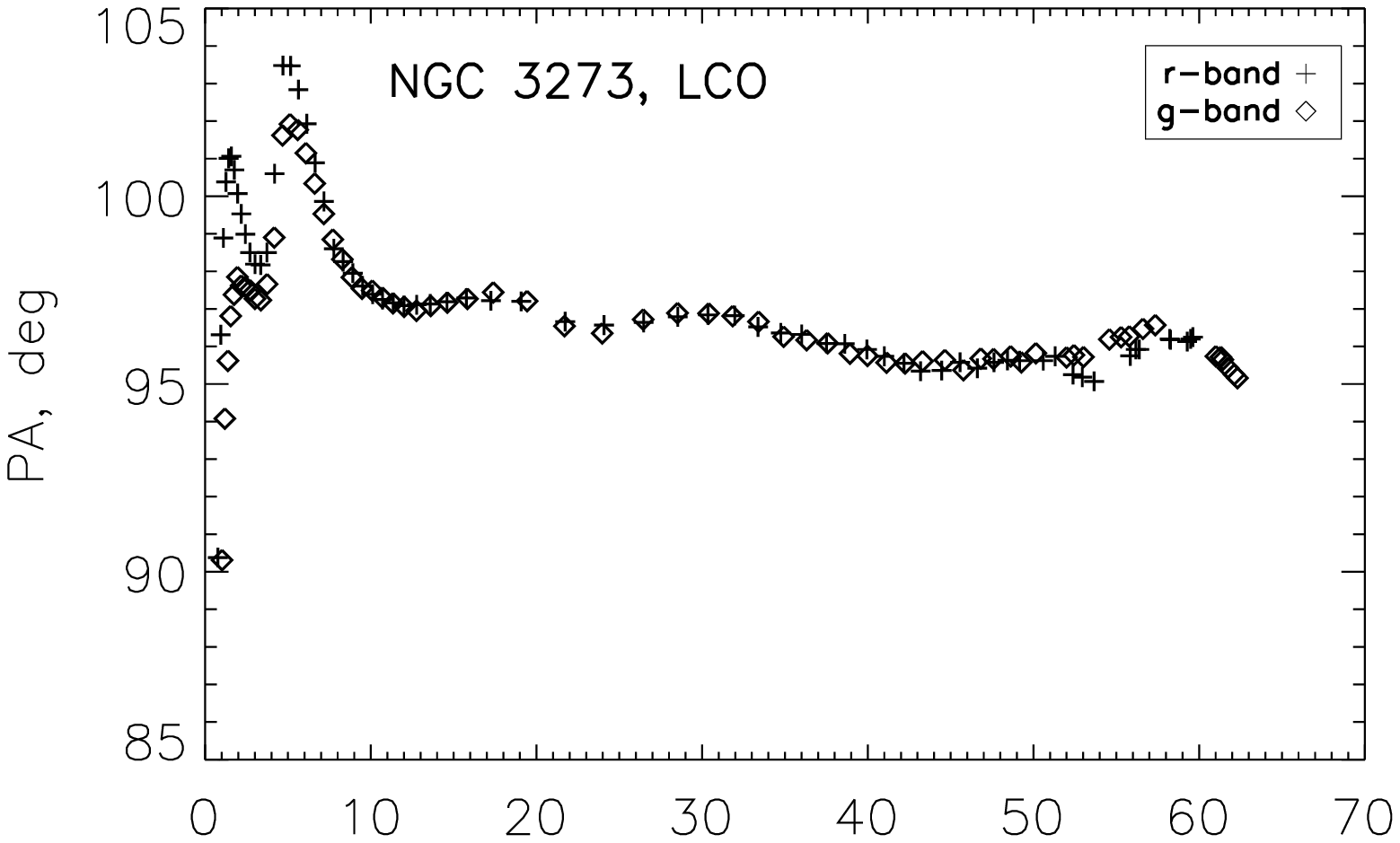} &
 \includegraphics[width=8cm]{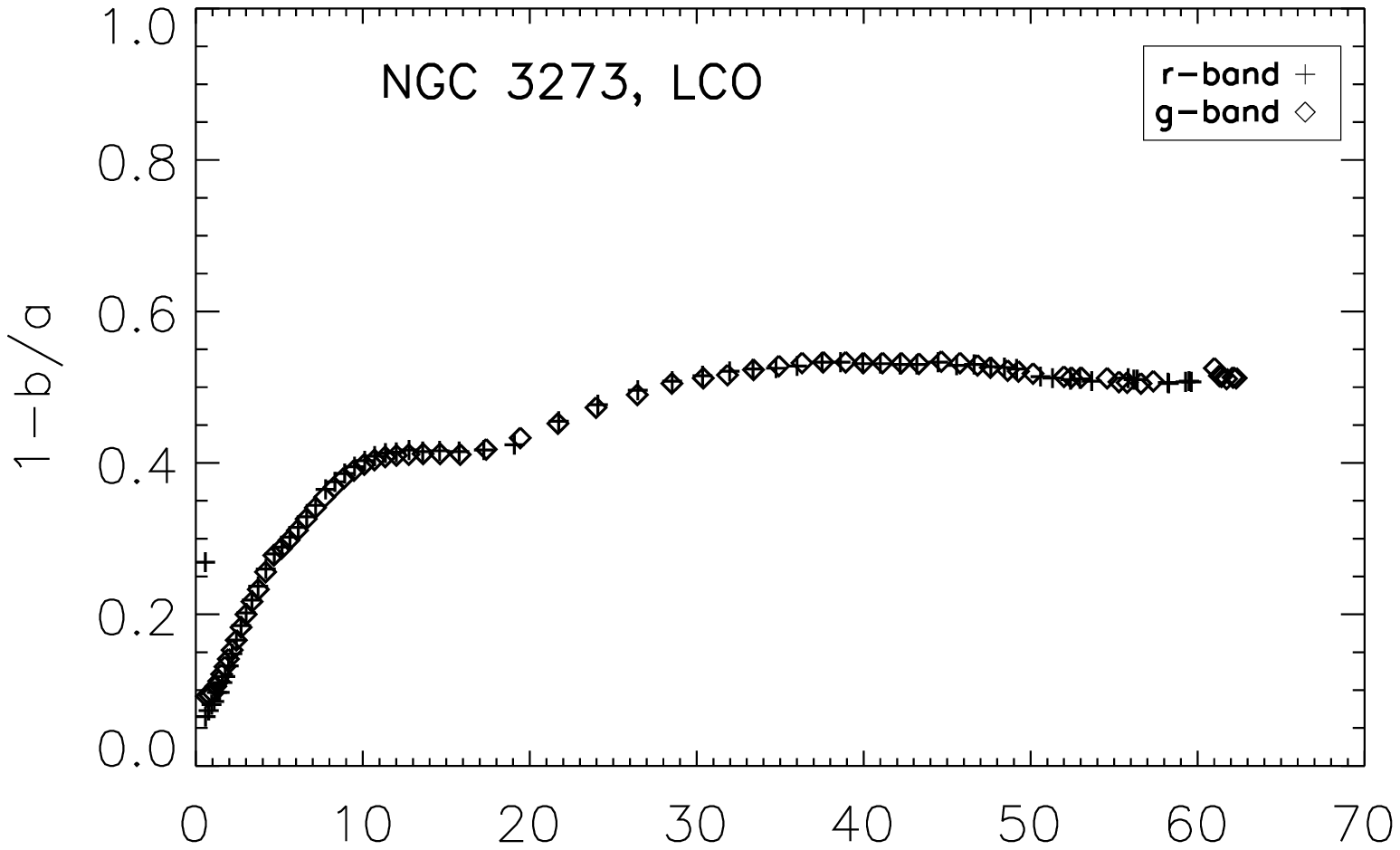} \\
 \end{tabular}
 \includegraphics[width=12cm]{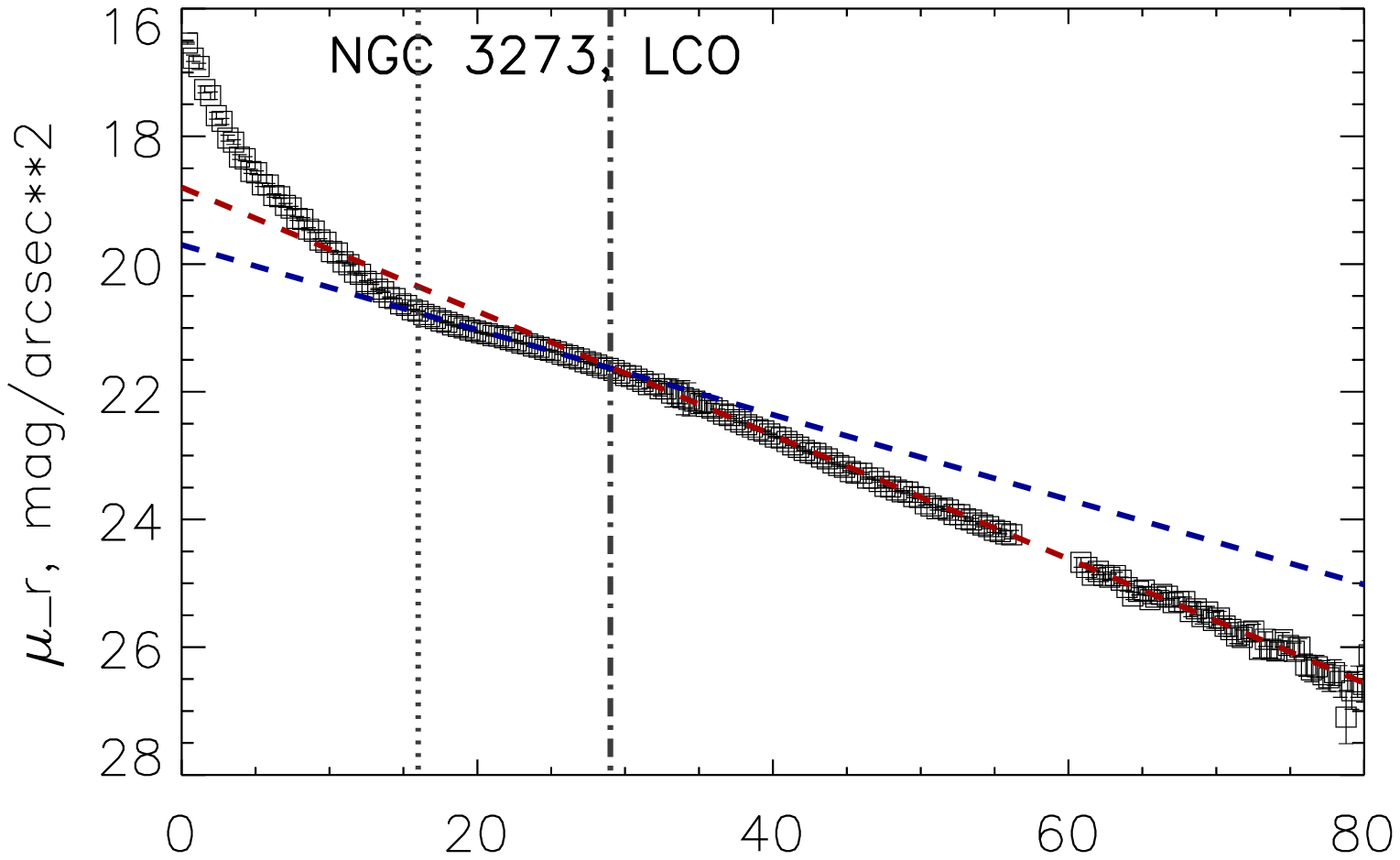} \\
\caption{An example of a truncated exponential surface-brightness profile (Type II):
NGC 3273. In the central part of the galaxy there is a rather shallow segment of the
surface-brightness profile -- a so called lens. The isophote ellipticity profile
has two plateau: in the range of $10^{\prime \prime}-20^{\prime \prime}$ at the level
of 0.4 and at $R>30^{\prime \prime}$ -- at the level of 0.5. The outer exponential law  
being extrapolated to the center goes above the lens 
surface-brightness profile -- a certain signature of the truncated surface-brightness profile.
 We show the boundary between the bulge and the disk by a vertical dotted line;
another vertical, dashed-dotted line marks the break radius.} 
\label{type2}
\end{figure*}

\begin{figure*}[p]
\centering
\begin{tabular}{c c}
 \includegraphics[width=8cm]{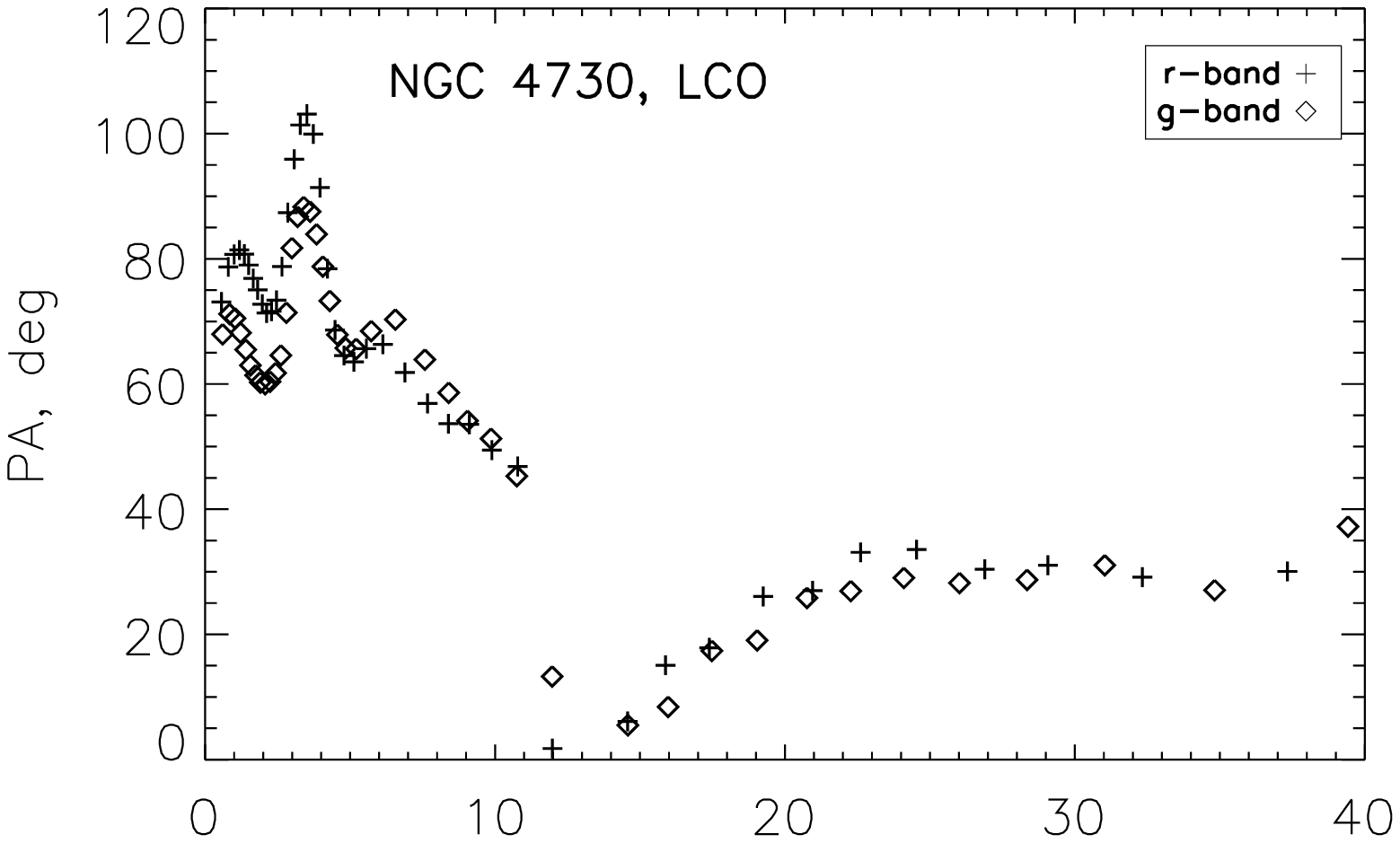} &
 \includegraphics[width=8cm]{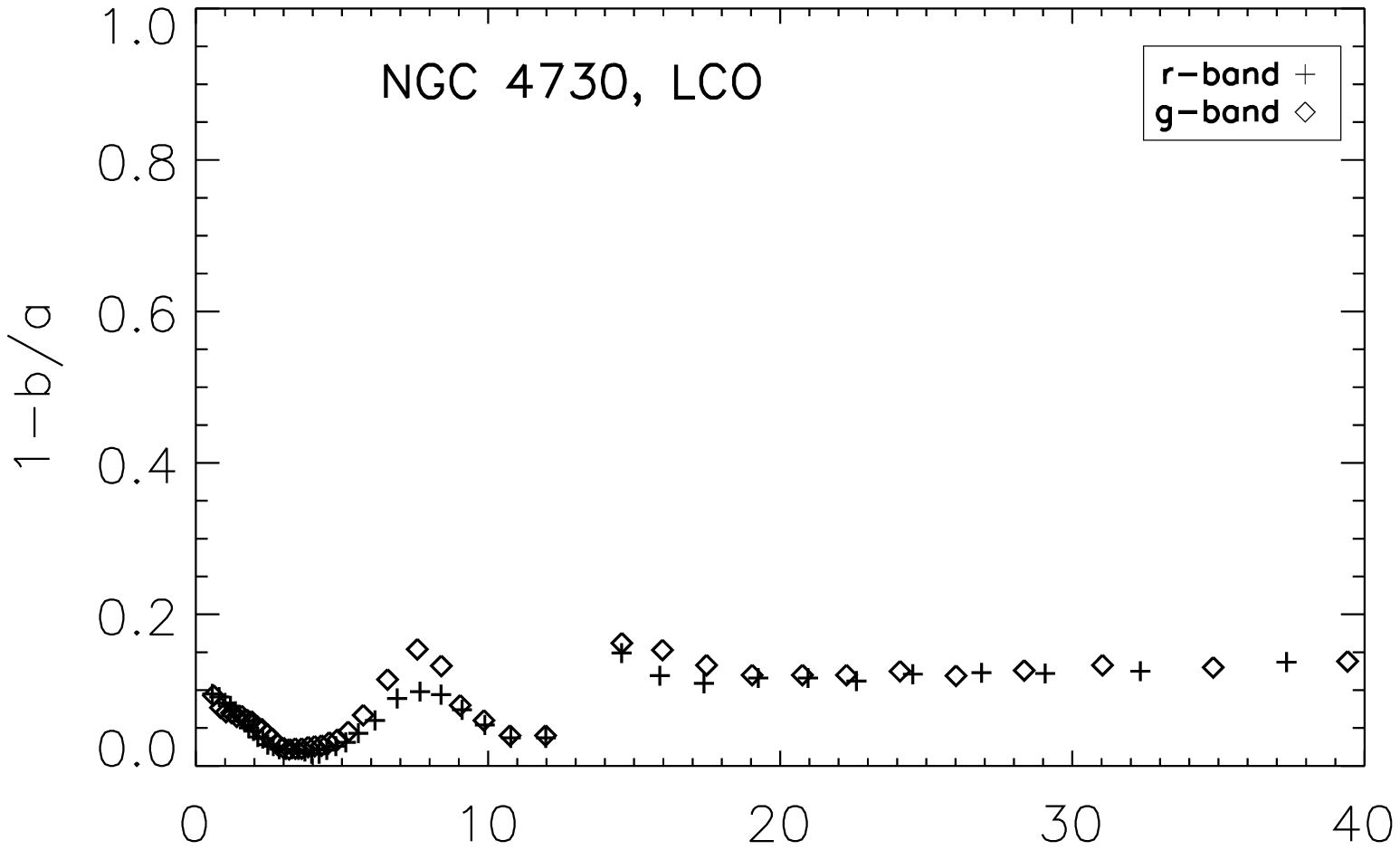} \\
 \end{tabular}
 \includegraphics[width=12cm]{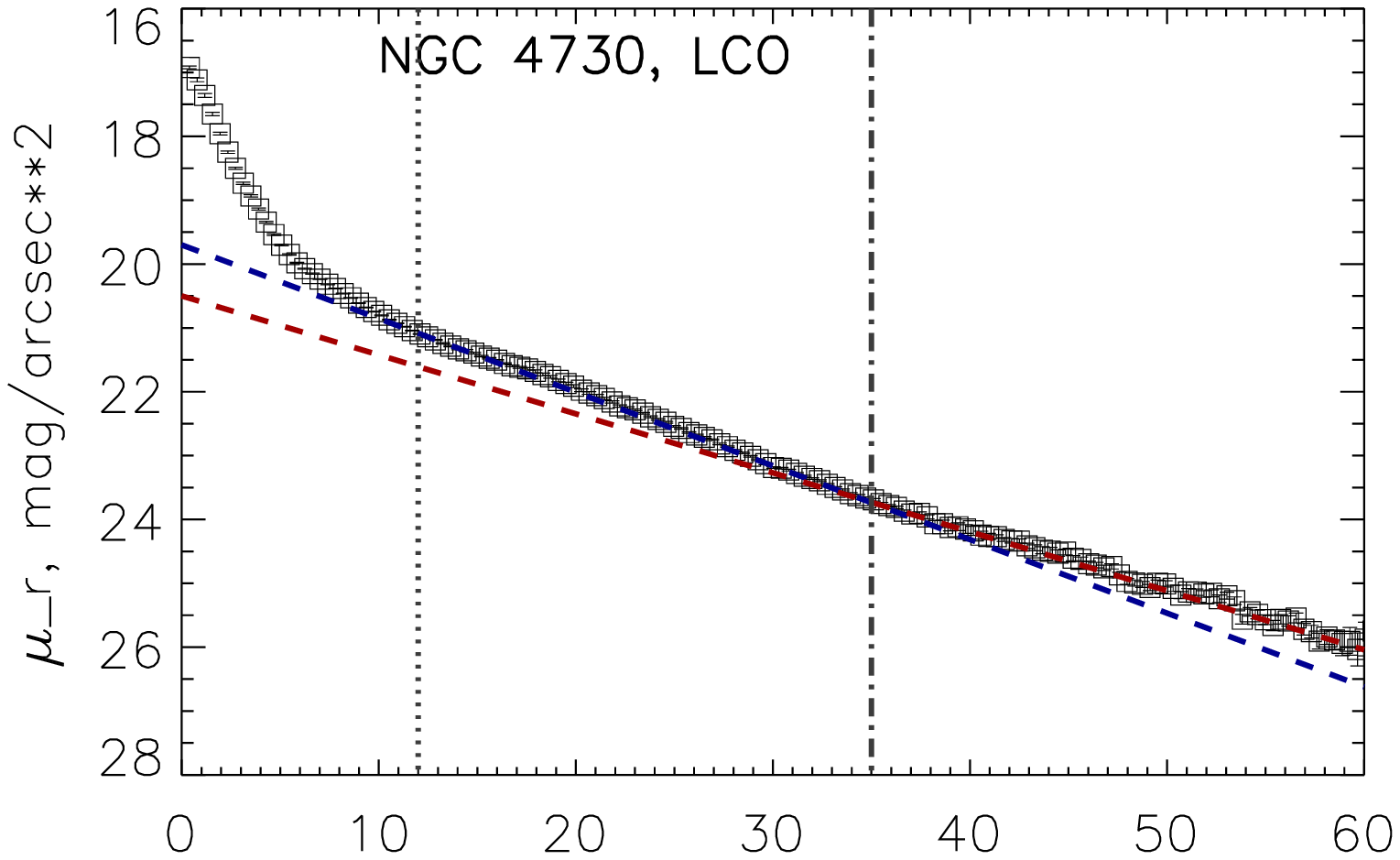} \\
\caption{An example of a two-tiered (antitruncated) exponential surface-brightness profile (Type III):
NGC 4730. The galaxy is seen nearly face-on, and in the center the isophotes are quite round.
However, starting from $R\approx 15^{\prime \prime}$ the ellipticity stays at $\epsilon \sim 0.15$,
and it is a disk-dominated  area. The disk surface-brightness profile can be fitted by two
exponential laws with different scalelengths, the outer scalelength being larger.
Two exponential laws meet at $\sim 35^{\prime \prime}$ -- it is a so called break radius, $R_{brk}$.
We show the boundary between the bulge and the disk by a vertical dotted line;
another vertical, dashed-dotted line marks the break radius.} 
\label{type3}
\end{figure*}

With flat-fielded and sky-subtracted images in hands, we have undertaken isophotal analysis for every
galaxy and have derived radial variations of the isophote ellipticities and major-axis position angle.
By assuming that large-scale stellar disks of our S0 galaxies are flat and have no warps, for every galaxy
we have found a radius where the isophote ellipticity stops to rise; we then suggest that the flat disks
dominate in the total surface brightness of the outer regions of the galaxies, starting from these radii outward.
To calculate the azimuthally averaged surface-brightness profiles of the disks,
we fixed the shape of the ellipses characterizing the round disk projection onto the sky plane, by taking
the isophote ellipticity and major-axis position angle just at these radii, and by moving outward we averaged the surface
brightnesses in the elliptical rings at every value of the radius. So the azimuthally averaged surface-brightness 
profiles of the disks have been derived. Then we fitted these disk surface-brightness profiles by an exponential
law starting from the outermost point exceeding the sky level by an rms sky-level scatter value. The quality of the fit
was recognized to be good if the rms scatter of the points around the fitting line was within typical errors of the
individual points. If we find an inner
radius where the azimuthally averaged surface-brightness profile starts to deviate systematically up or down from 
the fitted exponential law and if this radius is still within the disk-dominated radial range, we concluded that the
profile is not of the Type I, and fitted another exponential segment into the inner part of the disk azimuthally
averaged surface-brightness profile. As a result of this procedure, we have divided the total sample into three 
subsamples: the S0s with Type-I profiles, the S0s with Type-II profiles, and the S0s with Type-III profiles. 
Figures~\ref{type1}, \ref{type2}, and \ref{type3} demonstrate typical examples of all three types of the disk 
surface-brightness profiles.

The relative disk thicknesses have been calculated by our original method described in details by \citet{thickmeth};
 the careful testing of the method and determining the boundaries of its application will be presented later 
by \citet{chudfuture}.  Briefly, we followed the consideration of the projection effects on an ellipsoid
with the axes $a_1=a_2>a_3$ made by \citet{hubble26}. If we observe an intrinsically round, infinitely thin disk
projected onto the sky plane under the inclination $i$, we see an ellipse with the axis ratio of
$b/a=\cos i$. If the disk is not infinitely thin and can be imagined as an oblate ellisoid with the
vertical-to-radial axis ratio of $q$, then $\cos ^2 i =\frac{(b/a)^2 - q^2}{1-q^2}$. The latter equation
allows us to calculate the relative thickness $q$, if we have the possibility to determine independently 
the inclination $i$ and the isophote axis ratio $b/a$. The latter parameter is provided by the isophote
analysis. The former parameter can be obtained directly from the 2D surface photometry if we deal with
exponential-profile disks. The exponential scalelength for a given galactic disk can be used as a standard rule, 
and its visible variations with the azimuth, from $h$ along the major axis to $h \cos i$ along the minor axis,
follow a pure cosine law and provide an independent estimate of the inclination $i$. We divide the whole
galaxy image into 18 sectors, with 20-degree opening angle each, and calculate surface-brightness profiles within
all of them. These surface-brightness profiles are fitted by exponential laws, and the on-plane azimuthal distribution
of the 18 (projected) scalelengths obtained in such a way is approximated by an ellipse. Just this ellipse has an axis ratio
equal to $\cos i$ so giving us a possibility to determine the disk inclination.  

\section{The results}

\begin{figure*}[p]
\centering
\begin{tabular}{c c c}
 \includegraphics[width=5cm]{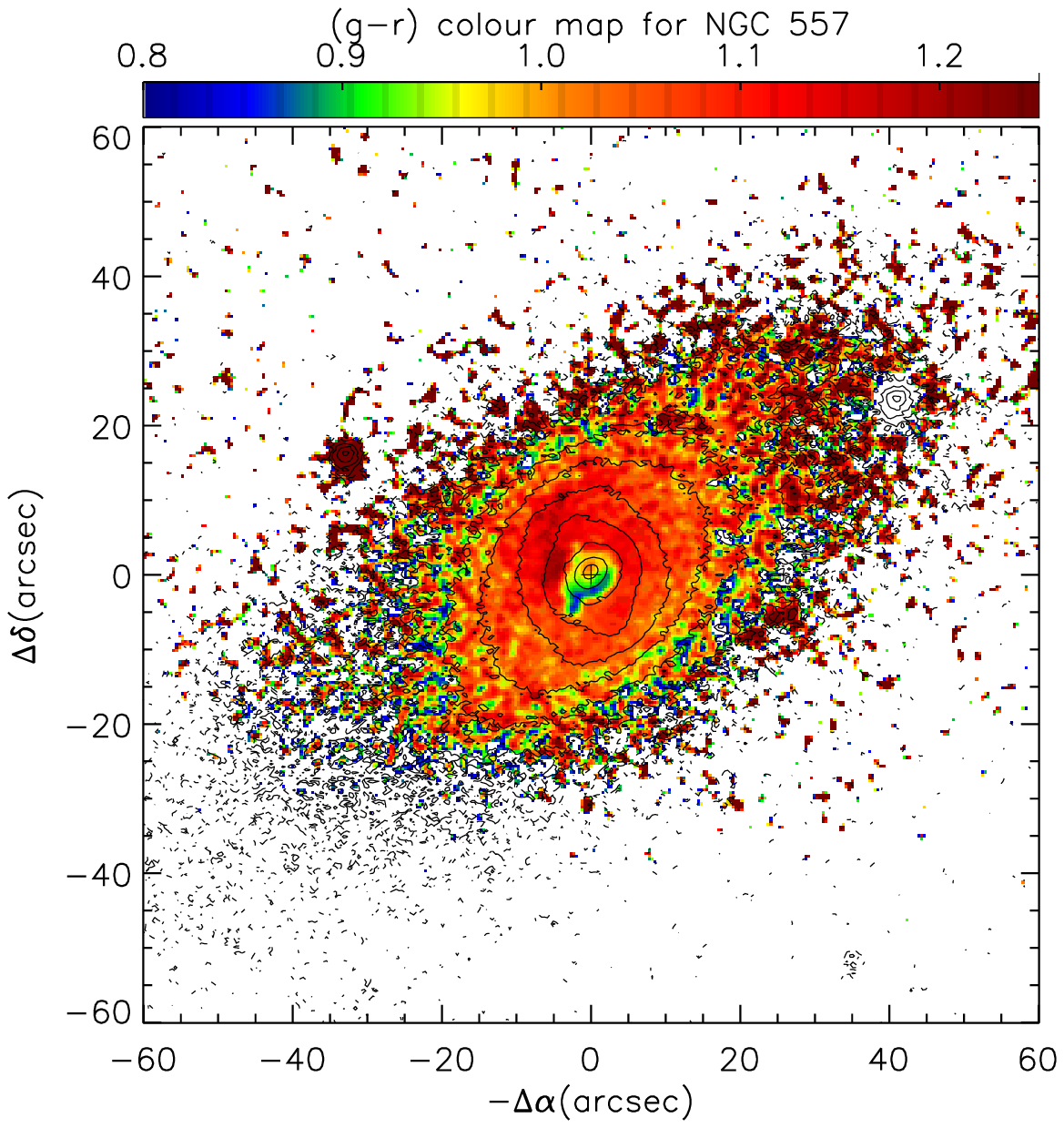} &
 \includegraphics[width=5cm]{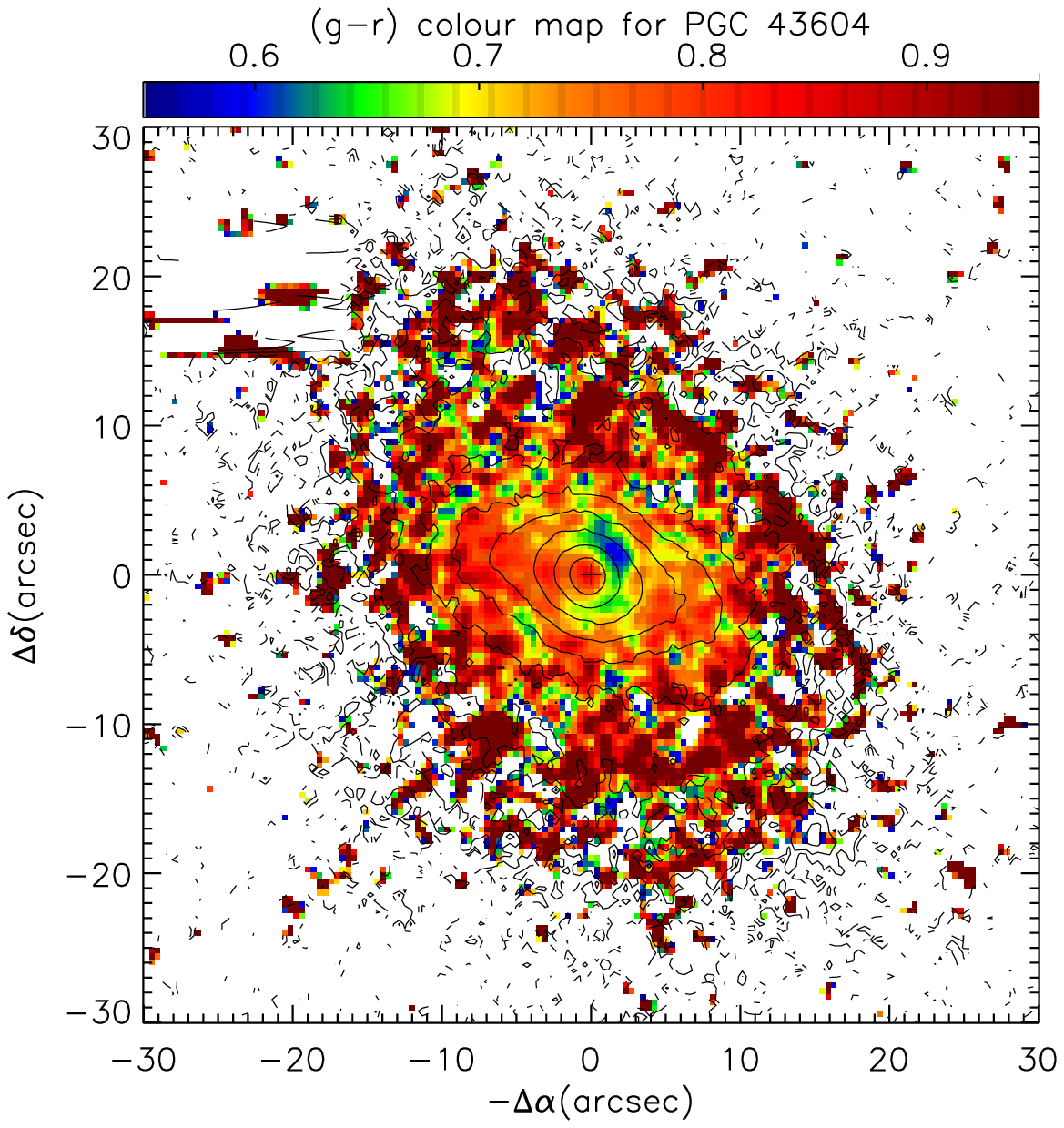} &
 \includegraphics[width=5cm]{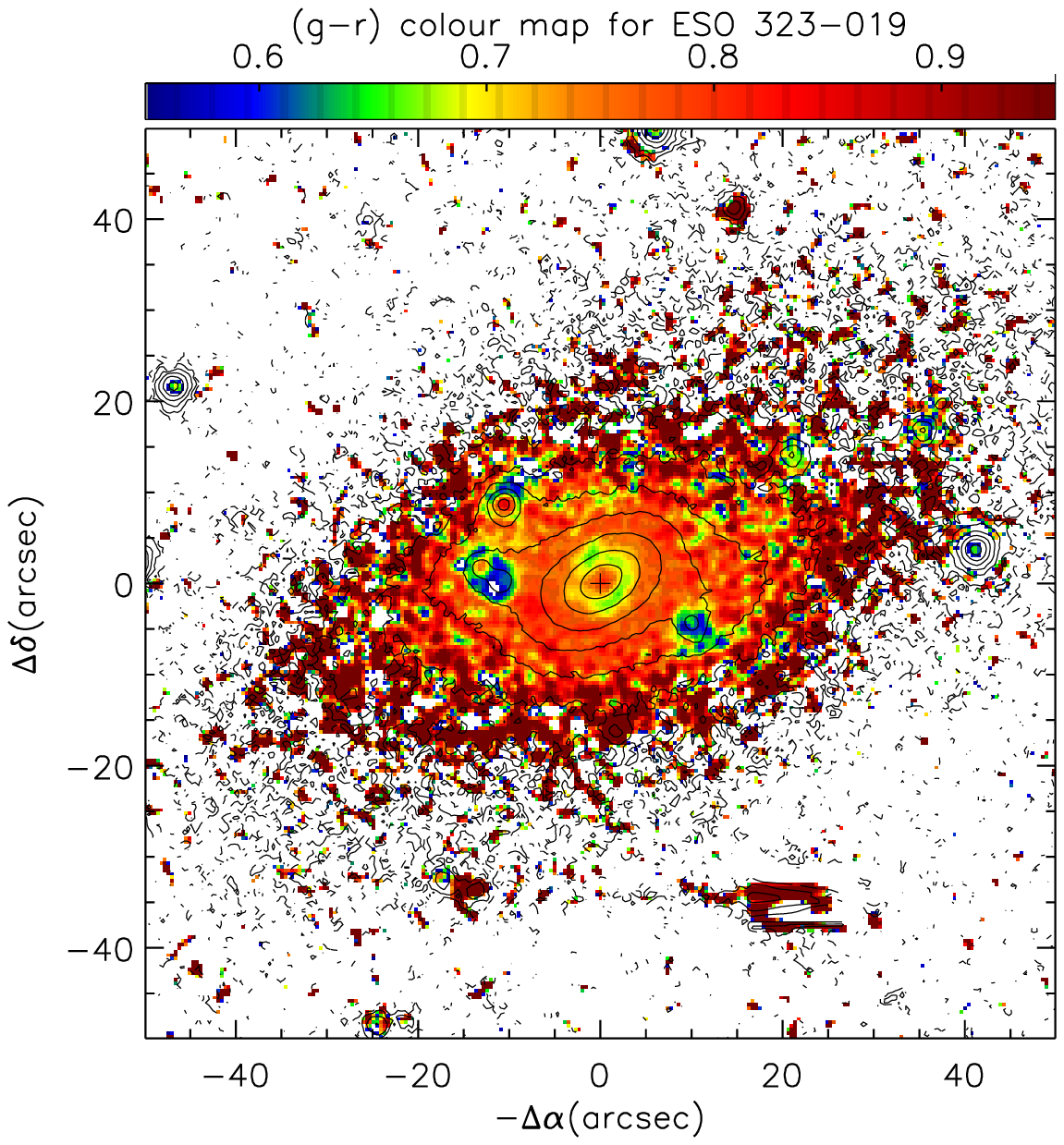} \\
\includegraphics[width=5cm]{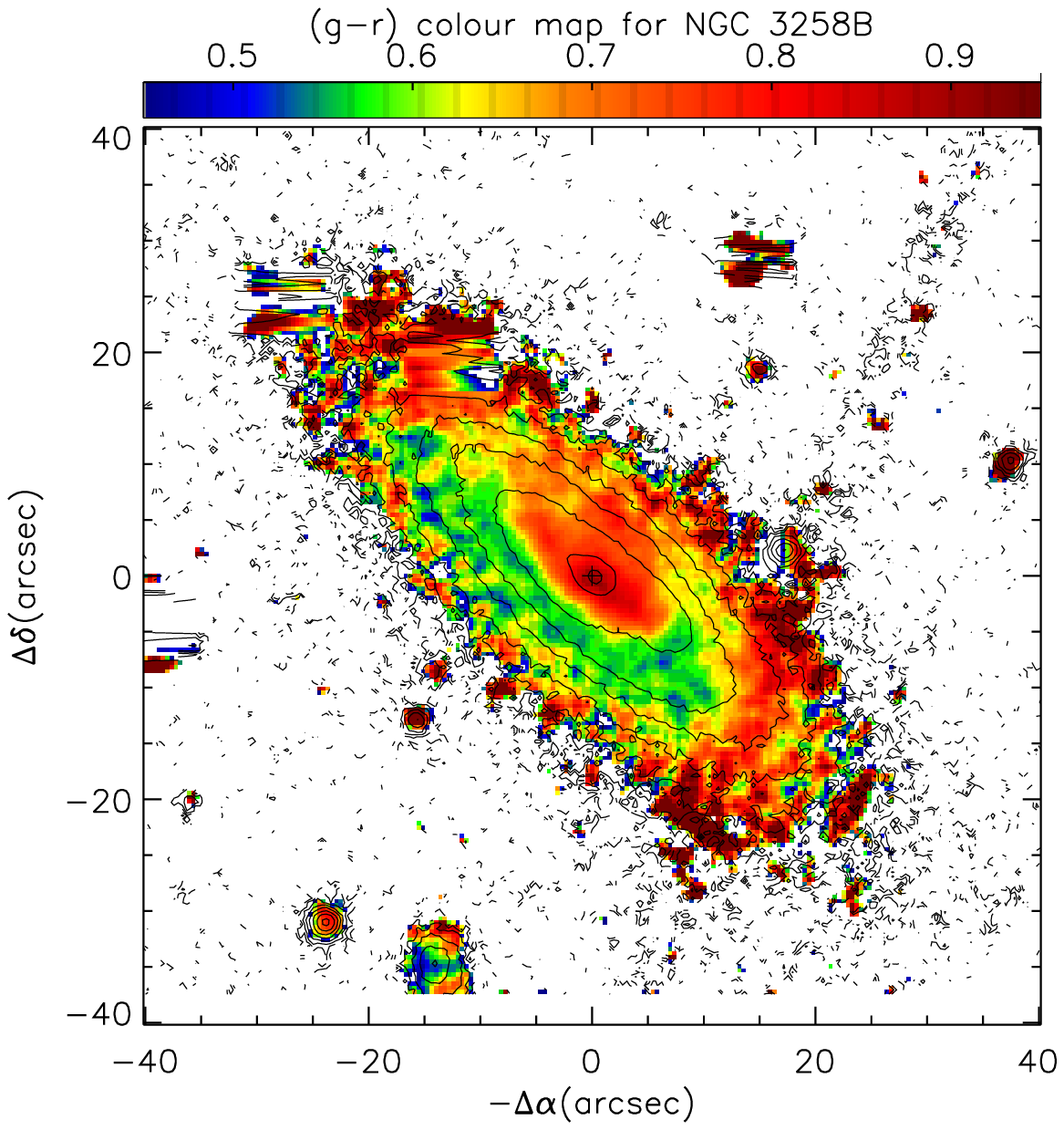} &
 \includegraphics[width=5cm]{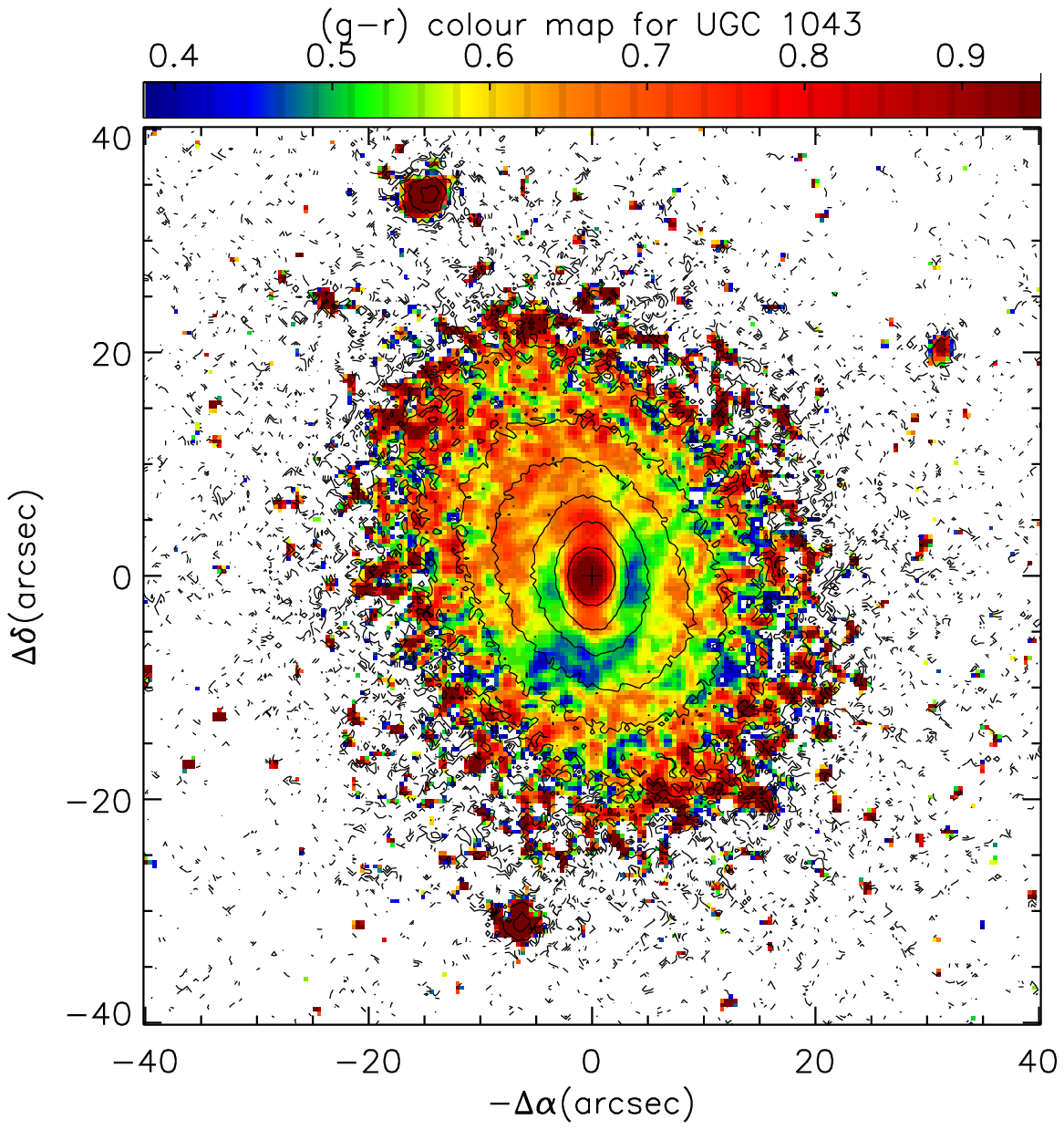} &
 \includegraphics[width=5cm]{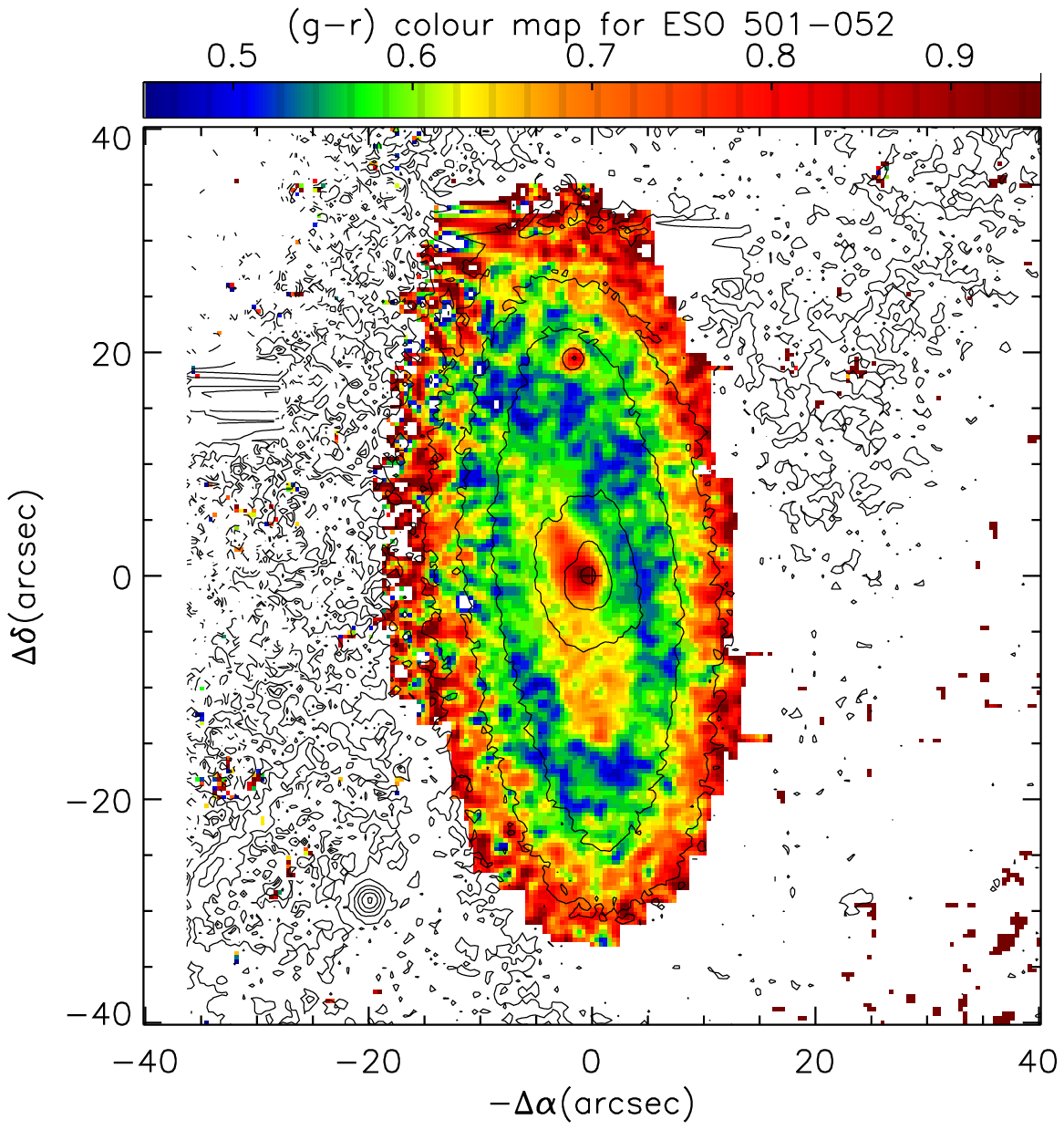} \\
\includegraphics[width=5cm]{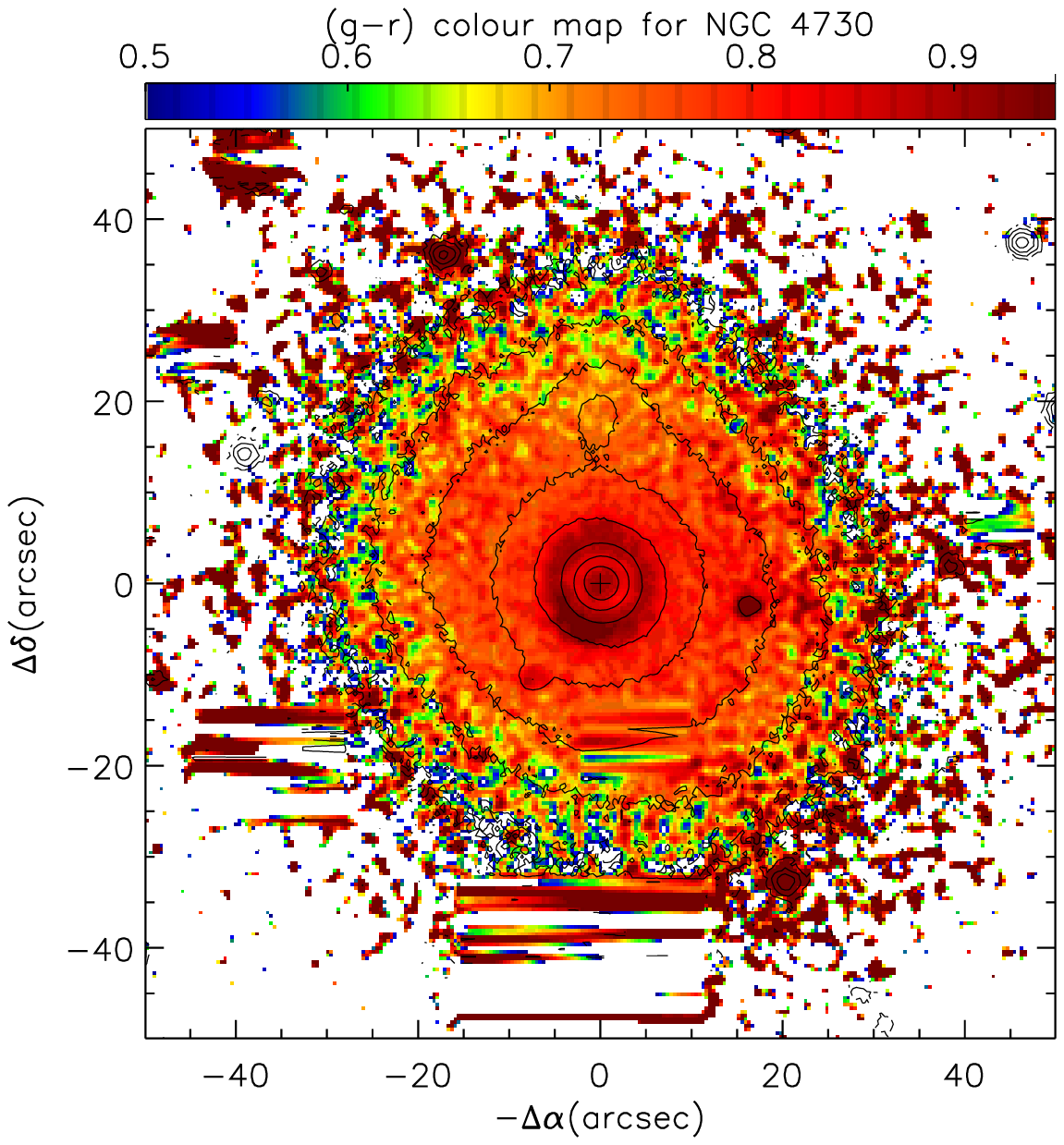} &
 \includegraphics[width=5cm]{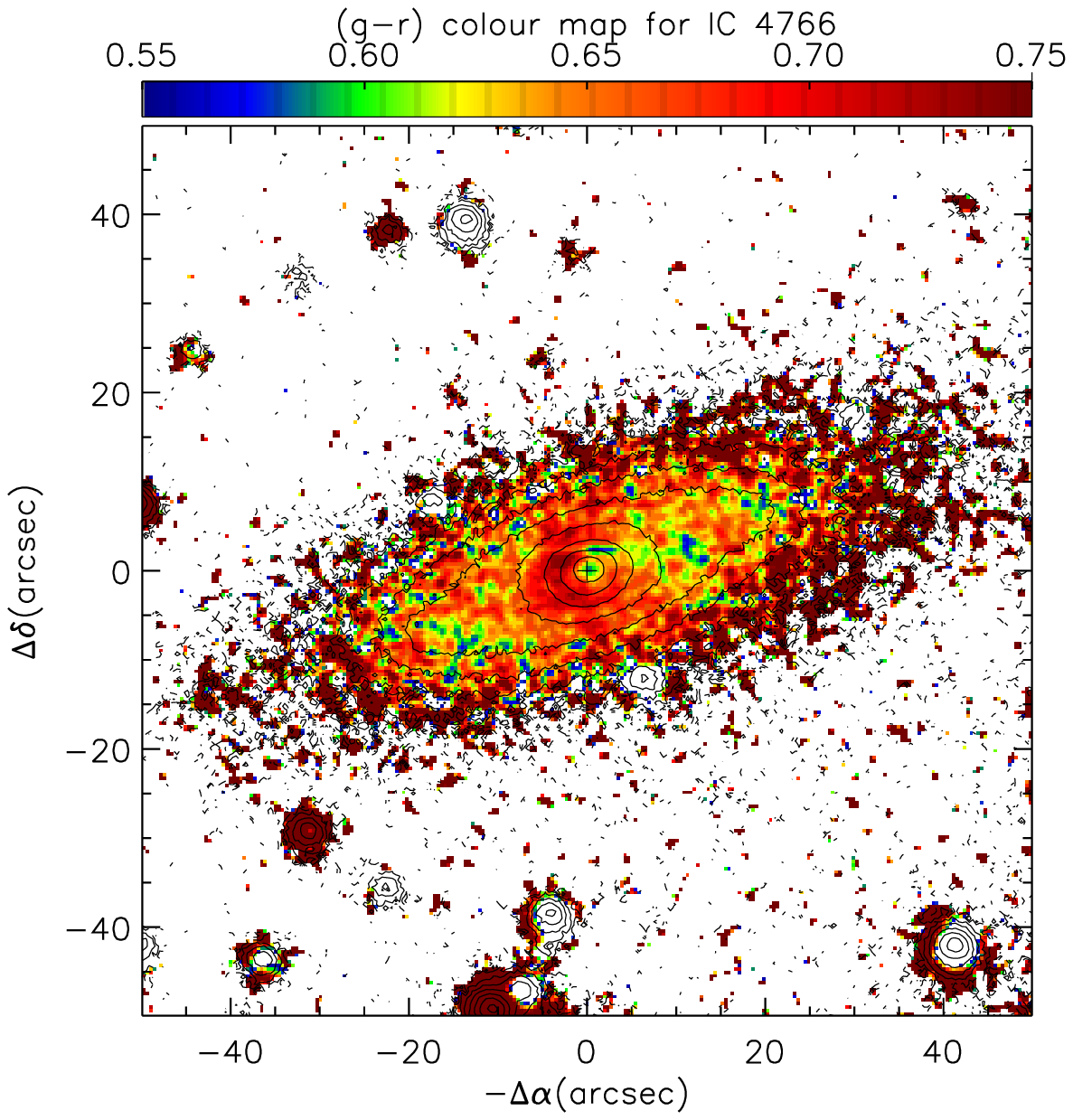} &
 \includegraphics[width=5cm]{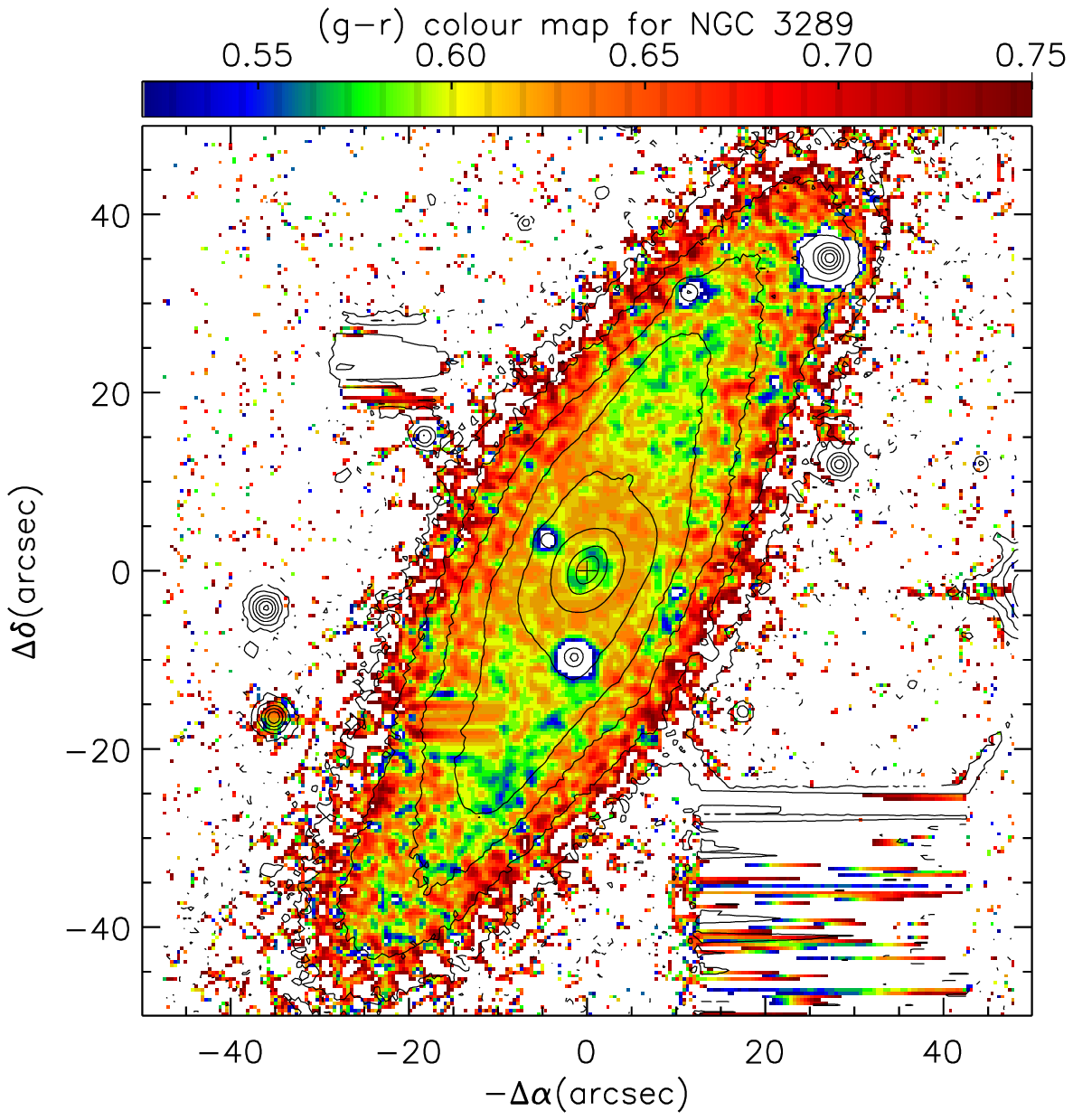} \\
\end{tabular}
\caption{Some examples of the color structures: {\it upper row} -- circumnuclear blue rings, {\it middle
row} -- large-scale blue rings, {\it bottom row, two left plots} -- circumnuclear red rings. NGC~3289
({\it the right plot in the bottom row}) demonstrates a rather complex system of blue and reddish rings.} 
\label{ringcol}
\end{figure*}

Table~\ref{listgal} lists all 60 galaxies which have been analyzed, with mentioning the type of the disk
surface-brightness profile derived and also containing notes about some structure details and color features 
which have been seen in the color maps. The presence of a bar was recognized if the isophote ellipticity profile
had a distinct local maximum exceeding the outer ellipticity level prescribed by the disk orientation. The 
existence of a ring is derived from the visual inspection of the surface-brightness profiles; the rings
can be or not be distinguished by the color. The color distributions demonstrate mostly red centers and smoothly 
reddish stellar bodies, but the low-luminosity S0s can instead possess blue nuclei. 
Sometimes distinct color features such as blue rings of various sizes or circumnuclear red (dust)
rings can also be noted (Fig.~\ref{ringcol}).

\startlongtable
\begin{deluxetable}{lclrcrr}
\tablecolumns{7} 
\tablewidth{0pc} 
\tablecaption{The galaxies studied photometrically with the LCO network.\label{listgal}} 
\tablehead{ 
\colhead{Galaxy} & \colhead{Type\tablenotemark{a}}  & \colhead{$M_H$\tablenotemark{a}} & 
\colhead{$R_{25}$,$^{\prime \prime}$\tablenotemark{b}} 
& \colhead{Type of disk profile} & \colhead{Bar/ring?}   & \colhead{Color features}}
\startdata 
\sidehead{Abell 194}
NGC 541 & S0-? & --24.3 & 66  &  I  & minibar  & \nodata   \\
NGC 557 & SB0$^+$(rs)pec & --23.9 & 43  & III & ring   & a blue filament   \\
PGC 5216 & S0: & --22.35 & 25  &  III & \nodata  &  \nodata   \\
PGC 5313 & E3: & --22.3 & 16.5  & III &  ring  & \nodata   \\
PGC 5314 & S0/a & --22.4 & 18  &  III  & bar   & \nodata  \\
UGC 1003 & S0 & --22.9 &  24   & III  & \nodata  & \nodata  \\
UGC 1030 & E & --23.2 &  31  &  I  & \nodata    & blue disk   \\
UGC 1043 & E & --22.8 & 26  &   III  & ring & blue ring  \\
\sidehead{NGC 1550 group}
IC 366  &  E? (LEDA)  & --22.15 & 10 & III & \nodata   & \nodata   \\
UGC 3006 & SAO$^0$: & --23.7 & 52  &  I  & \nodata   & \nodata   \\
UGC 3008 & (R)SB(rs))$^+$  & --23.3 & 49 & III & Bar, ring & \nodata   \\
UGC 3011 & SA(rs)0$^0$: & --22.64 & 32  &  I   &  ring   & \nodata   \\
\sidehead{Fornax}
NGC 1351 & SA0$^-$pec: & --22.5 & 102 & III & \nodata    & \nodata     \\
NGC 1380B & SAB(s)0$^-$: & --21.1 & 60  & III & bar    &  \nodata   \\
NGC 1387 & SAB(s)0$^-$ & --23.44 & 91 & II &  bar   & \nodata     \\
\sidehead{Centaurus}
ESO 323-012 & SB0 & --22.86 &  27  &   I  & bar, ring & \nodata   \\
ESO 323-019 & E+ & --23.2 &  45  & III & bar, ring & blue nuclear ring \\ 
NGC 4677 & SB(s)0$^+$ & --23.5 &  44  &   III  & ring & \nodata  \\
NGC 4683 & SB(s)0$^-$ &  --23.4 & 43 & I  & bar &  \nodata    \\
NGC 4696B  & SA0$^-$ & --24.1 & 39.5  &  I  & \nodata  &  \nodata     \\
NGC 4730 & SA(r)0$^-$ & --23.8 & 38  & III & \nodata   &  red nuclear ring \\
NGC 4743 & SA0$^+$ & --23.0 &  52  &   I   & ring & \nodata \\
NGC 4744 & SB(s)0/a & --24.1 & 74 &  I  & bar, ring & dust lanes \\
PGC 43572 & S0 & --22.06 & 25.5  & I & \nodata  & blue nuclear polar ring \\
PGC 43604 & SB0 & --22.8 & 27 &  I  & Boxy bar & blue nuclear ring \\
PGC 43652 & S0  & --21.4 & 19  &  III & \nodata    & \nodata    \\ 
\sidehead{Hydra}
ESO 501-035 & SB(r)0$^0$ & --23.36 & 42  &   I   & ring  & \nodata   \\    
ESO 501-047 & SB0$^0$ &  --22.7 &  30 &  I  &  \nodata    & \nodata    \\
ESO 501-049 & SB(s)0  & --22.6 & 26 &   I   & ring    & \nodata \\
ESO 501-052 & (R)S0$^+$ & --22.2 & 37 & II & ring & blue ring   \\
LEDA 87329 & S0 & --21.04 & $<14$ &  III & \nodata & red nucleus \\
LEDA 141477 & E/S0 & --21.5 & 16.5 &   III  & \nodata    & \nodata     \\
NGC 3307 & SB(r)0/a pec & --22.6 & 29 & I & \nodata   & \nodata    \\
NGC 3308 & SAB(s)0$^-$ & --24.2 & 36 & III & \nodata    & \nodata   \\
NGC 3316 & SB(rs)0$^0$ & --23.96 & 29 &  III & \nodata    & \nodata  \\
PGC 31418 & S0 & --22.16 & 26(K) &  I  & \nodata     & \nodata     \\
PGC 31447 & S0  & --22.85 & 14  &  I  & \nodata   & \nodata   \\
PGC 31450 & SB(rs)0$^0$ & --22.3 & 20  &   III & \nodata   & \nodata  \\
PGC 31464 & S(rs)0 & --22.04 &  17  &   I  & ring & \nodata  \\
\sidehead{Antlia}
FS 72\tablenotemark{c} & S0 & --21.2 & 27 & III & bar, ring & red bar, blue disk \\
FS 80\tablenotemark{c} & dS0 & --21.6 & 24  & I & \nodata & \nodata \\
LEDA 83014 & S0 & --22.34 & 27  & III & bar & red bar, blue disk \\ 
NGC 3257 & SAB(s)0$^-$: & --22.5 & 32   &  III & bar  & a blue filament \\
NGC 3258A & SAB0$^+$: & --22.5 & 37  &   I   & ring & \nodata \\
NGC 3258B & SAB(r)0: & --20.85 & 32 &  III & bar & blue semi-ring \\
NGC 3273 & SA(r)0$^0$ & --23.6 & 57  &  II & \nodata    & \nodata    \\
NGC 3289 & SB(rs)0 & --23.35 & 60  & I & ring & reddish nuclear ring \\
\sidehead{Abell 3565}
ESO 383-030 & SAB(rs)0$^+$ & --23.0 & 42 &  III & rings & blue ring, red lanes \\
ESO 383-045 &  S0? & --23.9 & 39 &   III  & \nodata    & \nodata    \\
ESO 383-049 & E$+$? & --22.94 & 34.5  &   I & \nodata   &  \nodata \\
LEDA 183938 & SB(rl)0$^+$ & --22.4 & 31  &   I  & bar, ring &  \nodata  \\
\sidehead{Abell S0805}
ESO 104-002 & S0 & --22.5 & 32  &  III & \nodata   & blue disk   \\
IC 4749 &  S0? & --23.94 &  36  & I  & \nodata   & \nodata   \\
IC 4750 &  SAB(r)0$^+$ & --23.3 & 34.5  &  III & ring & \nodata   \\
IC 4766 & SA(r)0$^+$ & --23.54 & 37 & III & \nodata & red semi-ring \\
IC 4784 & S0? &  --24.7 & 45  & III & \nodata  & \nodata   \\
LEDA 93525 & E?(LEDA)  &  --23.56 & 35.5(K) &   I  & \nodata   &  \nodata   \\
PGC 62384 & SB0 & --22.45 &  20  &  I  & bars & blue nucleus  \\
PGC 62436 &  S0 & --21.65 & 21 &  III & \nodata   & \nodata  \\
PGC 62437 & S0 & --21.36 & 16 & I & \nodata & blue nucleus \\
\enddata 
\tablenotetext{a}{Mostly from NED; but some data taken from HyperLEDA \citep{hyperleda} are marked by `(LEDA)'.}
\tablenotetext{b}{Mostly the optical radii are taken from HyperLEDA \citep{hyperleda}; but some NIR radii taken from NED
are marked by `(K)'.}
\tablenotetext{c}{The designation of the galaxy is from the study of the Antlia cluster by \citet{antlia_cat}.}
\end{deluxetable}

The results of our analysis of the radial and vertical structures of the $r$-band images of the galactic disks 
are presented in the Table~\ref{s0type1} for the radial profiles of Type I, in the Table~\ref{s0type2} for the 
radial profiles of Type II, and in the Table~\ref{s0type3} for the radial profiles of Type III, correspondingly.
The radial surface-brightness profile parameters, namely, the central surface brightnesses $\mu _0$ and the exponential
scalelengths $h$, are obtained by fitting the profiles by exponential laws in the radius ranges noted in 
the Tables -- following \citet{freeman}, we have tried to explore the radius ranges which are at least twice larger
than the exponential scalelengths; in these Tables the $\mu _0$'s are not corrected for the extinction.
The relative thicknesses of some disks, $q$, are also presented in the Tables; they have been obtained only
for the high signal-to-noise images of the disks which are far from edge-on or face-on orientation -- so minus
20 galaxies of our sample. The relative thicknesses $q$ of the S0 galactic disks, characterizing the ratio
of the vertical and radial scalelengths, are presented in the Table~\ref{s0type1} for the single-scaled
exponential disks, in the Table~\ref{s0type2} for the inner part of the truncated disk in NGC~3273, and 
in the Table~\ref{s0type3} for the inner segments of the two-tiered antitruncated disks. Only one galaxy 
with a Type-II disk is suitable for disk thickness determination -- it is NGC 3273. As for other two, 
NGC 1387 is too round (face-on), and ESO 501-052 is too inclined (almost edge-on).

\begin{table*}
\caption{Parameters of the Type-I disks}
\label{s0type1}
\begin{flushleft}
\begin{tabular}{lccccc}
\hline\noalign{\smallskip}
Galaxy & Radius range,  & $\mu _{0,r}$,  & $h_r$, & $h_r$, & $q$ \\
   & $^{\prime \prime}$ & mag$/ \Box^{\prime \prime}$ & $^{\prime \prime}$ & kpc &  \\
\hline\noalign{\smallskip}
NGC 541 & 20--80 &  20.2 &  $21.21\pm 0.01$ & $7.020\pm 0.003$ & \\
UGC 1030  & 15--35 & 20.4 & $8.78\pm 0.13$ & $2.91\pm 0.04$ & 0.375 \\
UGC 3006 & 25--58 & 19.5 & $14.01\pm 0.01$ & $3.334\pm 0.002$ &  0.22 \\
UGC 3011 &  17--42 & 19.1 & $9.49\pm 0.09$ & $2.26\pm 0.02$ &  0.42 \\
ESO 323-012 & 13--30 & 19.0 & $9.29\pm 0.14$ & $1.97\pm 0.03$ &   \\
NGC 4683 & 23--44 & 20.9 & $14.41\pm 0.19$ & $3.05\pm 0.04$ & 0.35 \\
NGC 4696B & 18--50 & 19.7 & $11.84\pm 0.01$ & $2.510\pm 0.002$ & 0.46 \\
NGC 4743 & 16--35 & 19.1 & $9.94\pm 0.11$ & $2.11\pm 0.02$ & 0.34 \\
NGC 4744 & 45--85 & 20.9 &  $24.90\pm 0.03$ & $5.279\pm 0.006$ & 0.28 \\
PGC 43572 & 12--24 & 20.0 & $6.42\pm 0.32$ & $1.36\pm 0.07$ &  \\
PGC 43604 & 11--24 & 20.7 & $8.99\pm 0.35$ & $1.91\pm 0.07$ &  \\
ESO 501-035 & 25--50 & 20.8 & $15.27\pm 0.32$ & $4.15\pm 0.09$ &  \\
ESO 501-047 & 8--25 & 19.3 & $6.64\pm 0.04$ & $1.806\pm 0.011$ & 0.36 \\
ESO 501-049 & 11--31 & 20.1 & $8.38\pm 0.08$ & $2.28\pm 0.02$ & 0.31 \\
NGC 3307 & 8--24 & 19.1 & $5.93\pm 0.06$ & $1.61\pm 0.02$  & 0.31 \\
PGC 31418 & 12--24 & 19.6 & $5.25\pm 0.16$ & $1.43\pm 0.04$ & 0.31  \\
PGC 31447 & 10--25 & 19.1 & $5.84\pm 0.11$ & $1.59\pm 0.03$ & 0.31 \\
PGC 31464 & 5--16 &  19.7 & $4.94\pm 0.05$ & $1.34\pm 0.01$  & 0.45  \\
NGC 3258A & 9--32 &  18.9 & $7.19\pm 0.01$ & $1.474\pm 0.002$ & 0.27 \\
NGC 3289 & 30--55 &  18.8 & $13.40\pm 0.01$ & $2.747\pm 0.002$  & 0.16 \\
Antlia: FS80 & 10--28 & 18.8 & $6.22\pm 0.04$ & $1.275\pm 0.008$ & 0.25 \\
ESO 383-049 &  6--37 &  18.7 & $6.910\pm 0.003$ & $1.810\pm 0.001$ & 0.13 \\
LEDA 183938 & 22--32 &  20.9 & $8.15\pm 0.42$ & $2.14\pm 0.11$ & 0.27 \\
LEDA 93525 & 12--24 & 19.9 & $7.63\pm 0.11$ & $2.23\pm 0.03$ &   \\
PGC 62384 & 9--22 & 20.2 & $5.83\pm 0.08$ & $1.70\pm 0.02$ &   \\
PGC 62437 & 4--12 & 20.5 & $4.64\pm 0.04$ & $1.355\pm 0.012$  &  \\
IC 4749 & 9--37 & 21.5$^a$ & $11.77\pm 0.37^a$ & $3.44\pm 0.11^a$ &  \\
\hline
\multicolumn{6}{l}{$^a$\rule{0pt}{11pt}\footnotesize
The $r$-image was badly guided so here we use the $g$-band image.}\\
\end{tabular}
\end{flushleft}
\end{table*}

\begin{table*}
\scriptsize
\caption{Parameters of the Type-II disks}
\label{s0type2}
\begin{flushleft}
\begin{tabular}{lccccccccccc}
\tablewidth{0pt} 
\hline\noalign{\smallskip}
Galaxy & \multicolumn{5}{c}{Inner disk} &  &  & \multicolumn{4}{c}{Outer disk} \\
 & Range,  & $\mu _{0,r}$, & $h_r$,  
& $h_r$, & $q$ & $\mu _{brk}$, & $R_{brk}$, & Range, &  $\mu _{0,r}$, & $h_r$, 
& $h_r$,  \\
 & $^{\prime \prime}$ & mag$/ \Box ^{\prime \prime}$ & $^{\prime \prime}$ & kpc &  & mag$/ \Box ^{\prime \prime}$ &
$^{\prime \prime}$ & $^{\prime \prime}$ & mag$/ \Box ^{\prime \prime}$ & $^{\prime \prime}$ & kpc \\
\hline\noalign{\smallskip}
NGC 3273 & 16--30 & 19.7 & $16.33\pm 0.01$ & $3.35\pm 0.00$ & 0.13 & 21.7 & $29.1\pm 0.1$ & 35--70 & 18.8 & $11.15\pm 0.01$ & $2.29\pm 0.00$ \\
ESO 501-052 & 10--18 & 21.1 & $30.38\pm 1.19$ & $8.26\pm 0.32$ & & 21.8 & $17.9\pm 0.5$ & 20--33 & 20.1 & $11.84\pm 0.01$ & $3.22\pm 0.00$ \\
NGC 1387 & 30--85 & 19.9 & $24.74\pm 0.00$ & $2.18\pm 0.00$ & & 24.2 & $99.8\pm 1.7$ & 102--114 & 13.5 & $10.05\pm 0.10$ & $0.88\pm 0.01$ \\ 
\hline
\end{tabular}
\end{flushleft}
\end{table*}

\begin{table*}
\scriptsize
\caption{Parameters of the Type-III disks}
\label{s0type3}
\begin{flushleft}
\begin{tabular}{lccccccccccc}
\hline\noalign{\smallskip}
Galaxy & \multicolumn{5}{c}{Inner disk} &  &  \multicolumn{4}{c}{Outer disk} \\
 & Range, & $\mu _{0,r}$, & $h_r$, & $h_r$, & $q$ & $\mu _{brk}$ & $R_{brk}$ & Range, & $\mu _{0,r}$, & $h_r$,
& $h_r$, \\
& $^{\prime \prime}$ & mag$/ \Box ^{\prime \prime}$ & $^{\prime \prime}$ & kpc  & & mag$/ \Box ^{\prime \prime}$ &
$^{\prime \prime}$ & $^{\prime \prime}$ & mag$/ \Box ^{\prime \prime}$ & $^{\prime \prime}$ & kpc \\
\hline\noalign{\smallskip}
NGC 557 &  12--24 &  19.1  & $7.55\pm 0.08$  &  $2.50\pm 0.03$ & 0.54 & 22.8 & $25.9\pm 1.4$ & 30--47 & 21.3 & $18.42\pm 0.69$ & $6.1\pm 0.2$ \\
PGC 5216 &  7--14 &  19.3  & $5.42\pm 0.00$  &  $1.79\pm 0.00$ & 0.155 & 21.8 & $14.1\pm 1.8$ & 14--22 & 20.3 &  $8.38\pm 0.12$ & $2.77\pm 0.04$ \\
PGC 5313 & 8--15 & 18.1 & $3.56 \pm 0.03$ & $1.18\pm 0.01$ &  & 22.8 & $15.4\pm 0.7$ & 18--35 & 21.2 & $10.5 \pm 0.5$ & $3.48\pm 0.16$ \\
PGC 5314 & 9--15 & 19.9 & $6.75\pm 0.12$ & $2.23\pm 0.04$ & & 24.1 & $25.9\pm 1.7$ & 21--44 & 21.6 & $11.4 \pm 0.4$ & $3.77\pm 0.13$ \\
UGC 1003 & 10--20 &  18.7 &  $4.89\pm 0.04$ & $1.62\pm 0.01$ & 0.30 & 23.4 & $21.2\pm 2.2$ & 21--35 & 20.4 & $7.66\pm 0.30$ & $2.54\pm 0.10$ \\
UGC 1043 &  7--18 &  19.5 & $5.46\pm 0.04$  & $1.81\pm 0.01$ & 0.54 & 23.5 & $17.4\pm 2.1$ & 20--32 & 20.7 & $8.37\pm 0.25$ & $2.77\pm 0.08$ \\
IC 366 &  10--21 &   19.7 & $5.64\pm 0.08$  & $1.34\pm 0.02$ & 0.31  & 24.0 & $22.3\pm 2.4$ & 23--37 & 21.8 & $11.02\pm 0.91$ & $2.62\pm 0.22$ \\
UGC 3008 &  5--12 &  18.3 & $5.80\pm 0.11$ & $1.38\pm 0.03$ & 0.27 &  21.9 & $19.1\pm 0.4$ & 40--60 & 20.9 & $21.25\pm 0.58$ & $5.06\pm 0.14$ \\
NGC 1351 & 15--30 &  18.7 & $12.24\pm 0.01$ & $1.08\pm 0.00$ & 0.42 & 21.4 & $30.4\pm 2.7$ & 40--78 &  19.8 & $20.68\pm 0.02$ & $1.82\pm 0.00$ \\
NGC 1380B & 15--45 &  20.5 & $10.90\pm 0.00$ & $0.96\pm 0.00$ & 0.69 & 25.0 & $43.2\pm 7.6$ &  45--65 & 21.7 & $15.12\pm 0.55$ & $1.33\pm 0.05$ \\
ESO 323-019 &  13--28 & 19.2 & $8.62\pm 0.20$ & $1.83\pm 0.04$ &  & 23.0 & $30.0\pm 2.8$ & 35--55 & 21.2 & $18.3\pm 1.8$ & $3.88\pm 0.38$ \\
NGC 4677 & 30--45 & 19.3 &  $11.46\pm 0.01$ & $2.43\pm 0.00$ & & 23.5 & $44.2\pm 4.4$ & 50--80 &  21.4 & $22.96\pm 1.21$ & $4.87\pm 0.26$ \\
NGC 4730 & 12--36 & 19.7 & $9.41\pm 0.00$ & $2.00\pm 0.00$ & & 23.6 & $34.6\pm 9.8$ & 40--60 &  20.5 & $11.77\pm 0.47$ & $2.50\pm 0.10$  \\
PGC 43652 & 7--18 & 20.4 & $5.97\pm 0.07$ & $1.27\pm 0.02$ & 0.71 & 23.6 & $19.0\pm 10.4$ & 21--40 &  20.8 & $6.75\pm 0.35$ & $1.43\pm 0.07$ \\
LEDA 87329 & 6--16 & 19.2 & $3.94\pm 0.03$ & $1.07\pm 0.01$ &  & 23.4 & $15.2\pm 1.3$ & 19--25 & 20.9 & $6.64\pm 0.22$ & $1.81\pm 0.06$ \\
LEDA 141477 & 8--16 & 19.8 & $5.15\pm 0.10$ & $1.40\pm 0.03$ & 0.20 & 22.7 & $13.8\pm 1.1$ & 16--26 &  20.5 & $6.78\pm 0.04$ & $1.84\pm 0.01$ \\
NGC 3308 &  8--28 &  19.0 & $10.45\pm 0.01$ & $2.84\pm 0.00$ & 0.61 & 22.0 & $27.0\pm 3.0$ & 30--62 & 19.9 & $15.40\pm 0.01$ & $4.19\pm 0.00$ \\
NGC 3316 & 10--22 &  19.1 &  $7.43\pm 0.01$ & $2.02\pm 0.00$ & 0.56 & 23.4 & $29.2\pm 1.7$ & 40--60 &  22.0 & $23.13\pm 1.28$ & $6.29\pm 0.35$ \\
PGC 31450 & 9--18 & 19.6 &  $6.2\pm 0.1$ & $1.69\pm 0.03$ & 0.20 & 22.9 & $18.9\pm 1.8$ & 20--35 &  20.9 & $10.2\pm 0.4$ & $2.77\pm 0.11$ \\
Antlia: FS 72 & 2--13 & 18.5 & $5.04\pm 0.04$ & $1.03\pm 0.01$ & 0.13 & 21.5 & $13.9\pm 1.8$ & 15--30 & 19.4 & $7.20\pm 0.14$ & $1.48\pm 0.03$ \\
LEDA 83014 & 16--28 & 19.6 & $6.14\pm 0.14$ & $1.26\pm 0.03$ & 0.28 & 24.7 & $29.1\pm 0.4$ & 30--45 & 22.2 & $12.41\pm 0.24$ & $2.54\pm 0.05$ \\
NGC 3257 &  7--18 & 19.2 & $6.28\pm 0.04$ & $1.29\pm 0.01$ & 0.56 & 23.5 & $23.9\pm 1.5$ & 29--54 & 21.3 & $12.75\pm 0.35$ & $2.61\pm 0.07$ \\
NGC 3258B & 12--28 &  18.7 & $6.79\pm 0.06$ & $1.39\pm 0.01$ &  & 23.3 & $26.8\pm 4.9$ & 30--50 & 20.2 &  $10.45\pm 0.80$ & $2.14\pm 0.16$ \\
ESO 383-030 &  10--35 &  20.1 & $11.42\pm 0.01$ & $2.99\pm 0.00$ & 0.13 & 23.4 & $35.2\pm 4.7$ & 35--50 & 21.3 & $17.80\pm 0.52$ & $4.66\pm 0.14$ \\
ESO 383-045 & 8--28 & 18.7 & $7.93\pm 0.15$ & $2.08\pm 0.04$ & 0.27 & 23.9 & $37.7\pm 0.6$ & 45--80 & 22.5 & $30.1\pm 1.3$ & $7.89\pm 0.34$ \\
ESO 104-002 & 6--17 &  19.2 & $5.33\pm 0.04$ & $1.56\pm 0.01$ & 0.26 & 22.8 & $18.2\pm 0.9$ & 18--50 & 21.5 & $14.0\pm 0.4$ & $4.09\pm 0.12$ \\
IC 4750 & 16--30 & 19.3 & $7.63\pm 0.08$ &  $2.23\pm 0.02$ & 0.30 & 23.6 & $28.1\pm 5.6$ & 32--48 & 20.5 & $10.9\pm 0.7$ & $3.18\pm 0.20$ \\
IC 4766 & 18--32 & 19.0 & $8.03\pm 0.08$ & $2.34\pm 0.02$ &  & 23.3 & $32.5\pm 1.8$ & 32--54 & 21.1 & $15.4\pm 0.4$ & $4.50\pm 0.12$ \\
PGC 62436 & 6--16 & 19.5 & $4.36\pm 0.04$ & $1.27\pm 0.01$ & 0.28 & 23.7 & $16.0\pm 2.7$ & 18--26 & 20.7 & $6.24\pm 0.32$ & $1.82\pm 0.09$ \\
IC 4784 & 12--26 &  20.3$^a$ & $9.67\pm 0.01$ &  $2.82\pm 0.00$ &  & 23.3$^a$ & $24.1\pm 2.2$ &  28--48 & 21.4$^a$ 
& $16.28\pm 0.02$ & $4.75\pm 0.01$ \\
\hline
\multicolumn{11}{l}{$^a$\rule{0pt}{11pt}\footnotesize
The $r$-image was badly guided so here we use the $g$-band image.}\\
\end{tabular}
\end{flushleft}
\end{table*}

\section{Discussion}

\subsection{Radial structure of the S0 disks in the clusters and beyond}

\citet{erwin12}, by analyzing radial structure of 24 S0 galaxies -- members of the
Virgo cluster, have concluded that the statistics of the surface-brightness profile types
in the cluster differs significantly from that in the field: they have not found Type-II
profiles in the Virgo S0s at all while in the field a quarter of all S0s demonstrate
truncated stellar disks \citep{erwin08,guti_erwin}. They reported the following fractions of the profile types
in the Virgo: 46\%$\pm 10$\%\ of the Type I, 0\%$\pm 4$\%\ of the Type II, and the remaining
54\%\ of the Type III. Our results on 60 S0 galaxies in 8 southern clusters are: 27 S0s have the profiles
of Type I -- 45\%$\pm 6$\%, 3 S0s have the profiles of Type II -- 5\%$\pm 3$\%\ (the errors indicated
correspond to the root square of the binomial distribution variance). We 
conclude that our results are completely consistent with the statistics of the Virgo S0s
reported by \citet{erwin12} and also differ from the field statistics where 26\%$\pm 6$\%\ of
Type I and 28\%$\pm 6$\%\ of Type II are reported by \citet{erwin12}. The larger statistics
of the field S0 galaxies by \citet{s4gbreaks} obtained through the photometry in the NIR bands
gives around 40 percent both for the Type I and Type II. So we confirm the probable deficit of the
Type-II surface-brightness profiles in the lenticulars in clusters with respect to the field. If to compare our results 
with the results of the STAGES survey \citep{stages} where the difference of the disk profile types
between the cluster A901/902 and the field projected onto the cluster at lower photometric redshifts
has not been found, we must recognize disagreement concerning the proportion of the Type I first of all:
in our sample about half of all cluster S0s have one-scaled exponential disks while in the A901/902
cluster only 25 percent S0s with Type I disks are detected by \citet{stages}. Interestingly, all other
profile types demonstrate consistency of their fraction in dense environment including the absence
of the Type II. The explanation of the disagreement concerning the Type I may be the note of
\citet{stages} that 20 percent of their cluster S0s have no any exponential parts in their profiles
al all. Since we have selected our S0 sample just by checking if they have exponential pieces in
their outer surface-brightness profiles, there is an evident difference in our and STAGES approaches
to the S0 sample selection.

\begin{figure*}[p]
\centering
\begin{tabular}{c}
 \includegraphics[width=10cm]{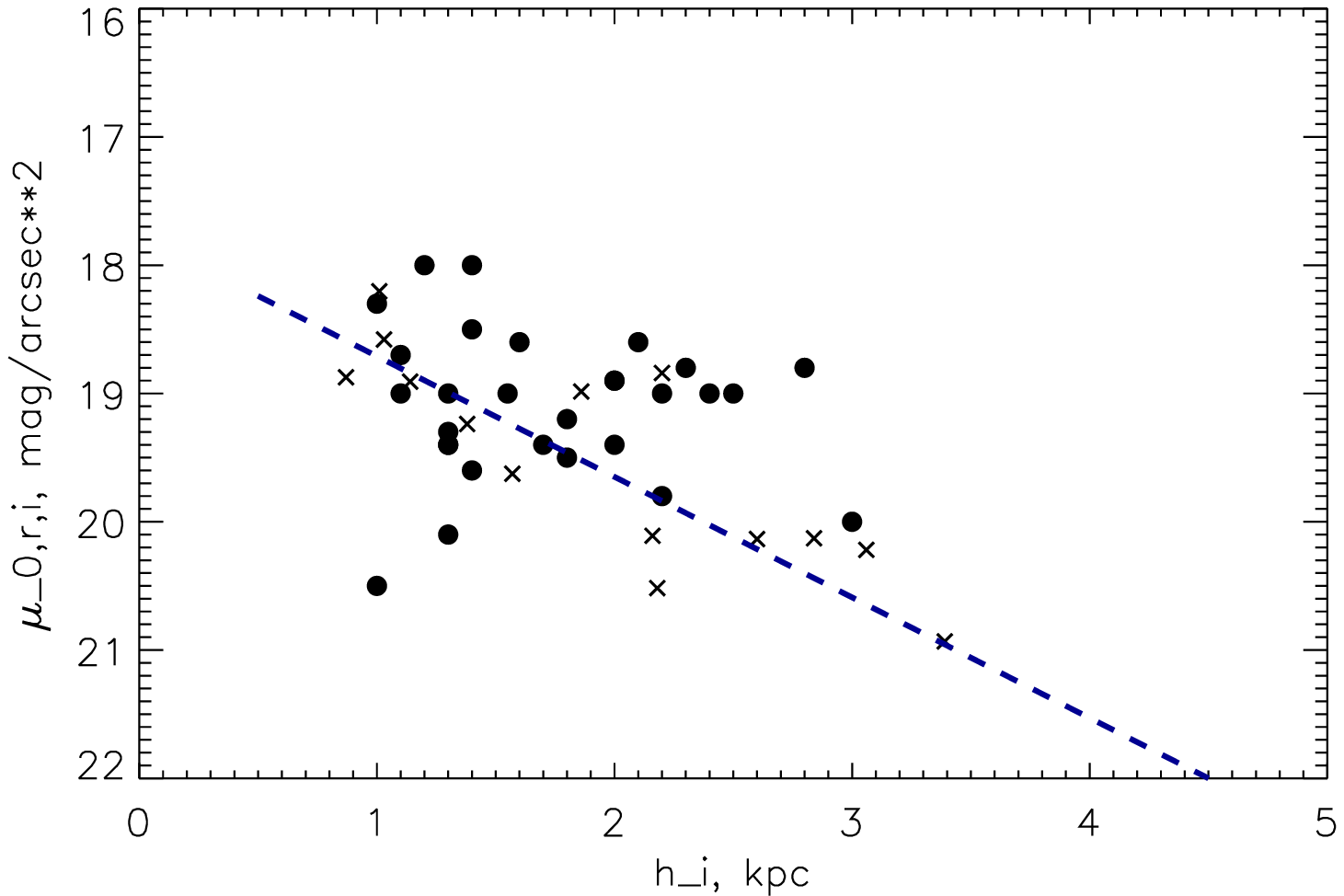} \\
 \includegraphics[width=10cm]{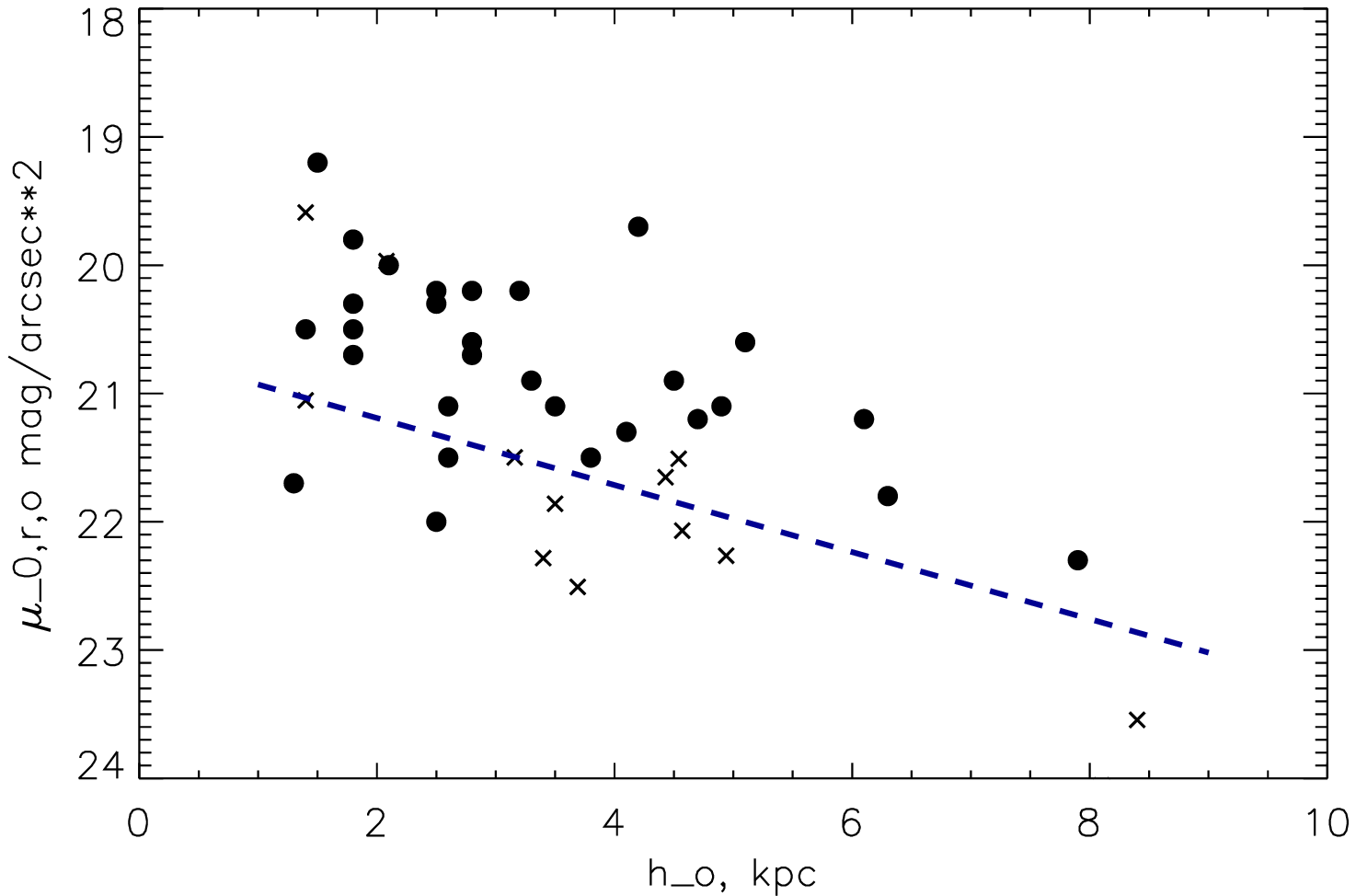} \\
 \includegraphics[width=10cm]{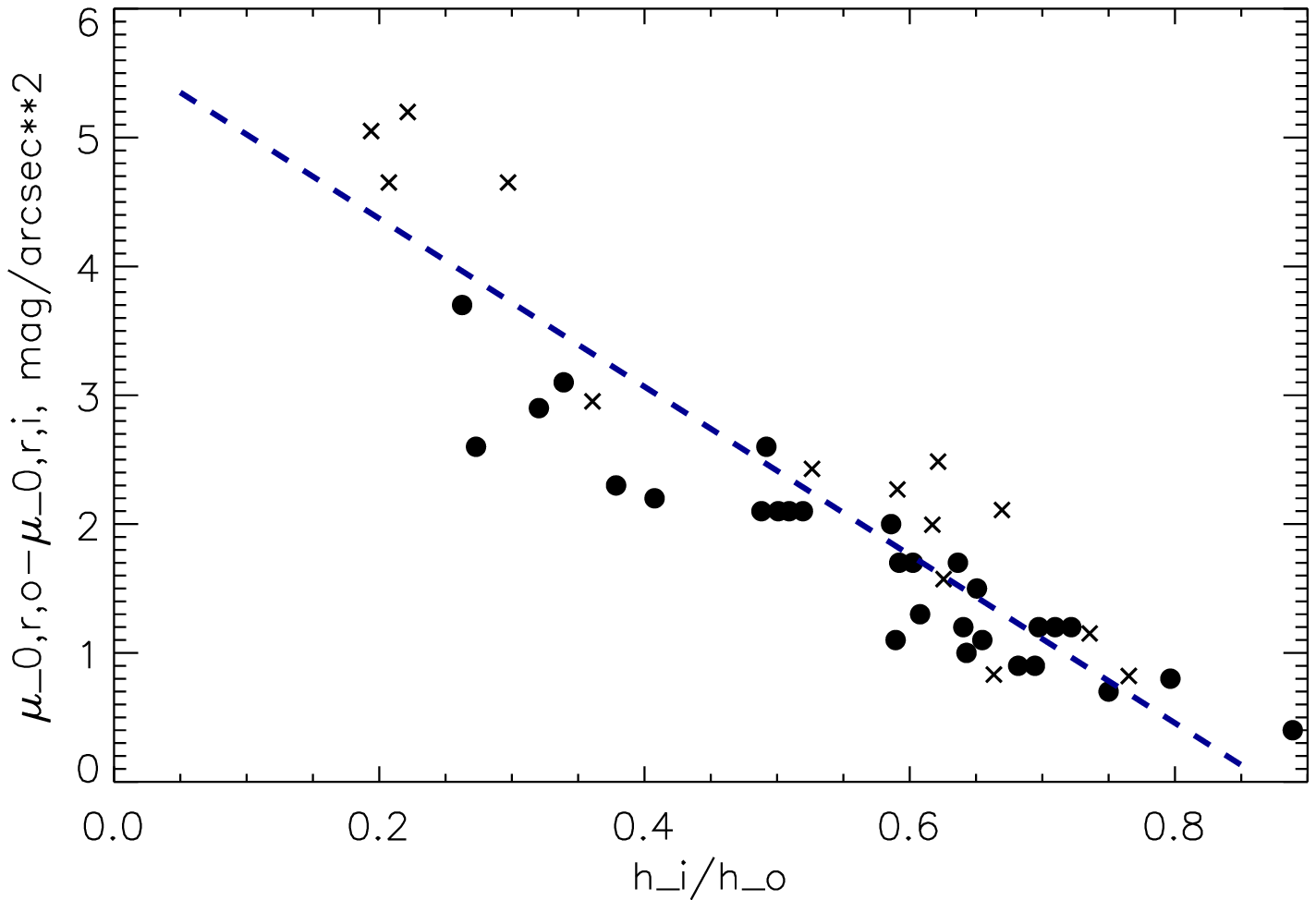} \\
\end{tabular}
\caption{Scaling relations for the two-tiered (antitruncated) exponential surface-brightness profiles (Type III)
in our sample galaxies (black points): {\it upper plot} -- the central surface brightness corrected for the Galactic
extinction according to NED vs exponential scalelength for the inner disks, $\mu _{0,i}$ vs $h_i$; {\it middle plot} --
the central surface brightness corrected for the Galactic extinction according to NED vs exponential scalelength 
for the outer disks, $\mu _{0,o}$ vs $h_o$; {\it bottom plot} -- the combined relation, $\mu _{0,o}-\mu _{0,i}$ 
versus $h_i/h_o$. By dashed blue straight lines we plot the scaling relations found by \citet{borlaff14} 
for the field S0 sample compiled over the works by \citet{erwin08} and by \citet{guti_erwin}, the crosses
are individual galaxies from \citet{erwin08} and \citet{guti_erwin} representing the field S0 population.} 
\label{type3comp}
\end{figure*}

As for the Type III, it seems probable that the environment does not affect the occurence
of the antitruncated profiles. The fraction of Type-III profiles in S0s remains almost
the same, half of all, in the field and in the clusters according to our results as well as to the
results by \citet{erwin12} and by \citet{stages}. With the profile parameters
derived by us for our cluster sample S0s, presented in the Table~\ref{s0type3}, we can
compare the scaling relations for the Type-III S0s in the clusters and in the field; we do it
in Fig.~\ref{type3comp}. The linear dependencies between the characteristics of the
inner-disk exponential fit, $\mu _{0,i}$ vs $h_i$, and between the characteristics of the outer-disk
exponential fits, $\mu _{0,o}$ vs $h_o$, as well as between $\mu _{0,o}-\mu _{0,i}$ and $h_i/h_o$,
were found by \citet{borlaff14} for the field sample compiled over the works by \citet{erwin08}
and by \citet{guti_erwin}. To compare their scaling relations with our data, we shift their linear dependencies 
by $+0.22$~mag, to transform the $R$-band into the $r$-band \citep{sdsscal}; and the individual disk central surface
brightnesses for the comparison sample of the field S0s (crosses in Fig.~\ref{type3comp}) are transformed into the $r$-band 
by using the formula (1) from \citet{guti_erwin} giving the similar shift. After that, one can see that the inner 
disks look almost the same in the clusters and in the field (Fig.~\ref{type3comp}, upper plot). The outer segments of 
the Type-III profiles of the S0s in the clusters demonstrate the correlation between $\mu _0,o$ and $h_o$
with the same slope as the S0s in the field but are slightly shifted to higher central surface
brightnesses (Fig.~\ref{type3comp}, middle plot). And the best
correlation among all found by \citet{borlaff14}, $\mu _0,o-\mu _0,i$ vs $h_i/h_o$, is
strictly the same in the field and in the clusters (Fig.~\ref{type3comp}, bottom plot).

\subsection{Vertical structure of the S0 disks in clusters and in the field}

The novel point of our photometric analysis is individual estimates of the stellar disk
relative thicknesses expressed in the terms of \citet{hubble26}'s $q$  -- see the Table~\ref{s0type1}, the Table~\ref{s0type2}, and 
the Table~\ref{s0type3}.

\begin{figure*}[t]
\centering
\begin{tabular}{c c}
 \includegraphics[width=8cm]{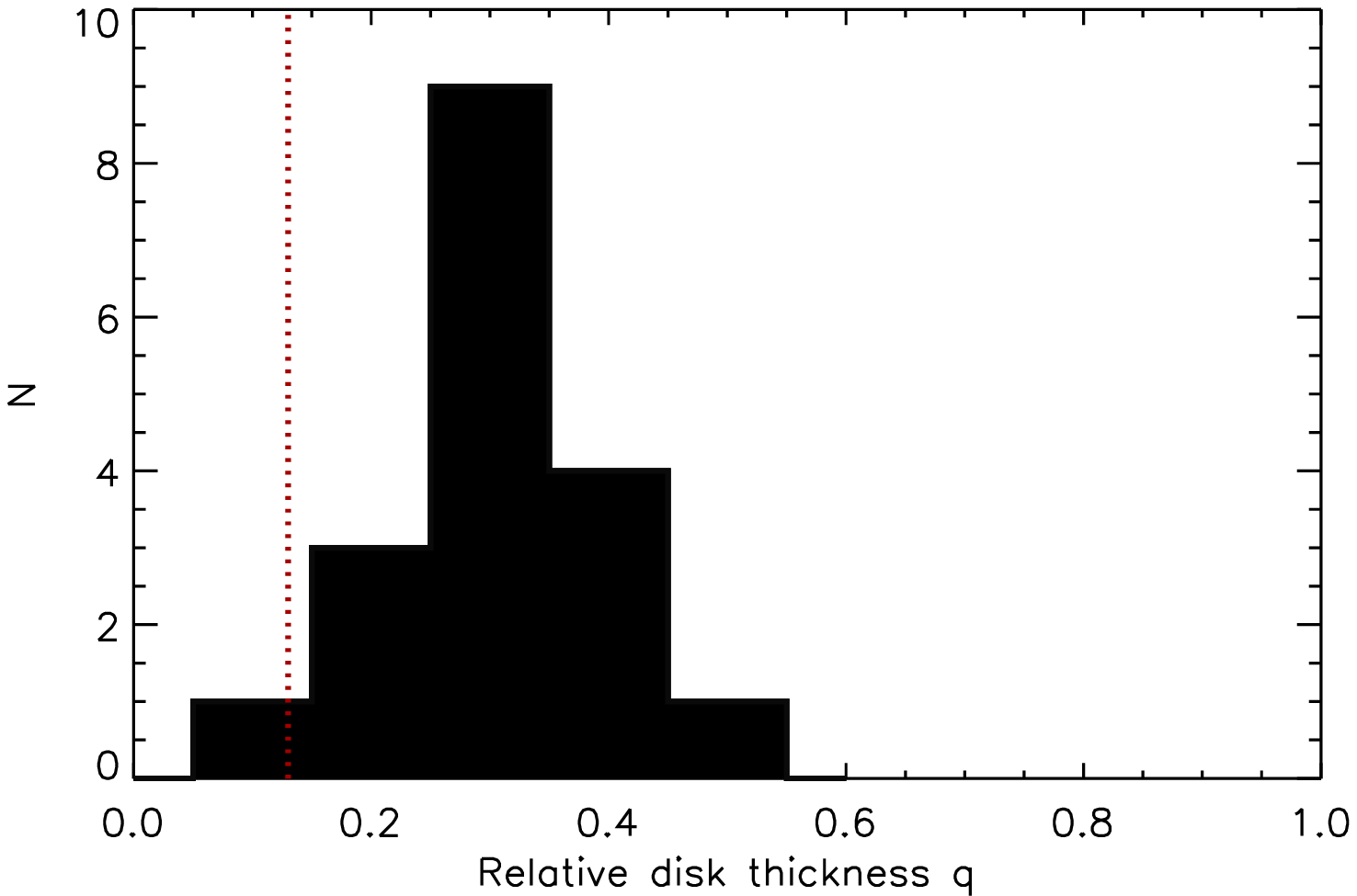} &
 \includegraphics[width=8cm]{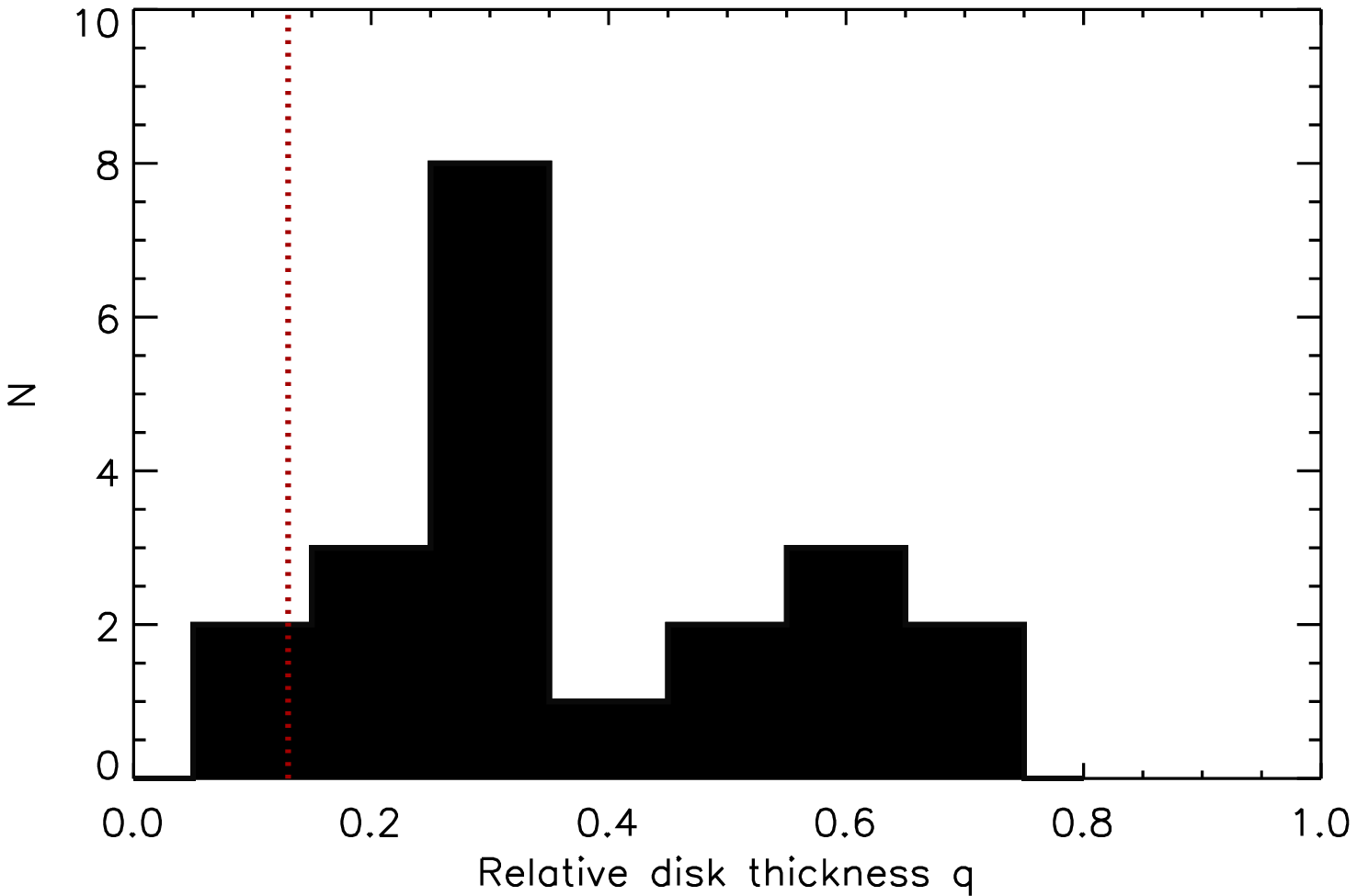} \\
\end{tabular}
\caption{The distributions of the disk relative thickness for the Type-I disks ({\it left}) and
for the inner segments of the Type-III disks ({\it right}). The red vertical dotted line shows the only
available thickness of the inner Type-II disk segment for NGC 3273.}
\label{qhisto}
\end{figure*}

Figure~\ref{qhisto} presents the distributions of the relative disk thicknesses for the Type-I disks ({\it left})
and for the inner segments of the Type-III disks ({\it right}). The Type-I distribution looks like a regular
normal law, rather narrow one. The distribution for the inner disks of Type-III profiles is obviously bimodal,
with two distinct maxima near $q=0.3$ and $q=0.6$. Indeed, we can easily mess the inner segments of the Type-III 
disks with {\it pseudobulges}: the latters have also exponential surface brightness profiles but must be 
{\it much thicker} than the disks -- because they are {\it bulges}. We have calculated the mean thicknesses
of the Type-I disks and of the {\it true} inner disks of Type-III -- those with $q<0.5$. We have obtained:
$\langle q \rangle = 0.31 \pm 0.02$ for the Type I disks and $\langle q \rangle = 0.25 \pm 0.02$ for the
inner disks of the Type-III profiles. Let us compare these estimates with similar estimates for the field
sample which we have presented in our first paper exploring our method of disk thickness measurement \citep{thickmeth}:
$\langle q \rangle = 0.46 \pm 0.05$ for the Type-I disks and 
$\langle q \rangle = 0.22 \pm 0.045$ for the inner disks of the Type-III profiles. We see again that the
Type-III disks, in particular their inner segments, are {\it the same} in the field and in the clusters 
while the Type-I disks look {\it thinner} in dense environments. Intuitively, one could expect the opposite trend: 
in dense environments the stellar disks had to be more dynamically heated due to frequent tidal interactions. However, 
it is not the first detected signature of more common accretion events in rarified environments \citep*{isomnras}.

Here we must note that after removing false Type-III disks where the inner segments are in fact pseudobulges,
the fraction of Type-III disks in the whole sample is diminished and the fraction of Type-I disks is probably increased.
But to compare the proportions of various disk types in the clusters and in the field, we must use initial
results because during the previous field S0 sample investigations there were no checking inner disk thickness. 

\subsection{Environment-driven evolution?}

Recently the first theoretical interpretation of the absence of Type-II profiles and prevalence of Type-I
profiles in cluster disk galaxies has been proposed. \citet{clarke_db} have simulated the process of
stellar disk growth in an isolated spiral galaxy and in the same galaxy after its infall into a cluster
with the Fornax-like parameters. They have found that when the infalling galaxy comes close to the cluster
center, at a distance of 20\%\ of the cluster virial radius, its Type-II surface density profile transforms 
into a pure single-scale exponential one -- or becomes a Type-I profile. The physical mechanism of this transformation
may be the effect of tidally driven transient spirals in the outer parts of the disk. Effective migration of disk 
stars into its outermost parts under the spiral density wave pressure removes any break of the surface density  
distribution at intermediate radii and produces purely exponential form of the disk profile in its final state. 
Interestingly, the galaxies suffer only radial-profile transformation, without significant disk thickening and/or 
dynamical heating. If really the cluster gravitational potential impact provides so effective
stellar radial migration in the galactic disks, it may explain not only the absence/rarity of the Type-II disks
in clusters, but also the brighter outer parts in the Type-III disks of the cluster S0s (Fig.~\ref{type3comp}).
Indeed, the effect of stellar migration outward must act not only on Type-II disks; it must affect all
infalling disk galaxies including Type-III ones. In the Type-III disks the effective stellar migration outward 
would strengthen the outer parts of the disks, and just this effect 
is found by us for the cluster S0s with the Type-III surface-brightness profiles in Fig.~\ref{type3comp}. 

\section{Conclusions}

We have analyzed the structure of the large-scale stellar disks in 60 lenticular galaxies -- members of
8 southern clusters. The parameters of the radial surface-brightness profiles have been determined,
and all the galaxies have been classified according to three types proposed by \citet{pohlen_trujillo}.
We confirm the result by \citet{erwin12} that the proportion of surface-brightness profile types
is significantly different in clusters and in the field: in the clusters the Type-II profiles are almost
absent while according to the literature data, in the field they constitute about one quarter of all lenticular galaxies. 
The Type-III profiles are equally represented in the clusters and in the field, and they follow similar scaling relations;
marginally we detect higher surface brightnesses for the outer segments of the Type-III disks in the
clusters. Also by applying our novel method \citep{thickmeth} we have determined the relative 
thicknesses (the vertical-to-radial scale ratios) for the stellar disks in 18 Type-I galaxies and for
21 inner segments of the Type-III disks. The relative thicknesses of the Type-I disks seem to be in average
smaller in our galaxies belonging to the clusters than it has been found by us earlier for the field Type-I S0s 
\citep{thickmeth}; perhaps it is due to the pollution of the subsample of the cluster Type-I disks by former
Type-II disks which may be re-structured during the galaxy infall into clusters \citep{clarke_db}.
Among the inner segments of the Type-III disks we have found 7 pseudobulges, with the average relative thickness
of $q=0.6$; since it is a third of our Type-III disk thicknesses determined, we note that a real fraction of Type-III disks
must be lower than it is thought up to now: some of the disks classified as Type-III ones may represent a combination
of a pseudobulge and a Type-I disk. The remaining 14 inner segments of our Type-III disks have on average the same relative thickness, 
$q=0.25 \pm 0.02$, as the Type-III S0 galaxies in the field.

\acknowledgments

This work is based on the imaging data obtained with the LCO robotic telescope network.
We acknowledge the usage of the HyperLEDA database (http://leda.univ-lyon1.fr). This research
has made use of the NASA/IPAC Extragalactic Database (NED) which is operated by the Jet
Propulsion Laboratory, California Institute of Technology, under contract with the National Aeronautics
and Space Administration. For the purpose of our photometric calibration we have used SDSS/DR9 \citep{sdss_dr9} and 
Pan-STARRS1 data \citep{panstarrs_1,panstarrs_2}.
Funding for the SDSS-III has been provided by the Alfred P. Sloan Foundation, the Participating Institutions,
the National Science Foundation, and the U.S. Department of Energy Office of Science. The SDSS-III Web site is http://www.sdss3.org/.
SDSS-III is managed by the Astrophysical Research Consortium for the Participating Institutions of the SDSS-III Collaboration 
including the University of Arizona, the Brazilian Participation Group, Brookhaven National Laboratory, Carnegie Mellon University,
University of Florida, the French Participation Group, the German Participation Group, Harvard University, the Instituto de 
Astrofisica de Canarias, the Michigan State/Notre Dame/JINA Participation Group, Johns Hopkins University, 
Lawrence Berkeley National Laboratory, Max Planck Institute for Astrophysics, Max Planck Institute for Extraterrestrial Physics, 
New Mexico State University, New York University, Ohio State University, Pennsylvania State University, University of Portsmouth, 
Princeton University, the Spanish Participation Group, University of Tokyo, University of Utah, Vanderbilt University, 
University of Virginia, University of Washington, and Yale University. 
The Pan-STARRS1 Surveys (PS1) and the PS1 public science archive have been made possible through contributions 
by the Institute for Astronomy, the University of Hawaii, the Pan-STARRS Project Office, the Max-Planck Society 
and its participating institutes, the Max Planck Institute for Astronomy, Heidelberg and the Max Planck Institute 
for Extraterrestrial Physics, Garching, The Johns Hopkins University, Durham University, the University of Edinburgh, 
the Queen's University Belfast, the Harvard-Smithsonian Center for Astrophysics, the Las Cumbres Observatory Global 
Telescope Network Incorporated, the National Central University of Taiwan, the Space Telescope Science Institute, 
the National Aeronautics and Space Administration under Grant No. NNX08AR22G issued through the Planetary Science 
Division of the NASA Science Mission Directorate, the National Science Foundation Grant No. AST-1238877, the University 
of Maryland, Eotvos Lorand University (ELTE), the Los Alamos National Laboratory, and the Gordon and Betty Moore Foundation.

\appendix

\section{Atlas of the S0 galaxies studied photometrically}

\clearpage

\bigskip

\begin{figure*}
\centering
\includegraphics[width=\textwidth]{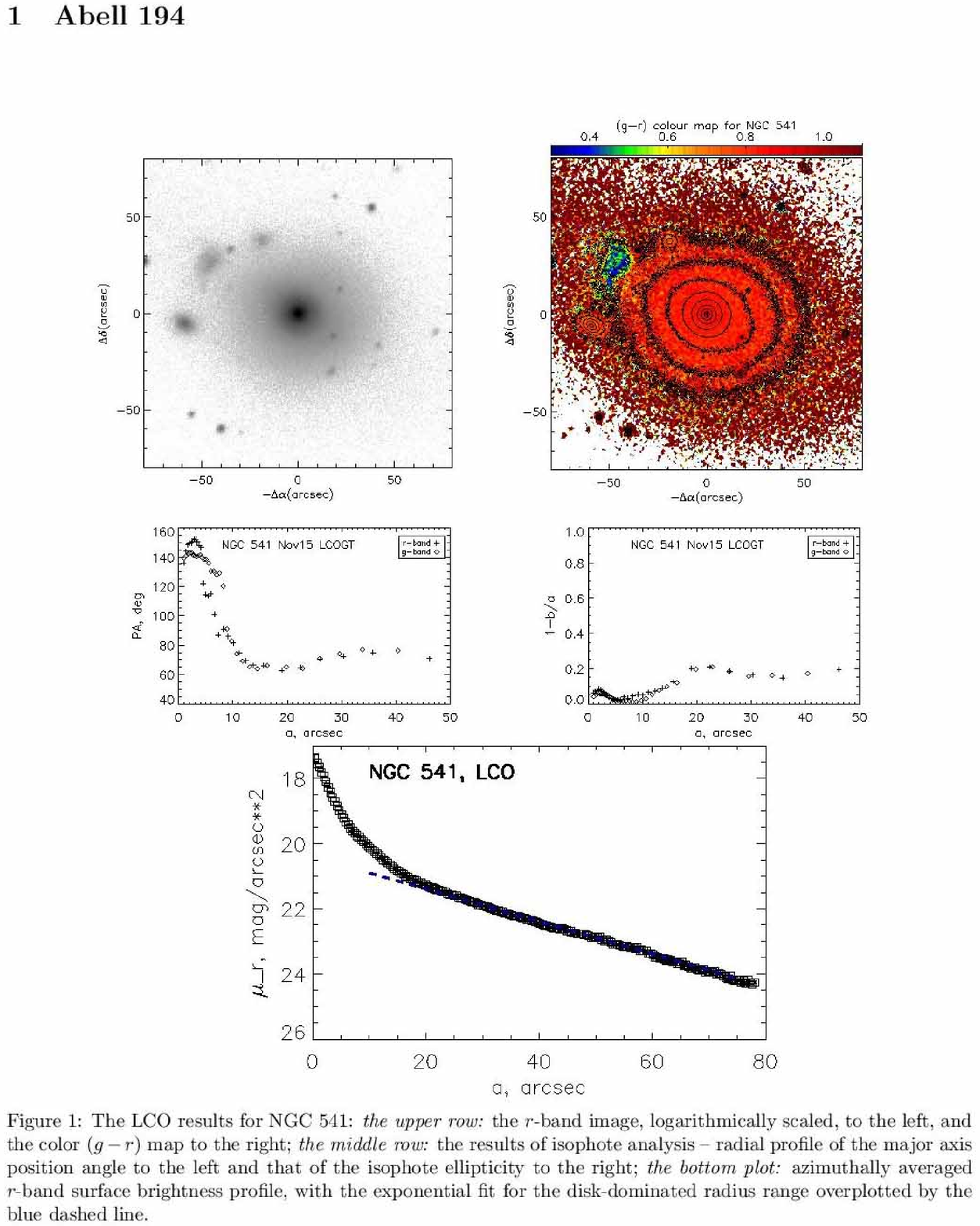}
\end{figure*}

\clearpage

\bigskip
\bigskip

\begin{figure*}
\centering
\includegraphics[width=\textwidth]{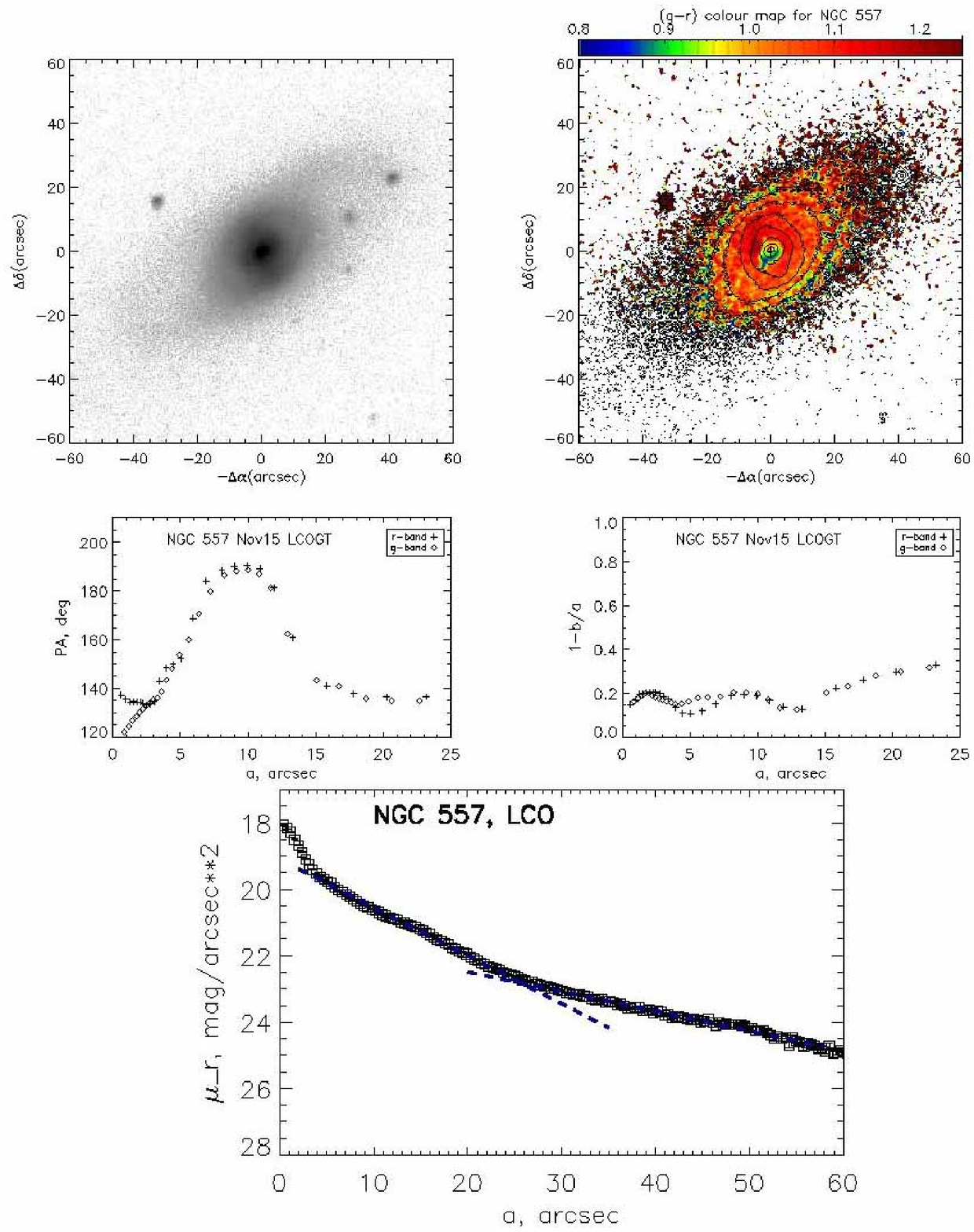}
\end{figure*}

\clearpage

\bigskip
\bigskip

\begin{figure*}
\centering
\includegraphics[width=\textwidth]{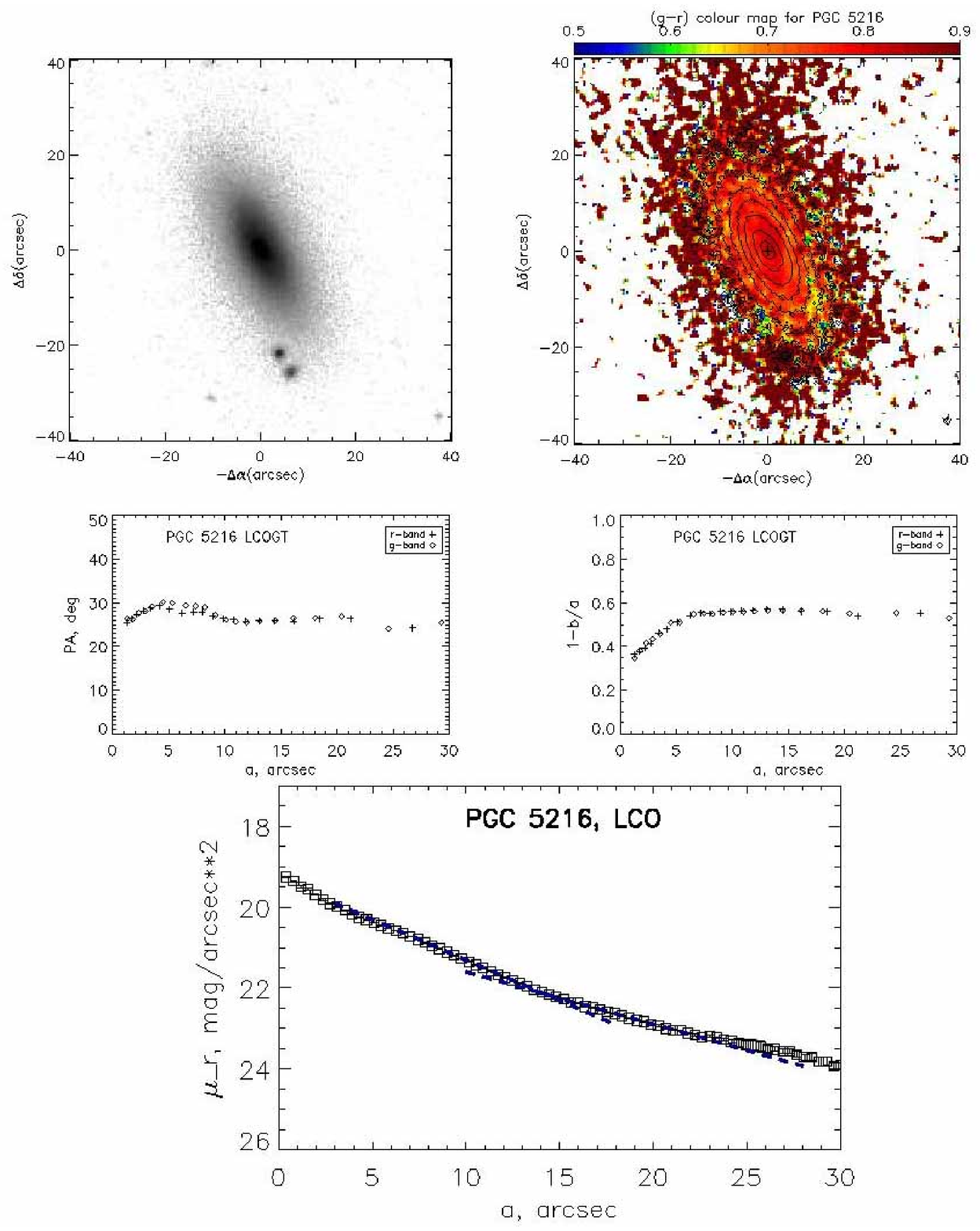}
\end{figure*}

\clearpage

\bigskip
\bigskip

\begin{figure*}
\centering
\includegraphics[width=\textwidth]{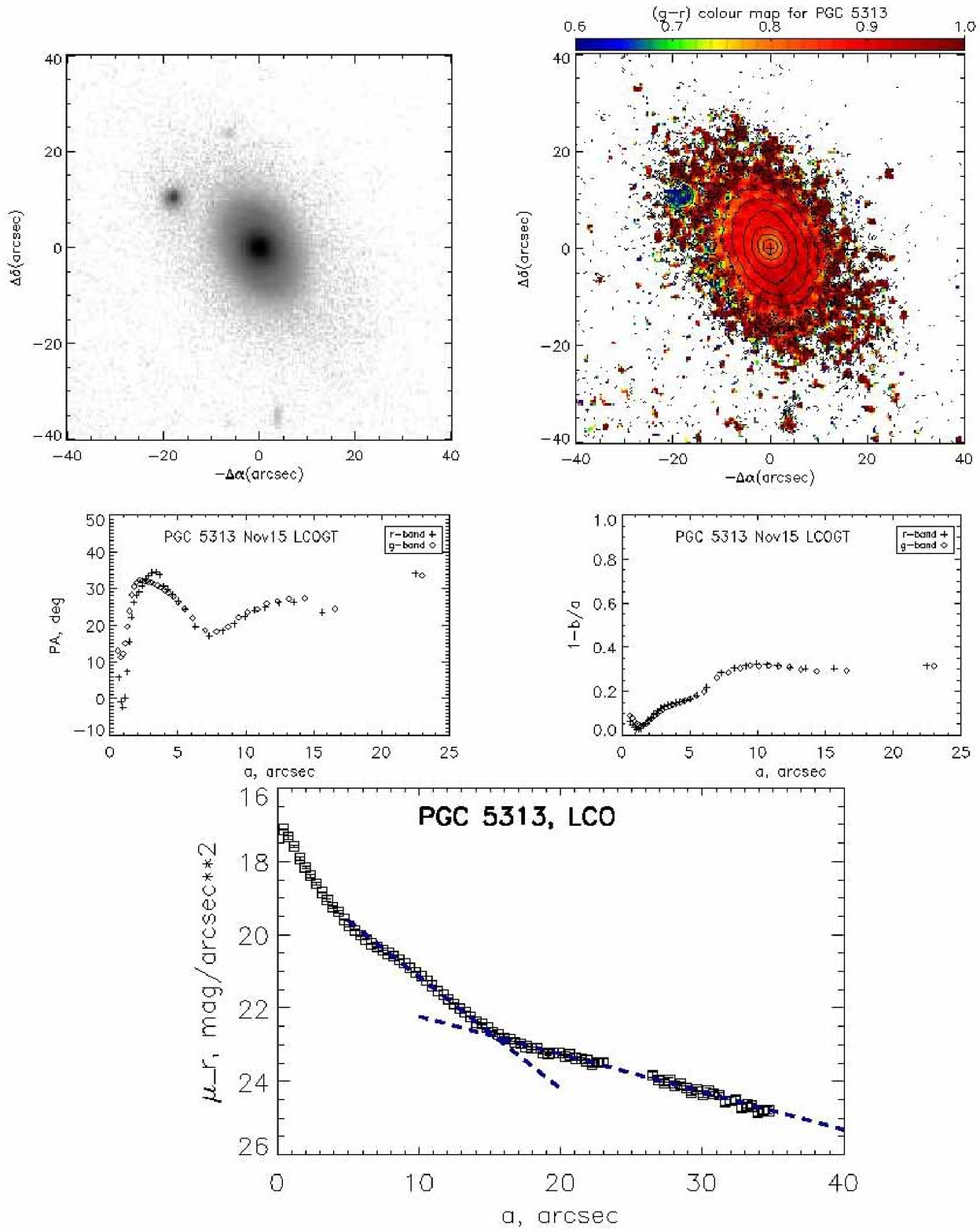}
\end{figure*}

\clearpage

\bigskip
\bigskip

\begin{figure*}
\centering
\includegraphics[width=\textwidth]{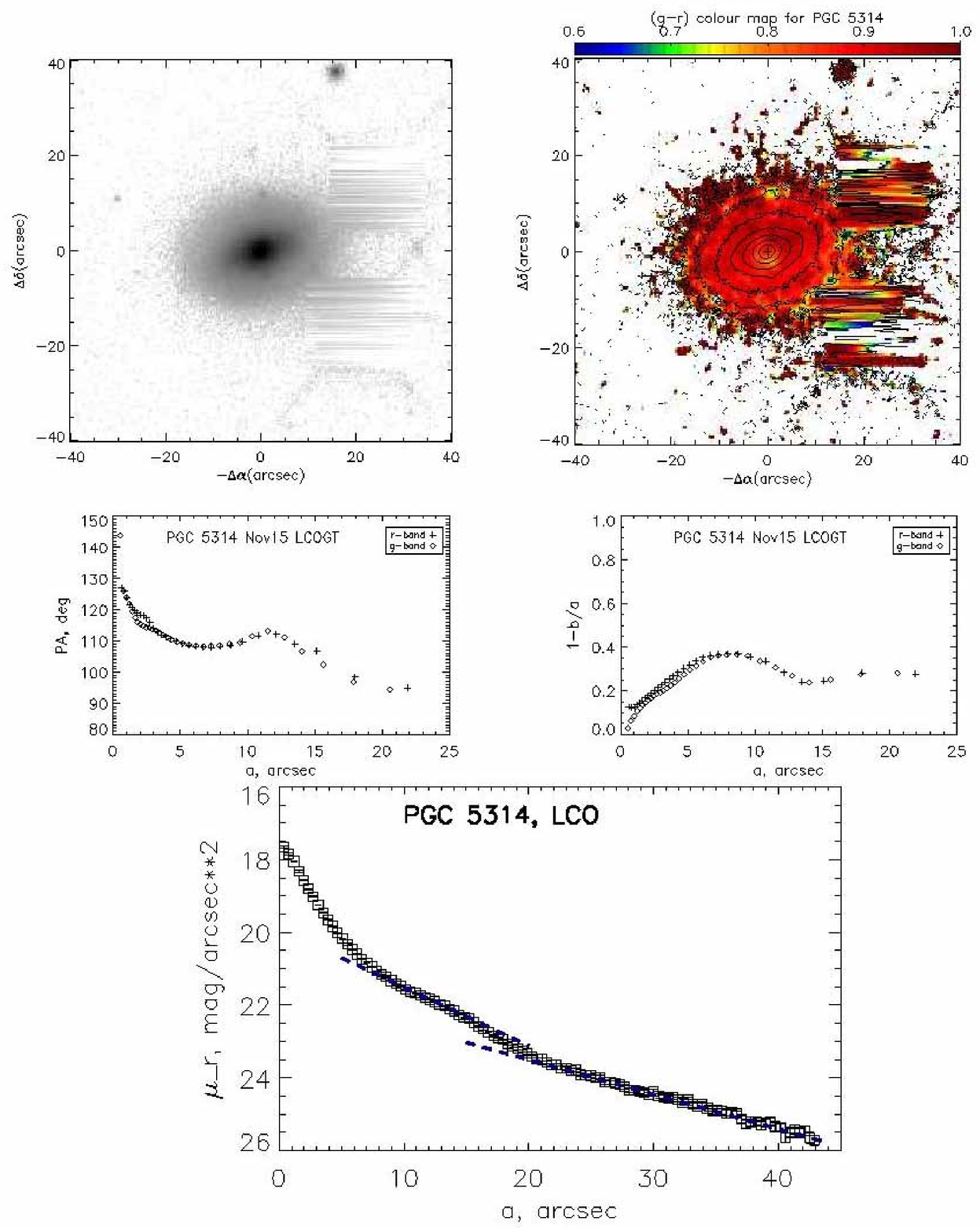}
\end{figure*}

\clearpage

\bigskip
\bigskip

\begin{figure*}
\centering
\includegraphics[width=\textwidth]{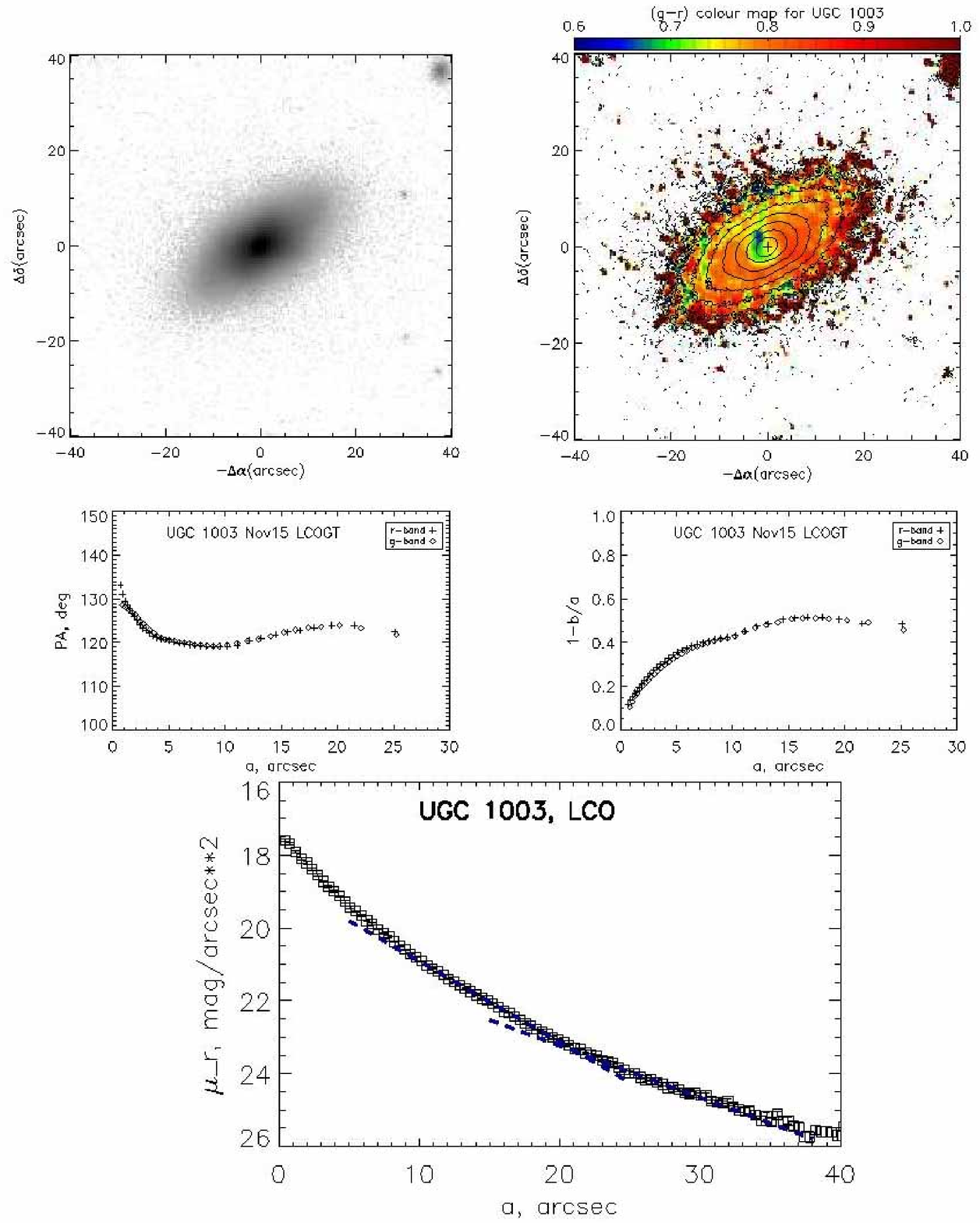}
\end{figure*}

\clearpage

\bigskip
\bigskip

\begin{figure*}
\centering
\includegraphics[width=\textwidth]{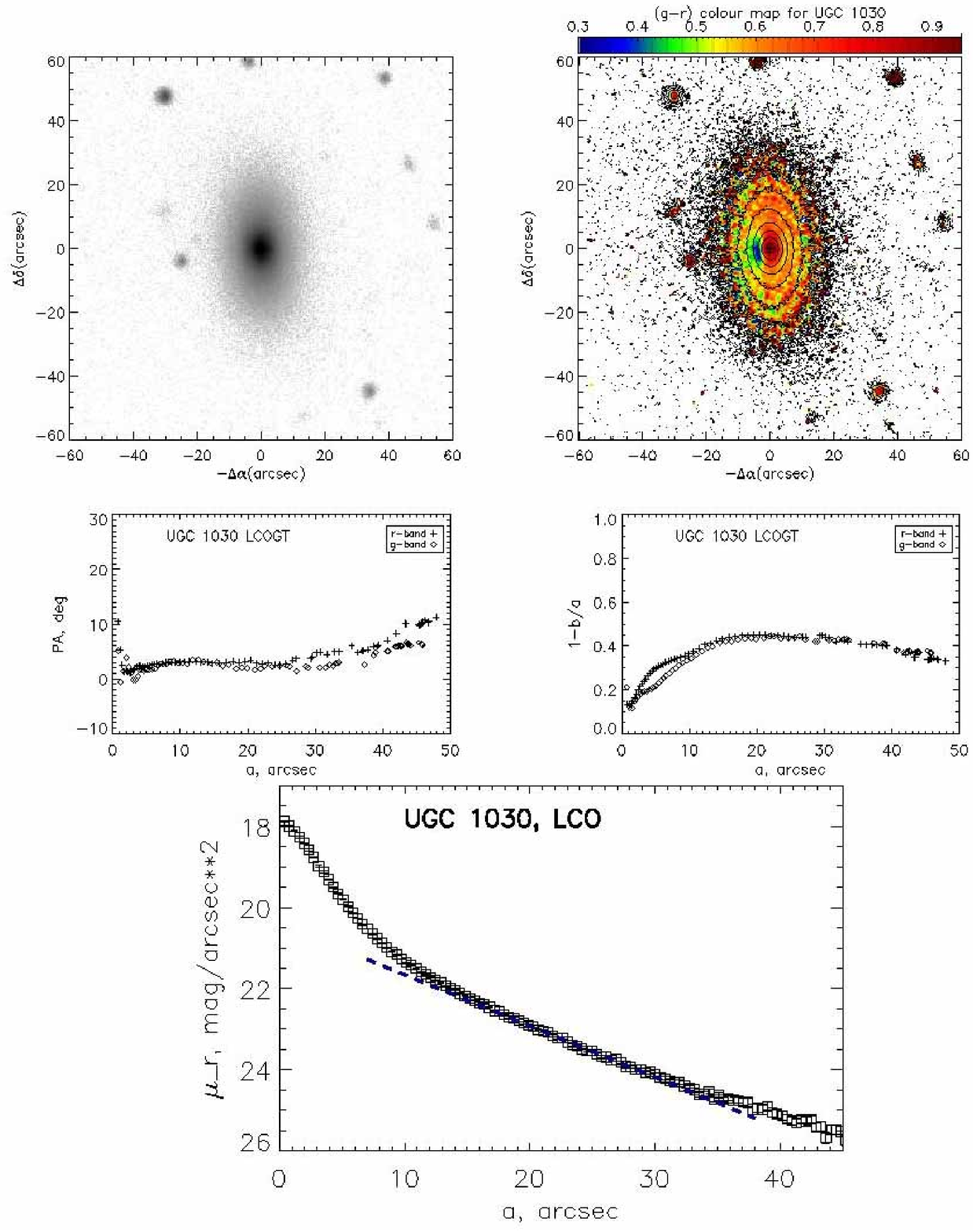}
\end{figure*}

\clearpage

\bigskip
\bigskip

\begin{figure*}
\centering
\includegraphics[width=\textwidth]{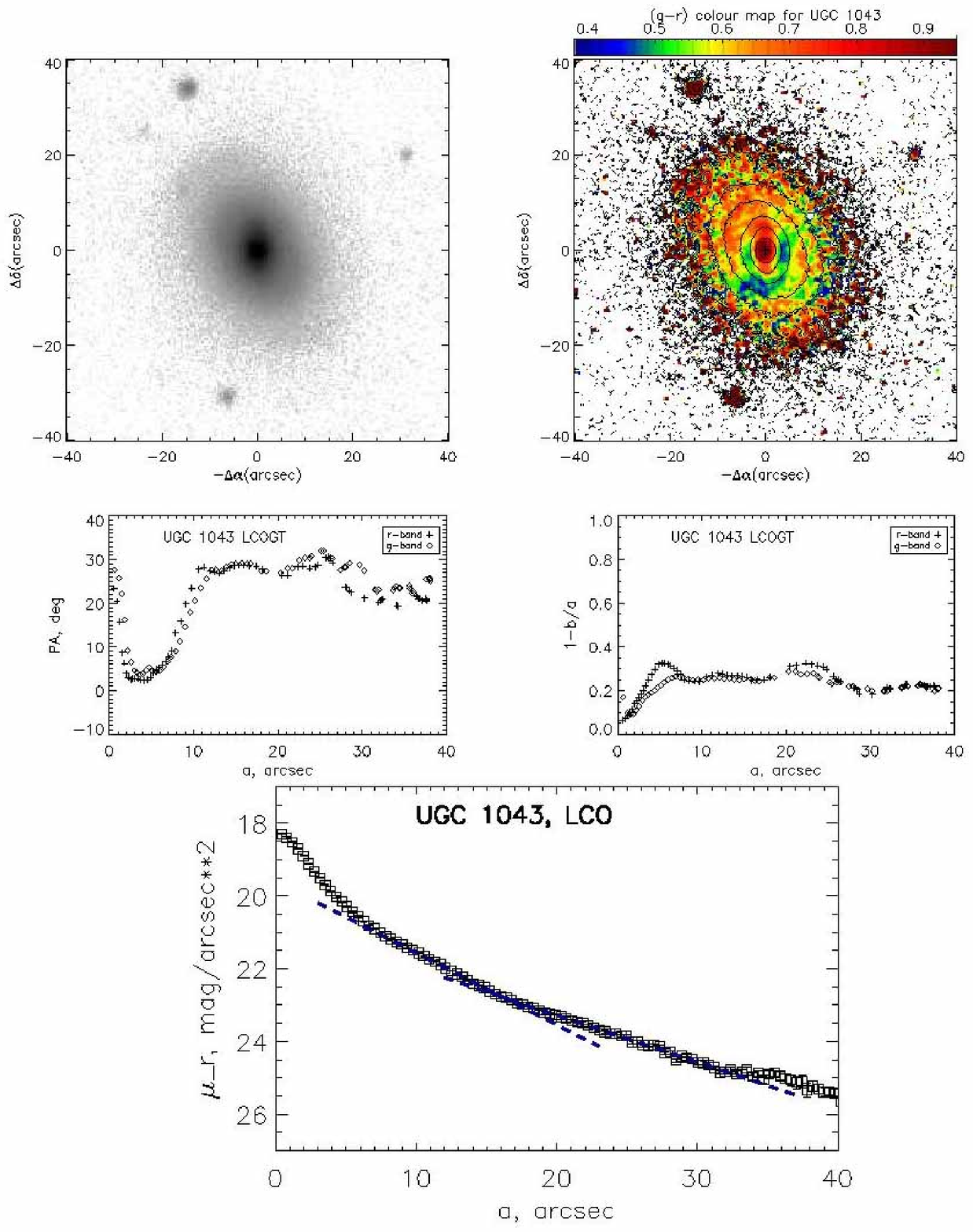}
\end{figure*}

\clearpage

\bigskip

\begin{figure*}
\centering
\includegraphics[width=\textwidth]{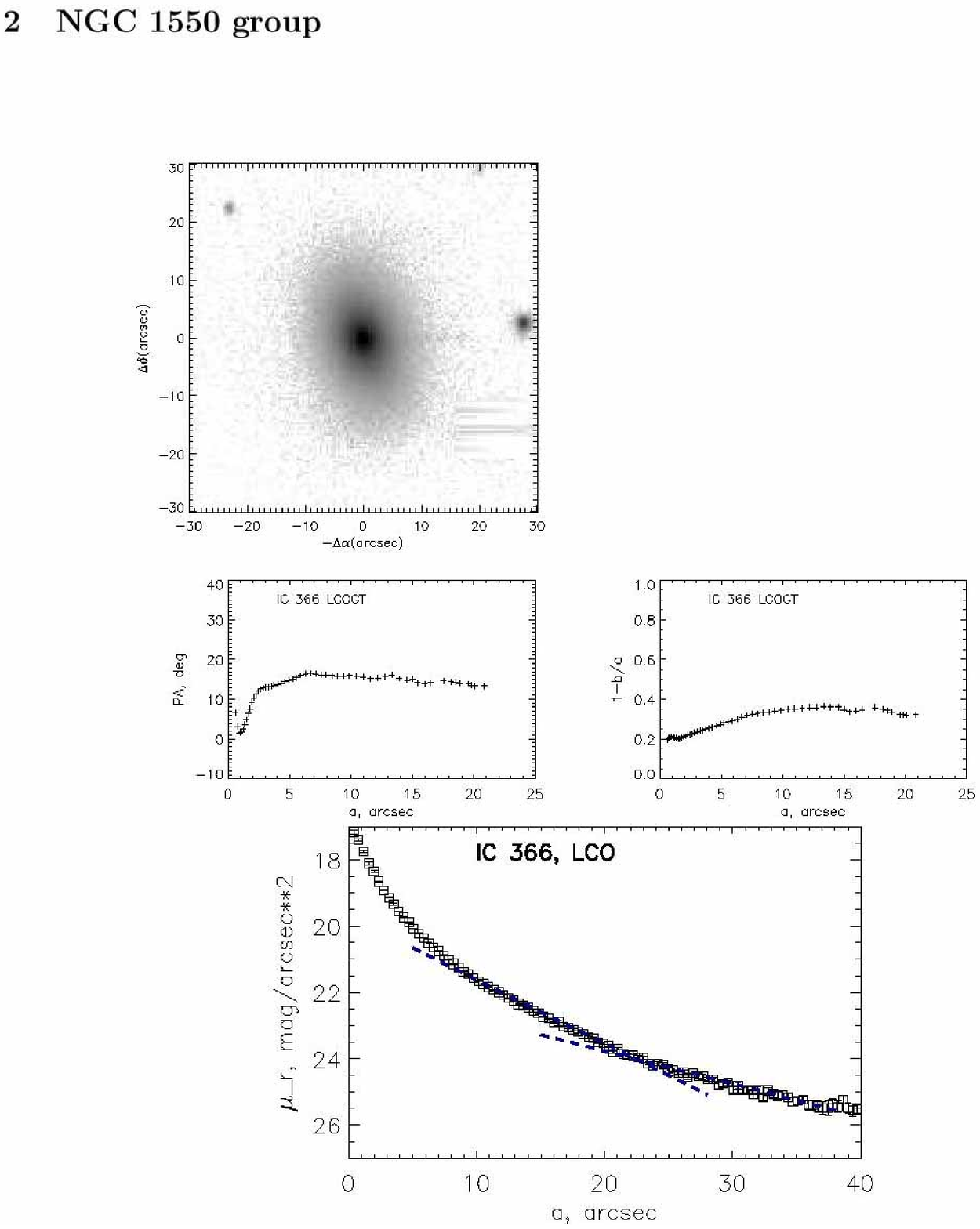}
\end{figure*}

\clearpage

\bigskip
\bigskip

\begin{figure*}
\centering
\includegraphics[width=\textwidth]{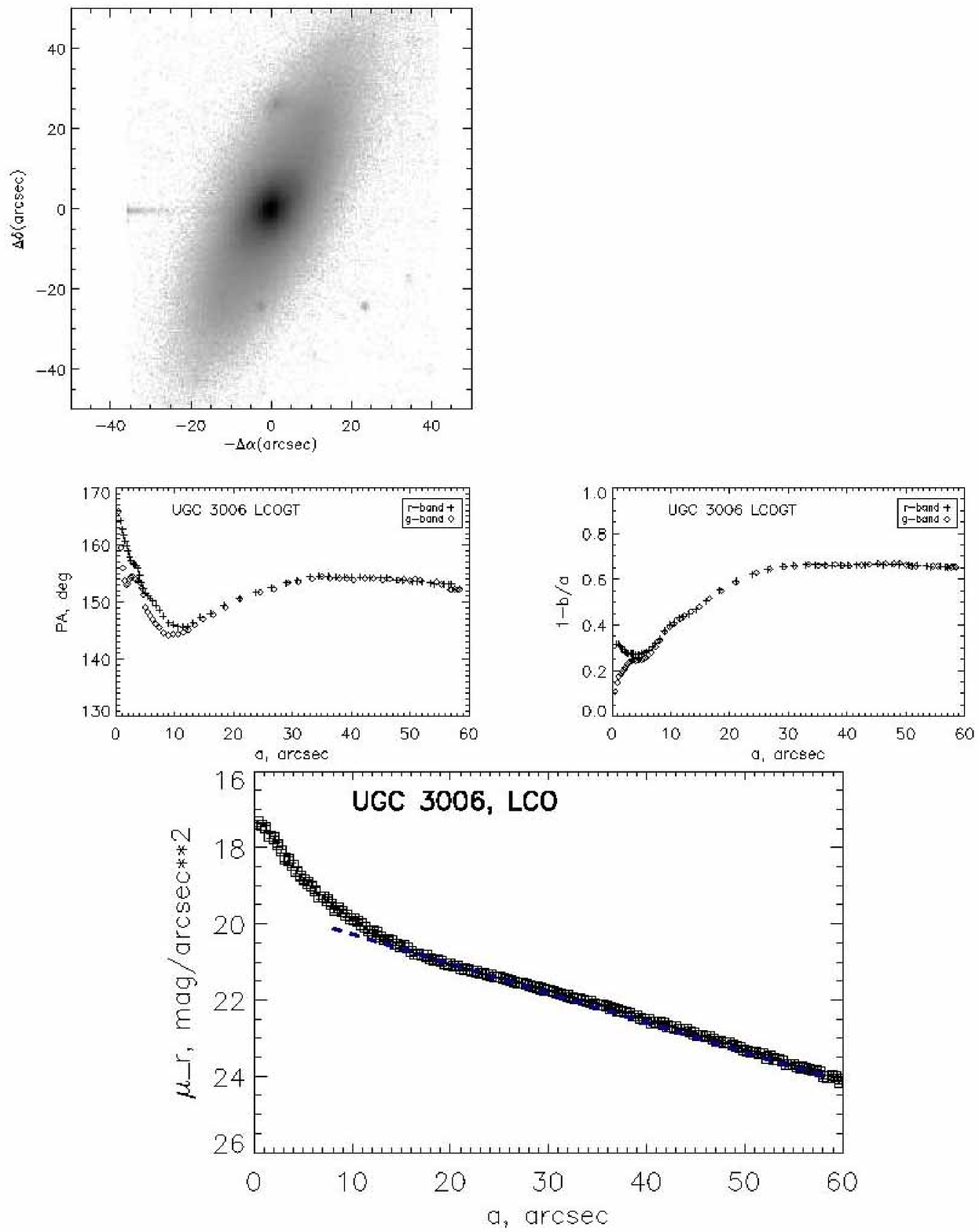}
\end{figure*}

\clearpage

\bigskip
\bigskip

\begin{figure*}
\centering
\includegraphics[width=\textwidth]{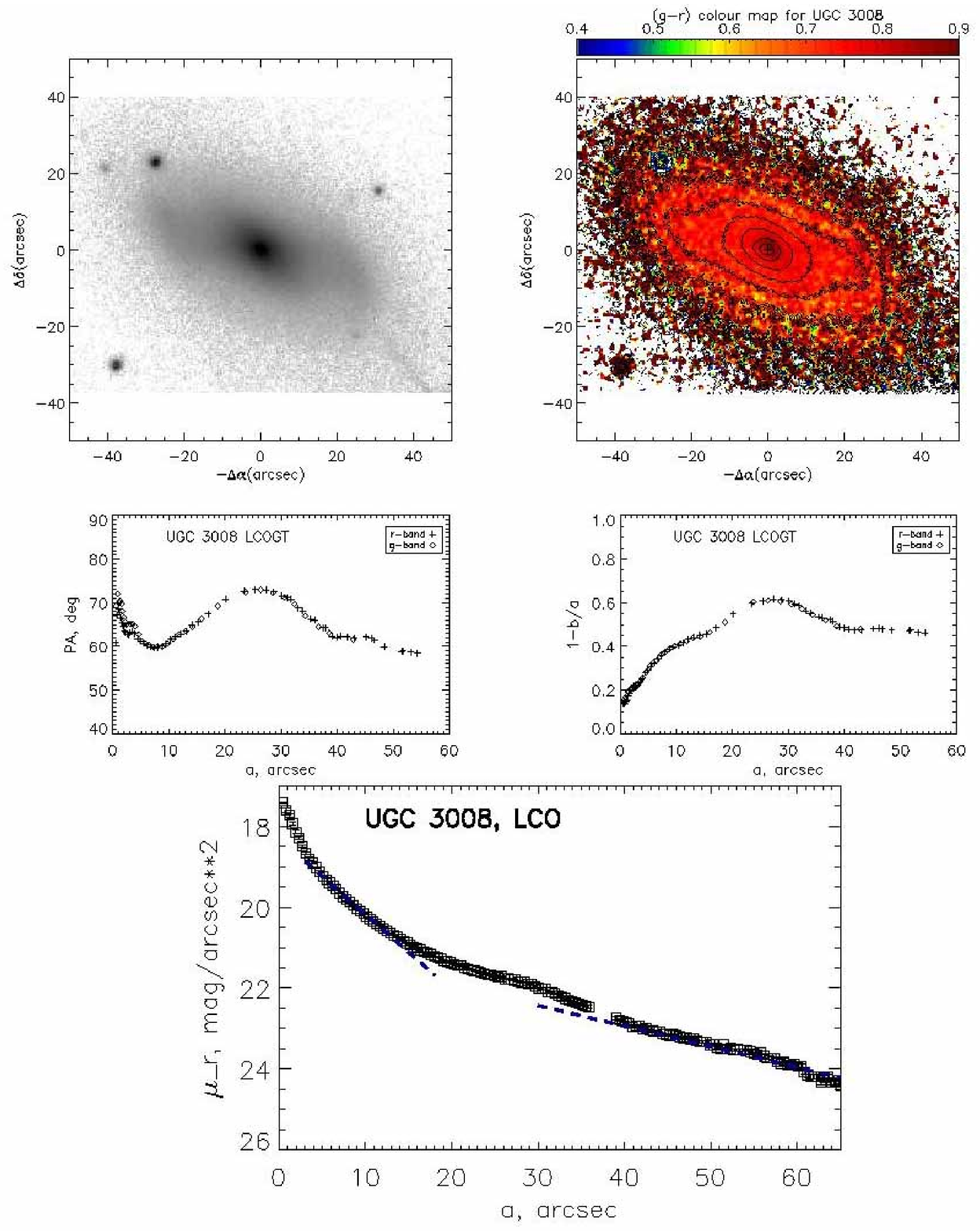}
\end{figure*}

\clearpage

\bigskip
\bigskip

\begin{figure*}
\centering
\includegraphics[width=\textwidth]{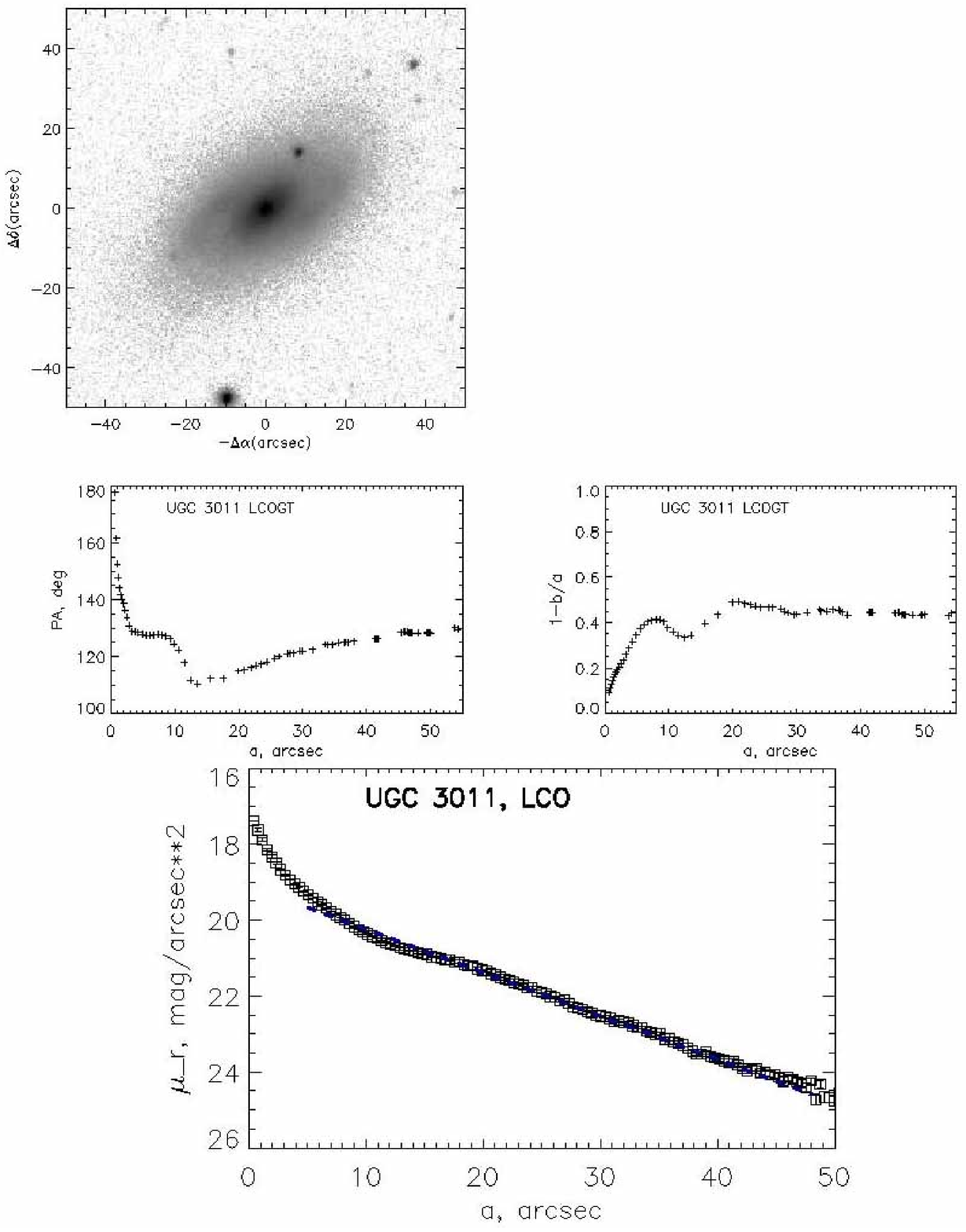}
\end{figure*}

\clearpage

\bigskip

\begin{figure*}
\centering
\includegraphics[width=\textwidth]{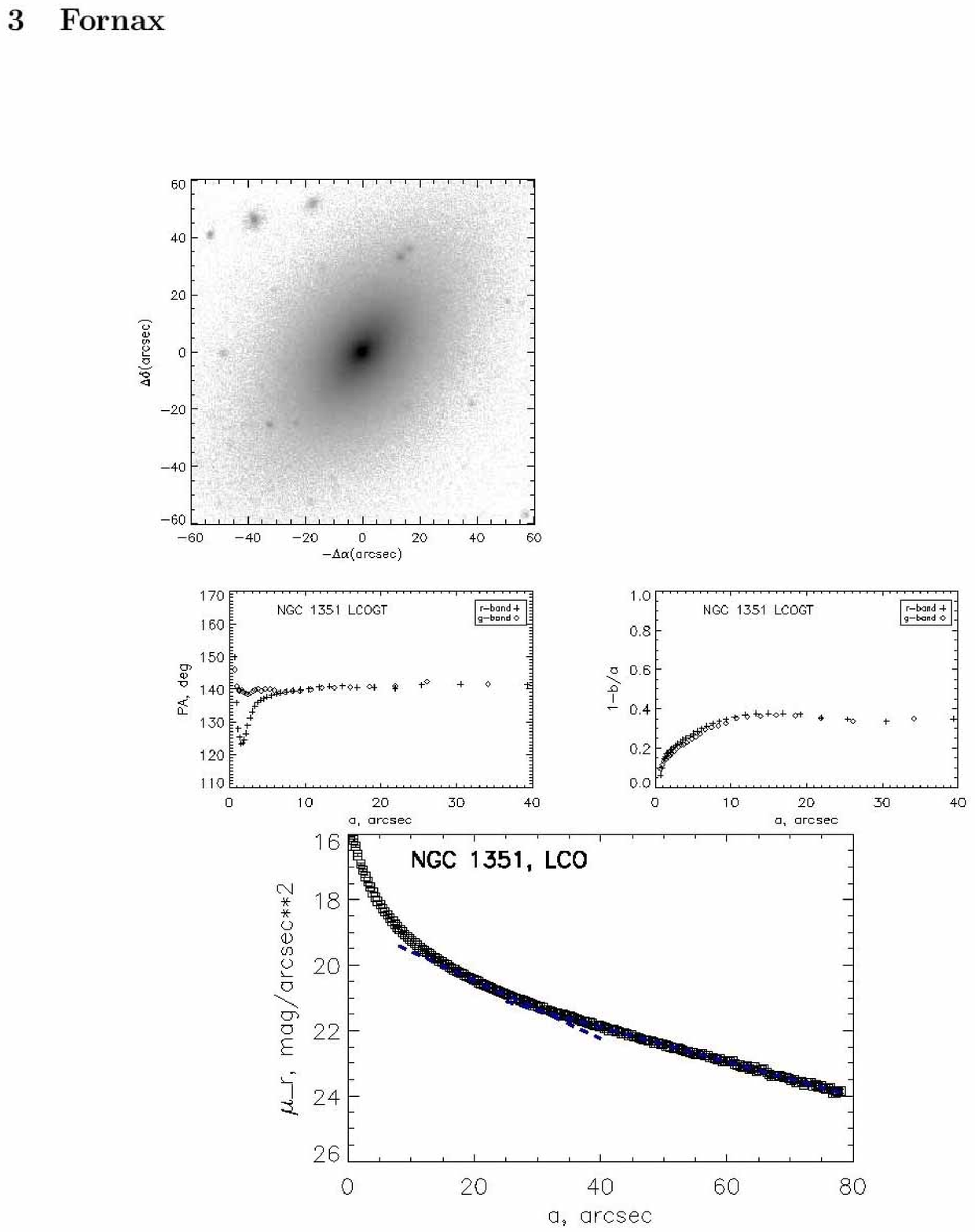}
\end{figure*}

\clearpage

\bigskip
\bigskip

\begin{figure*}
\centering
\includegraphics[width=\textwidth]{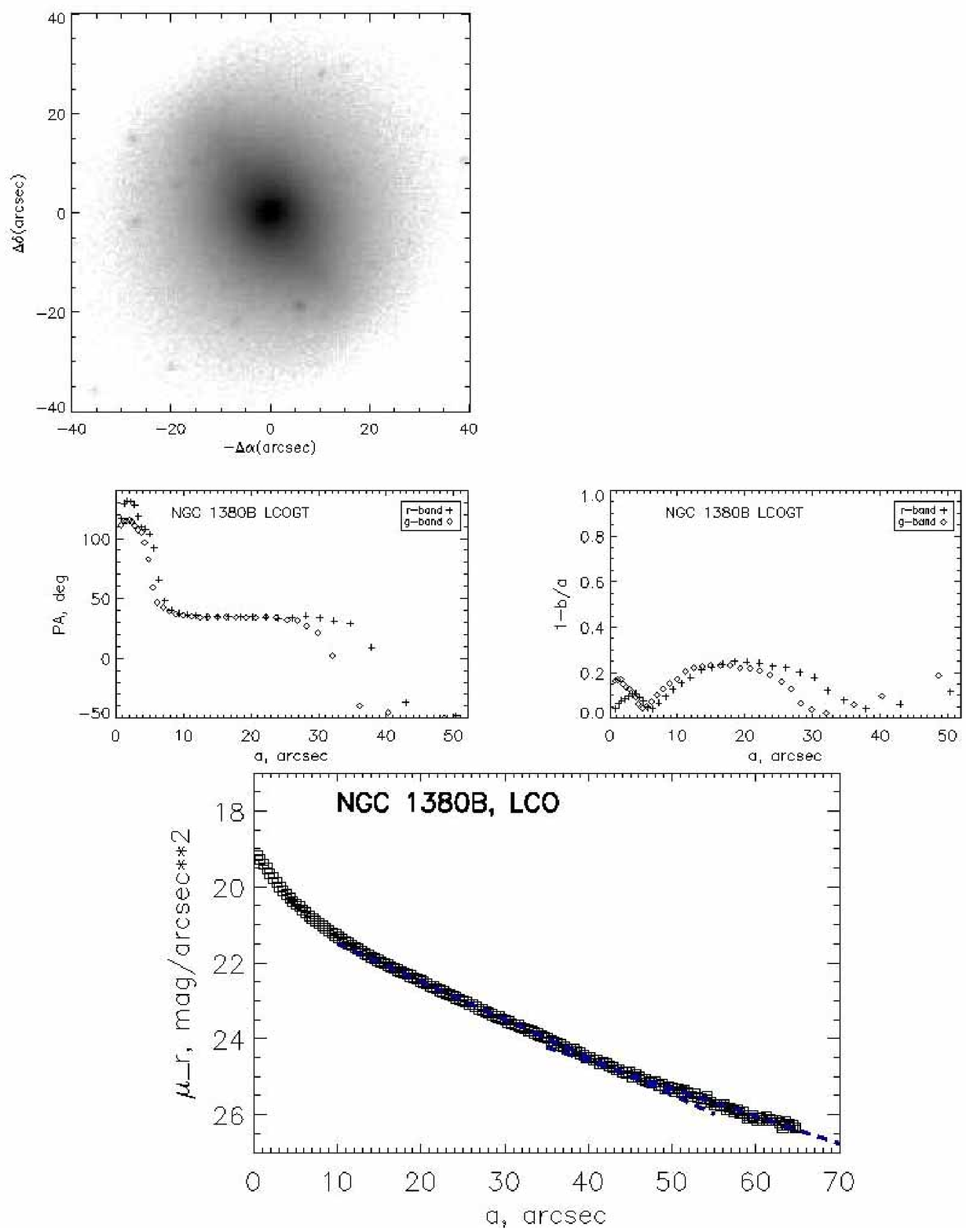}
\end{figure*}

\clearpage

\bigskip
\bigskip

\begin{figure*}
\centering
\includegraphics[width=\textwidth]{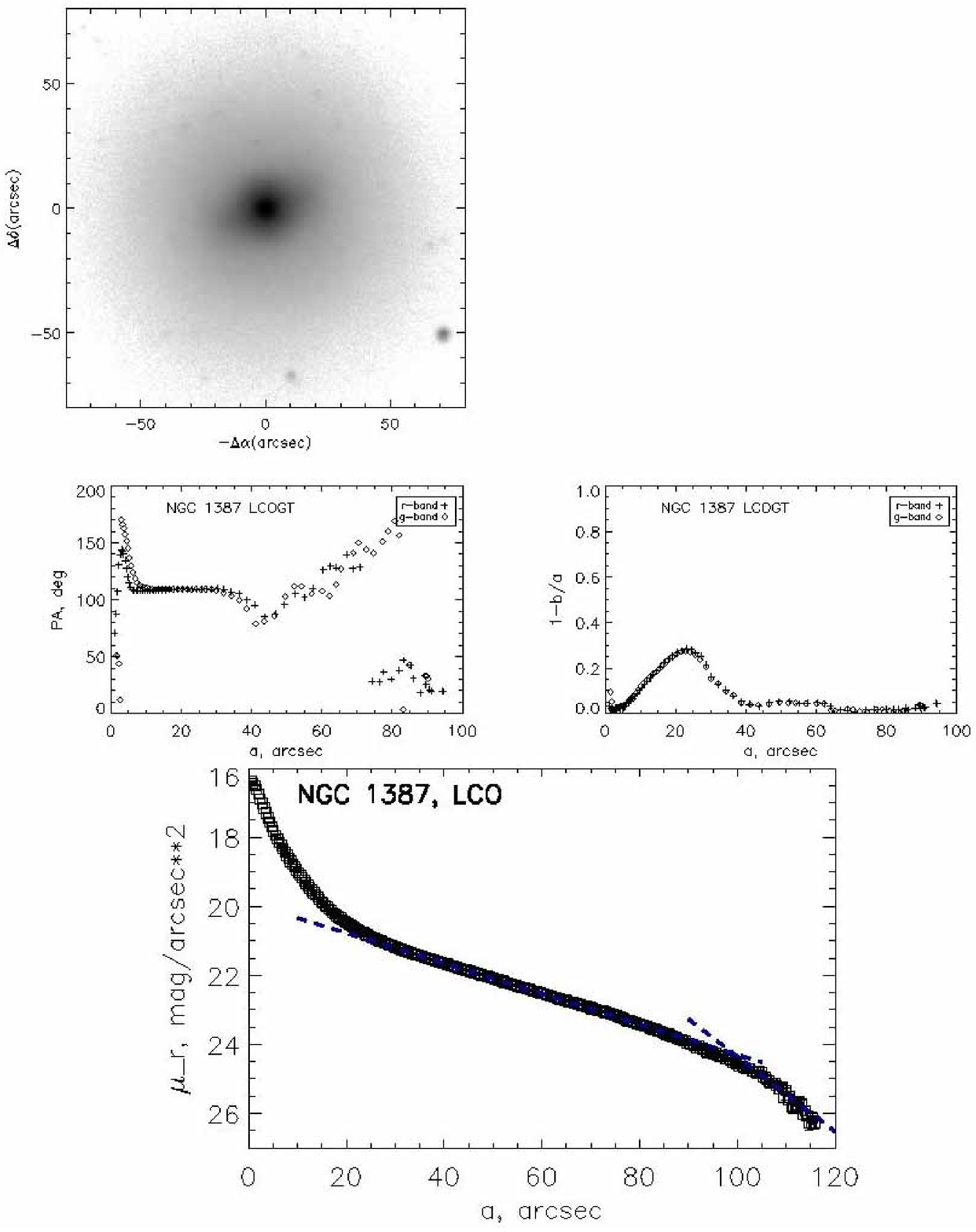}
\end{figure*}

\clearpage

\bigskip

\begin{figure*}
\centering
\includegraphics[width=\textwidth]{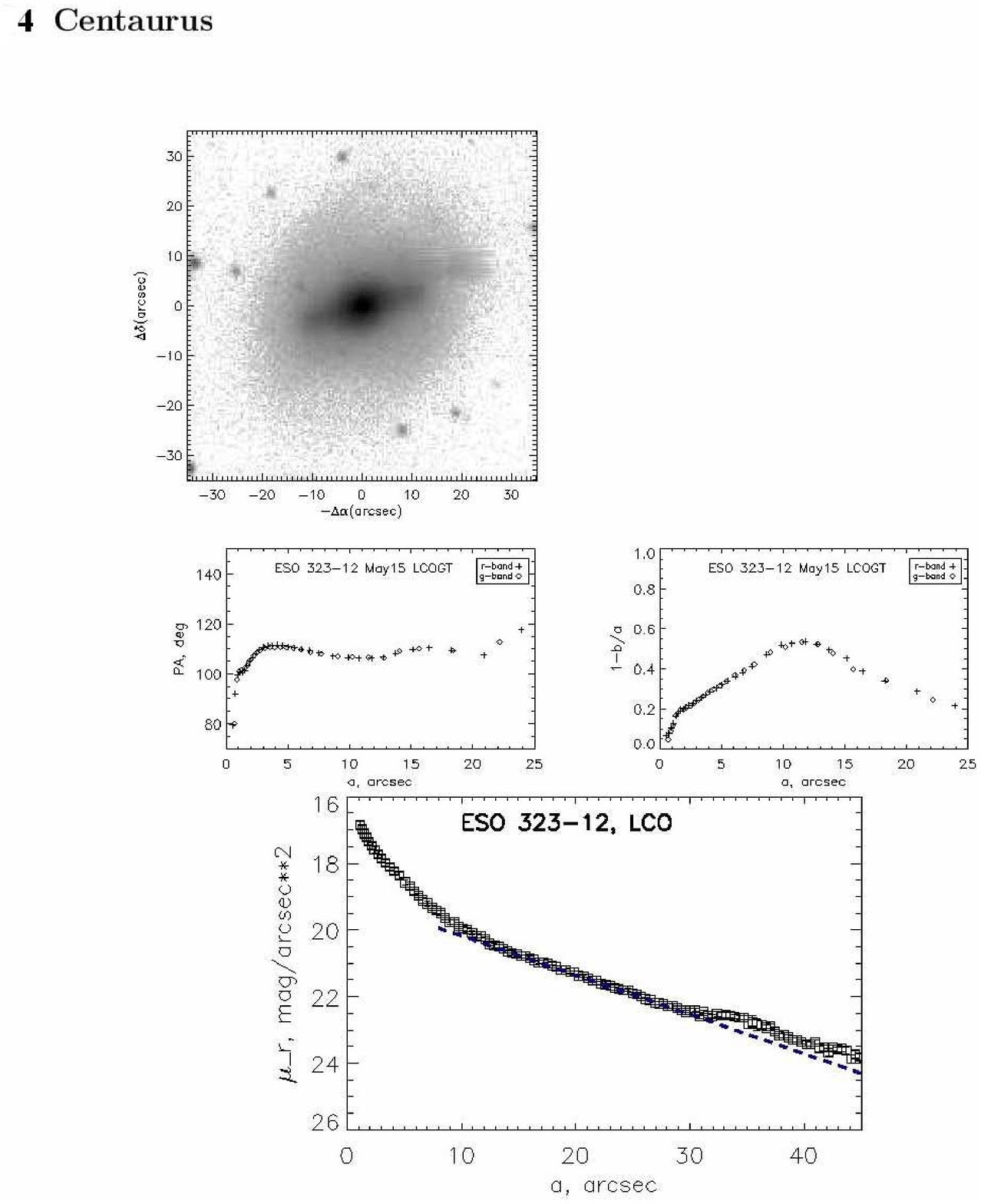}
\end{figure*}

\clearpage

\bigskip
\bigskip

\begin{figure*}
\centering
\includegraphics[width=\textwidth]{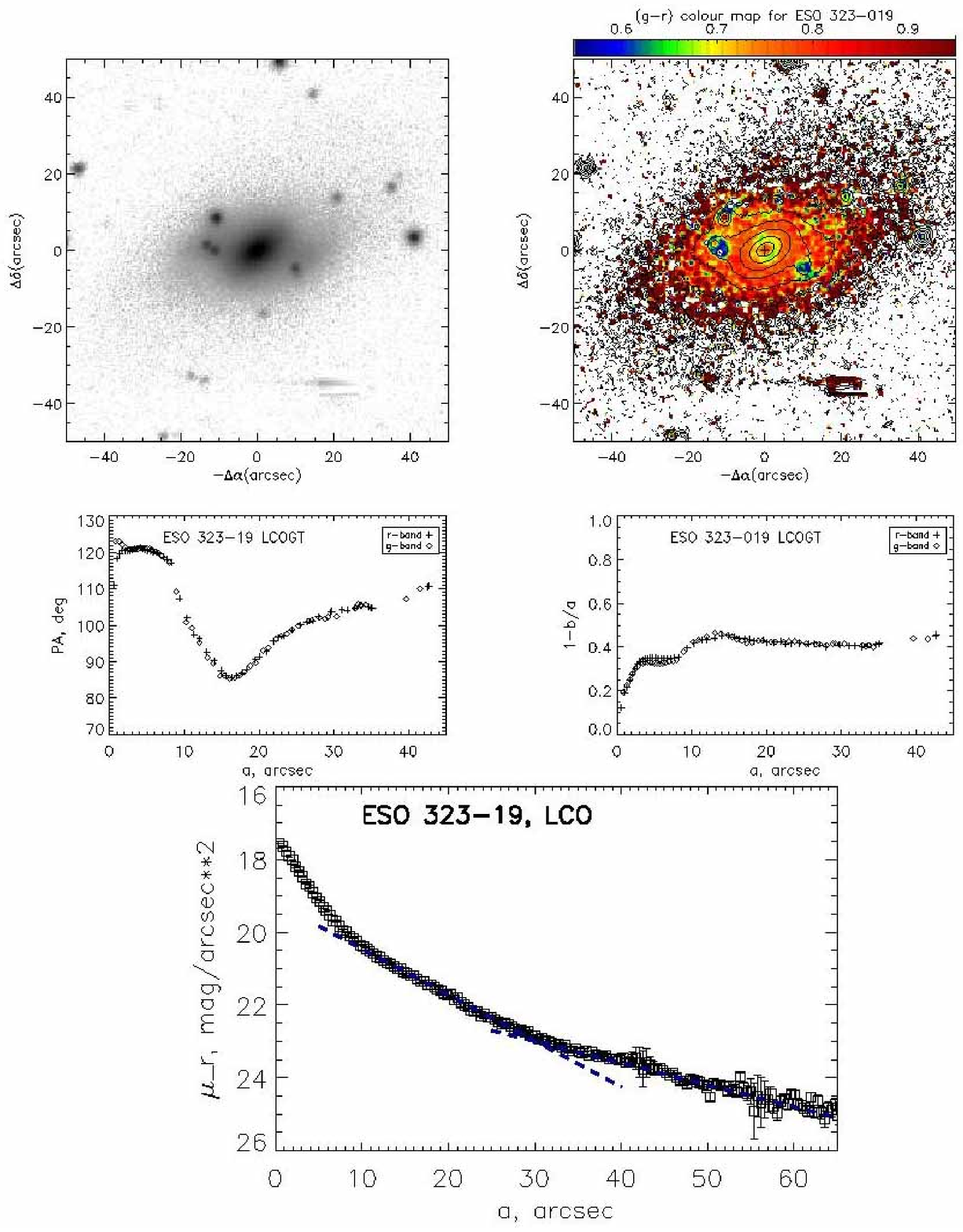}
\end{figure*}

\clearpage

\bigskip
\bigskip

\begin{figure*}
\centering
\includegraphics[width=\textwidth]{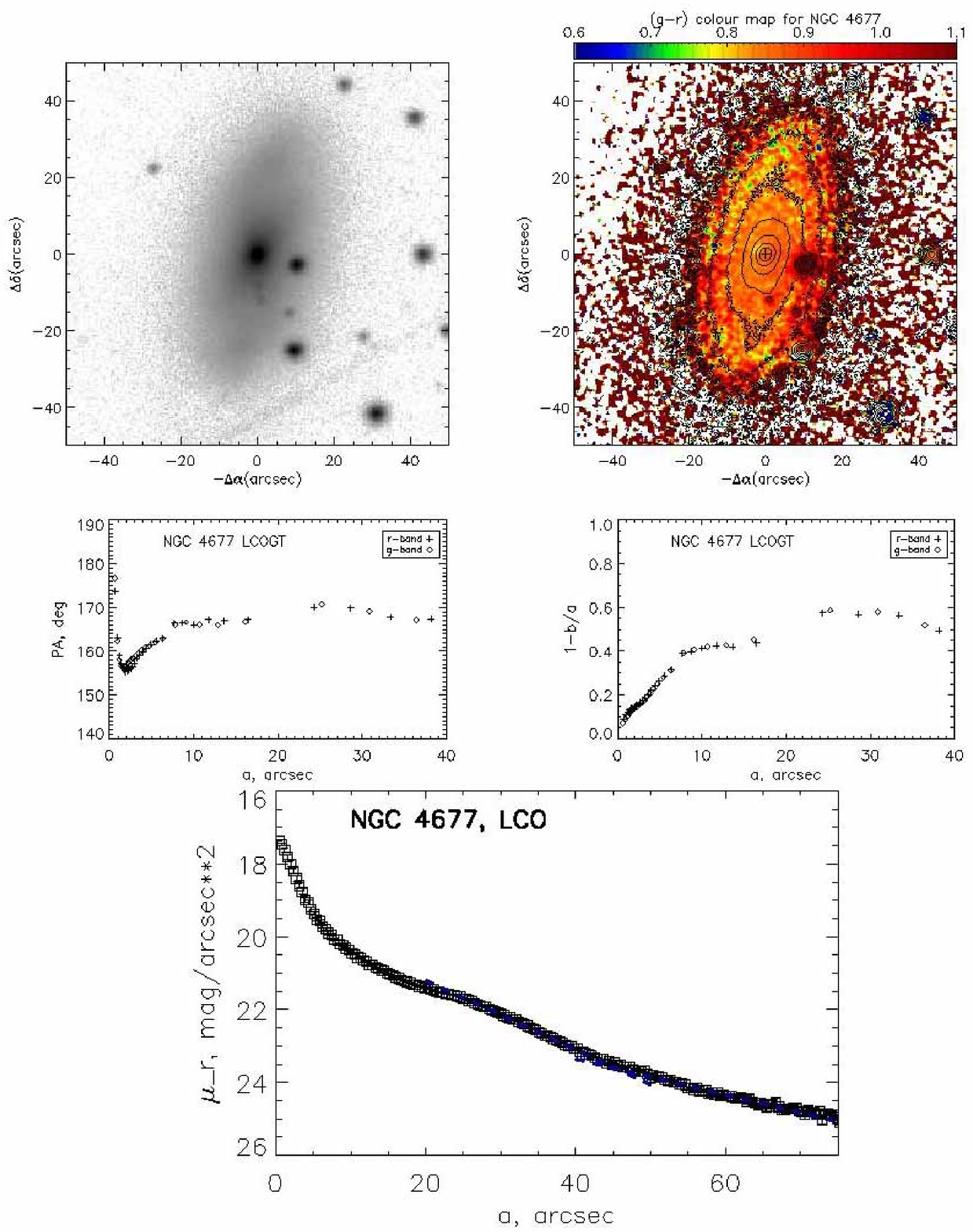}
\end{figure*}

\clearpage

\bigskip
\bigskip

\begin{figure*}
\centering
\includegraphics[width=\textwidth]{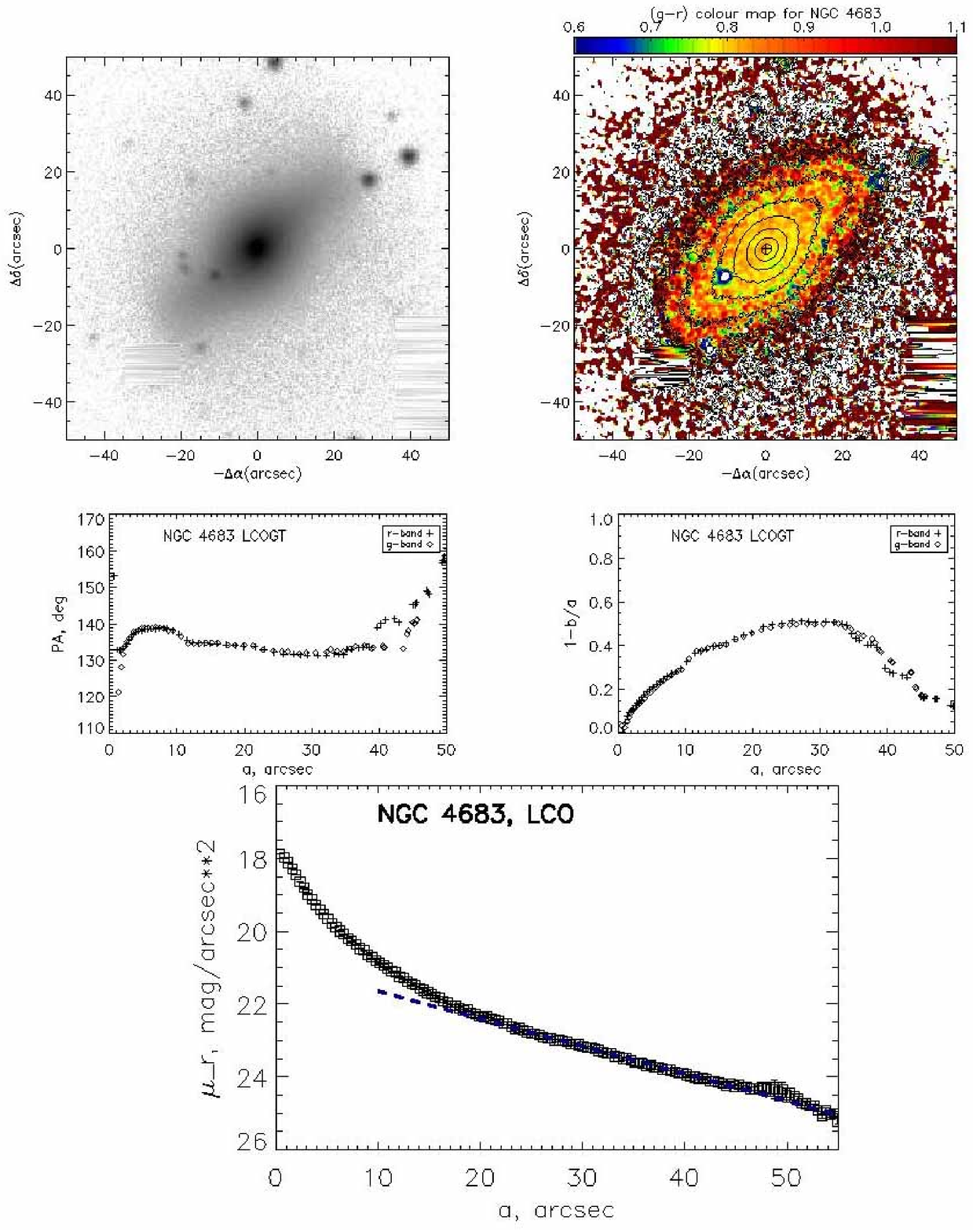}
\end{figure*}

\clearpage

\bigskip
\bigskip

\begin{figure*}
\centering
\includegraphics[width=\textwidth]{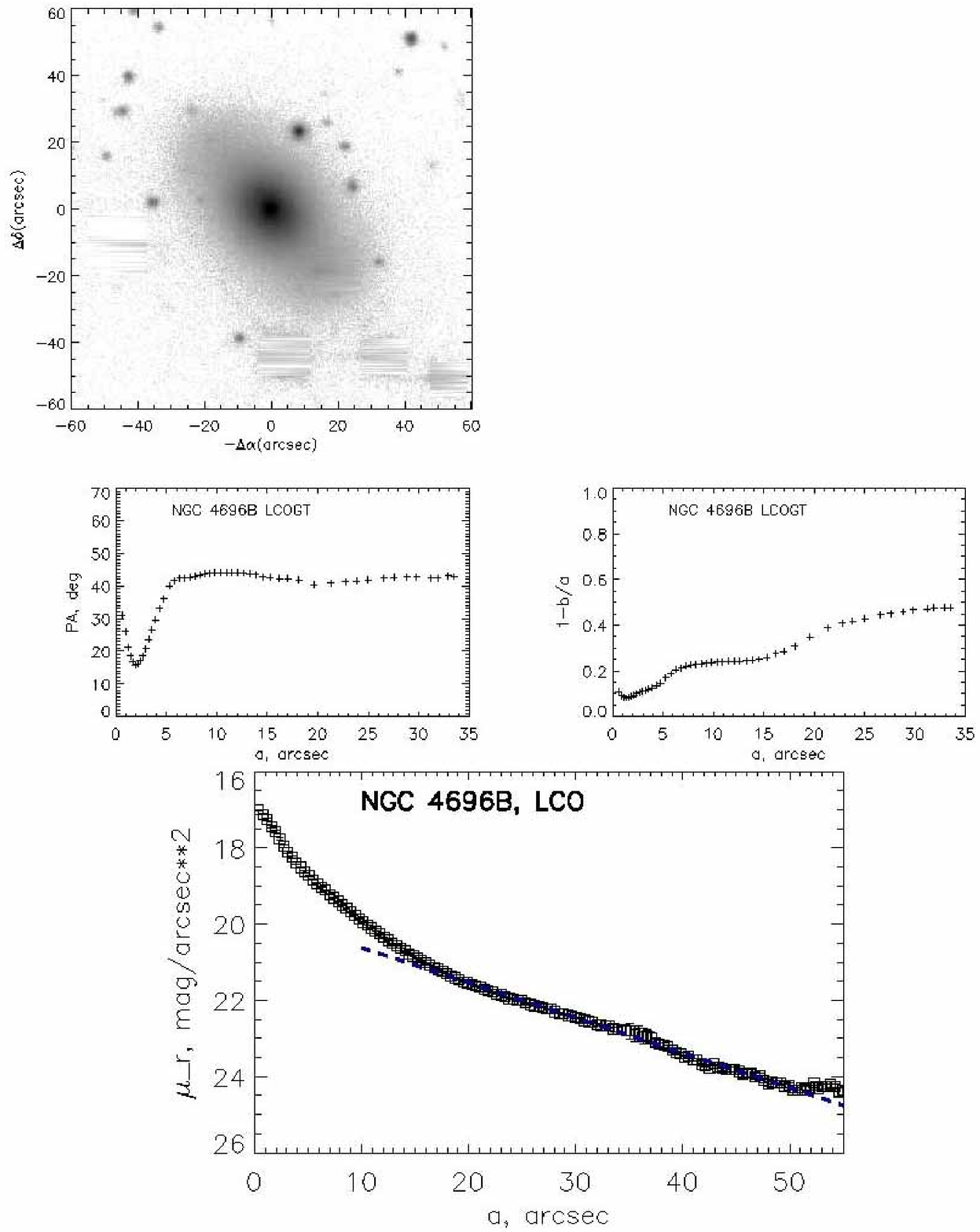}
\end{figure*}

\clearpage

\bigskip
\bigskip

\begin{figure*}
\centering
\includegraphics[width=\textwidth]{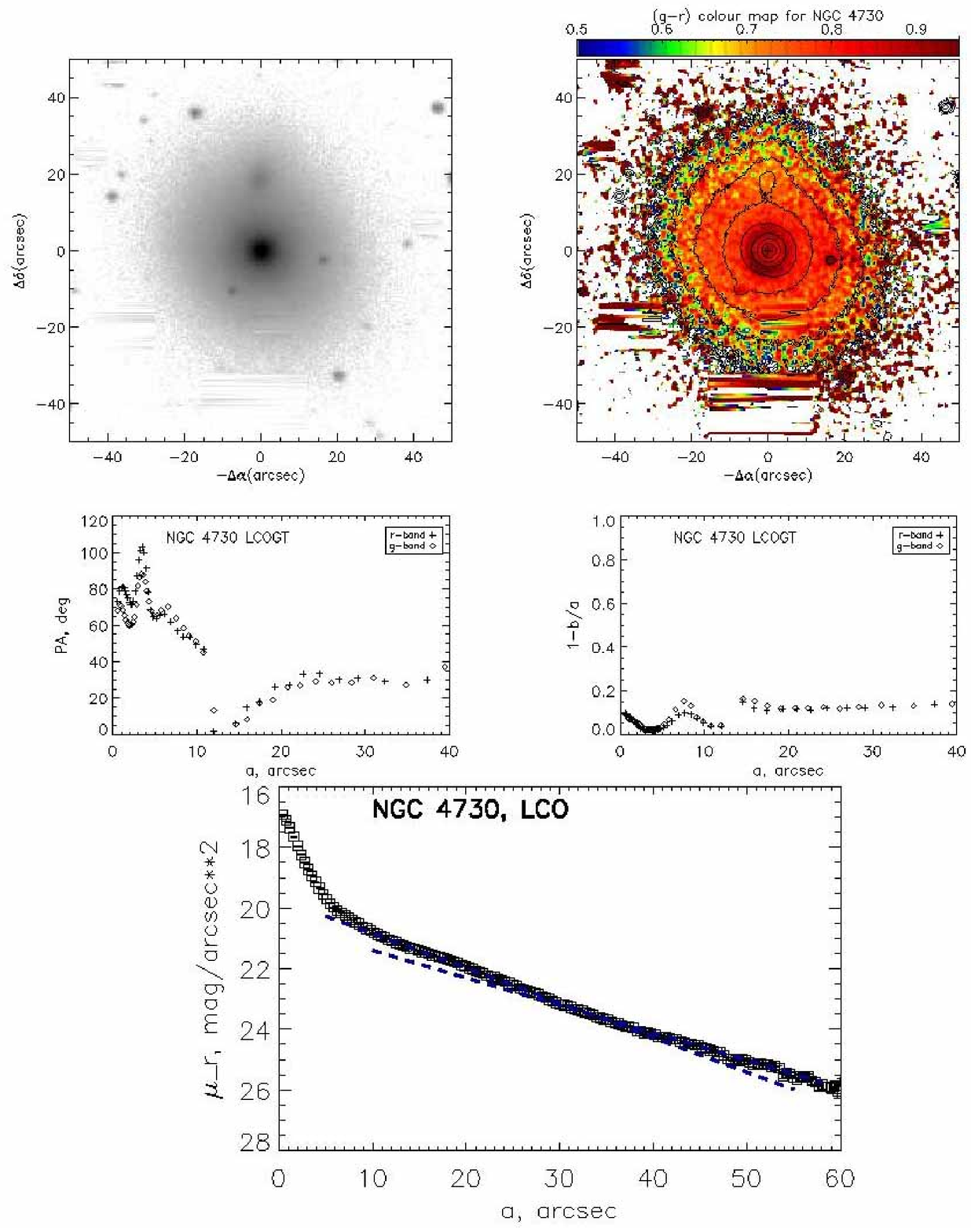}
\end{figure*}

\clearpage

\bigskip
\bigskip

\begin{figure*}
\centering
\includegraphics[width=\textwidth]{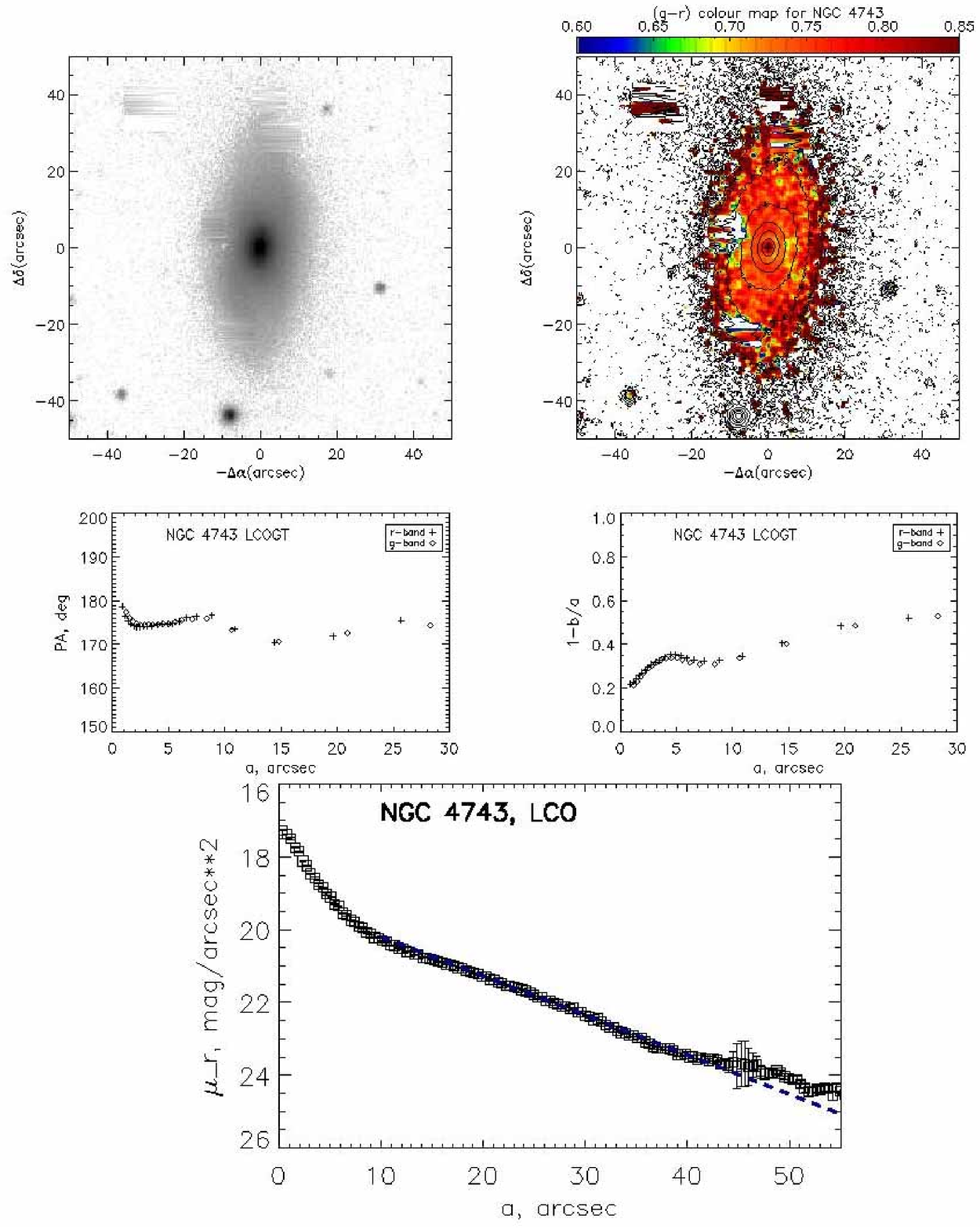}
\end{figure*}

\clearpage

\bigskip
\bigskip

\begin{figure*}
\centering
\includegraphics[width=\textwidth]{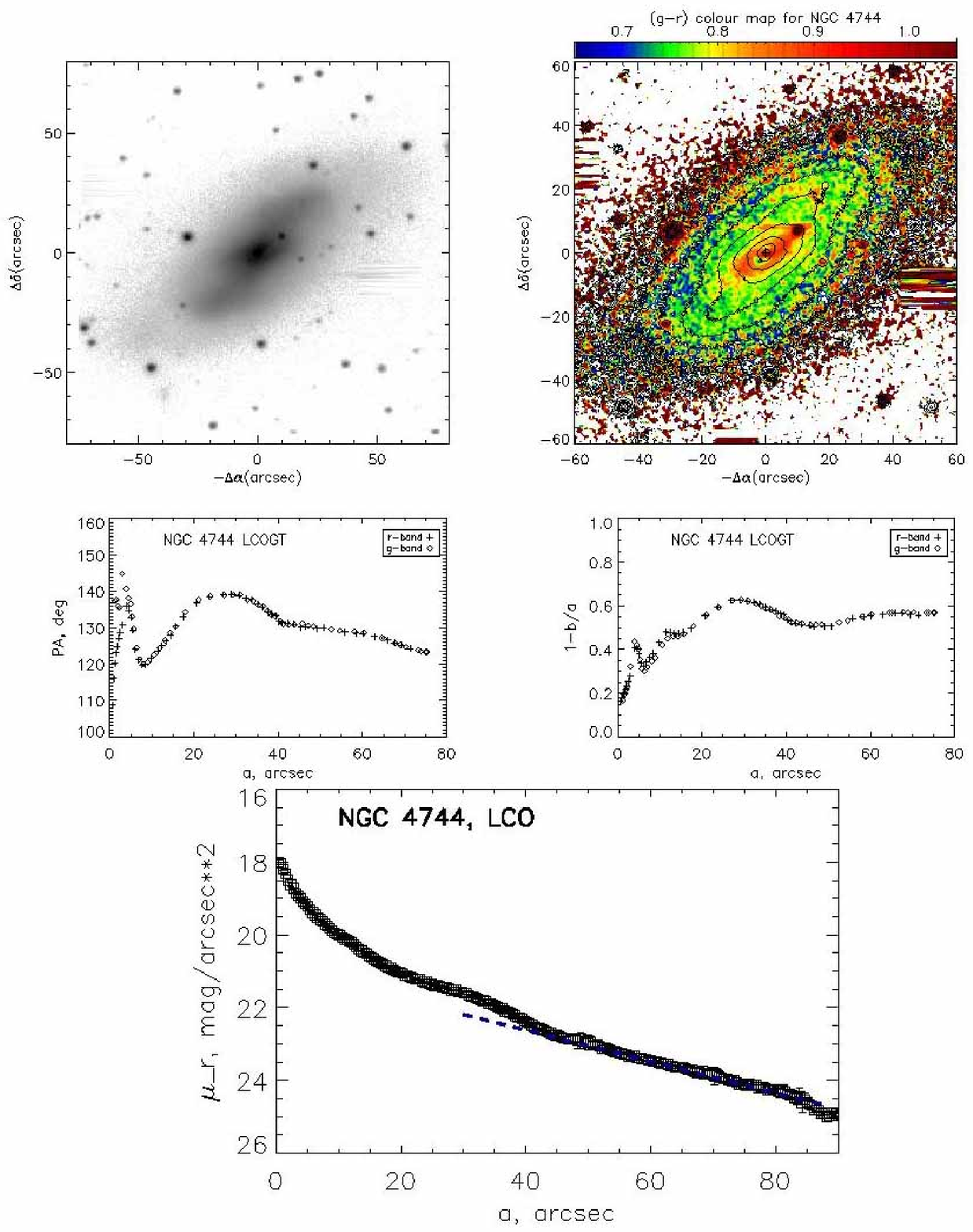}
\end{figure*}

\clearpage

\bigskip
\bigskip

\begin{figure*}
\centering
\includegraphics[width=\textwidth]{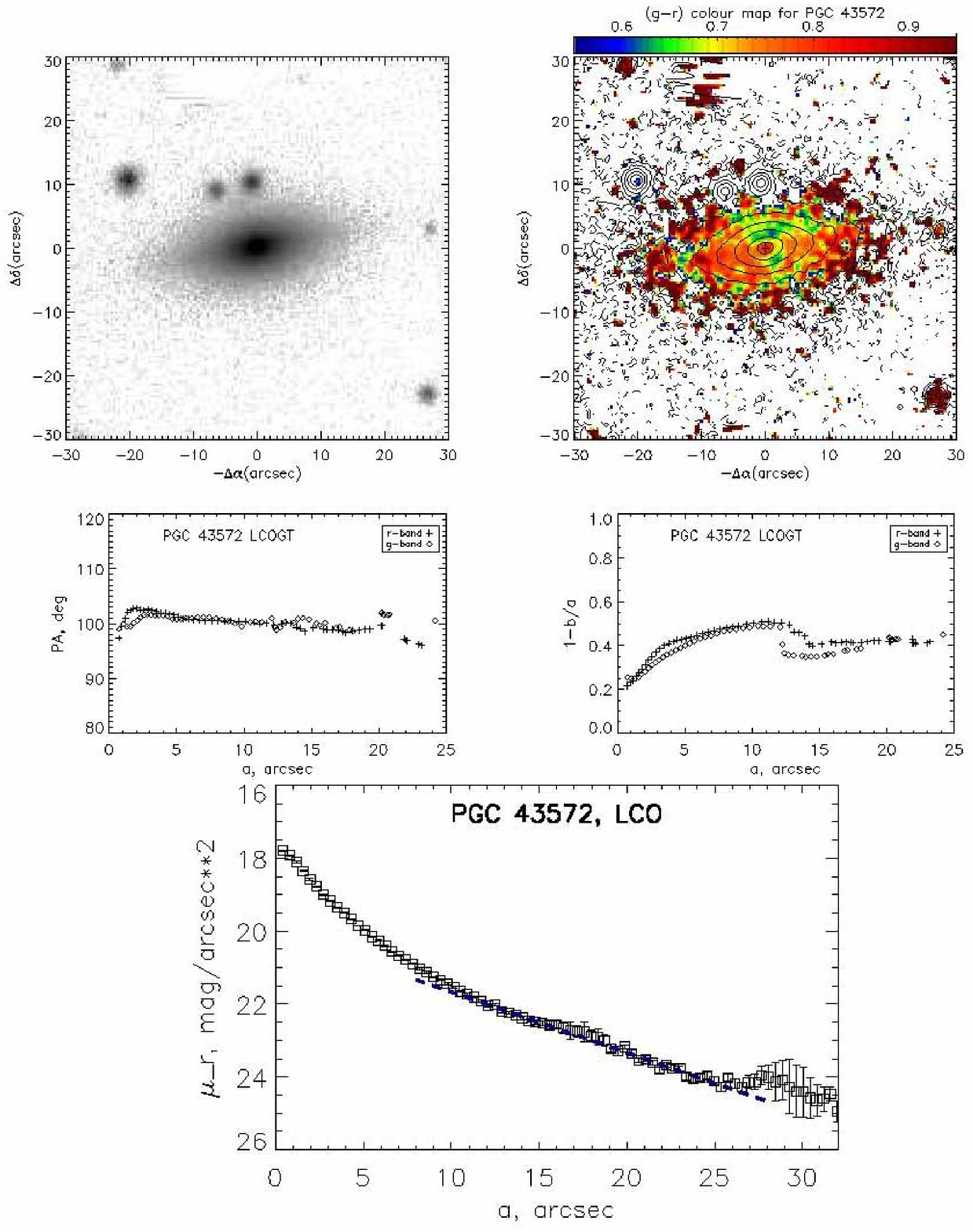}
\end{figure*}

\clearpage

\bigskip
\bigskip

\begin{figure*}
\centering
\includegraphics[width=\textwidth]{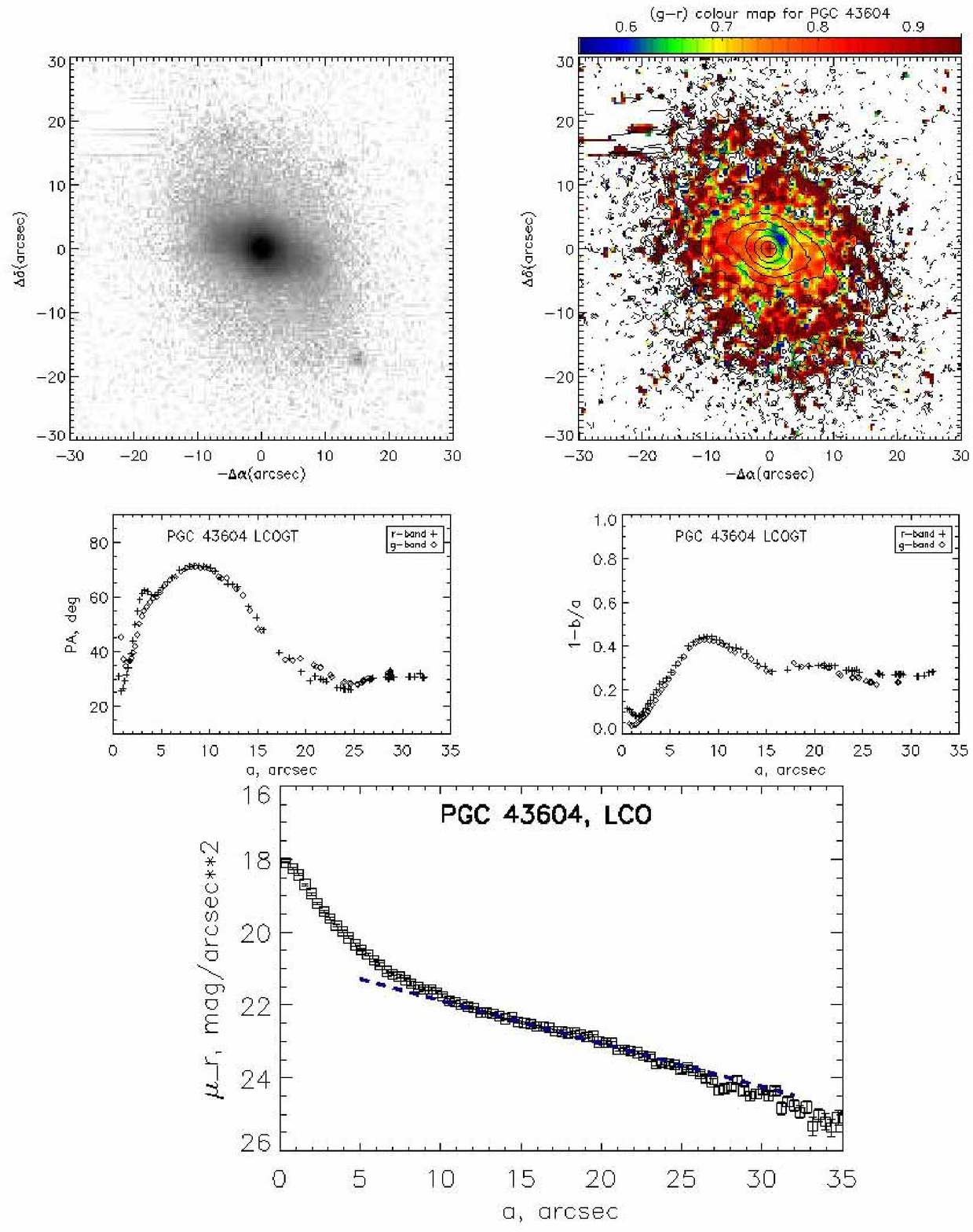}
\end{figure*}

\clearpage

\bigskip
\bigskip

\begin{figure*}
\centering
\includegraphics[width=\textwidth]{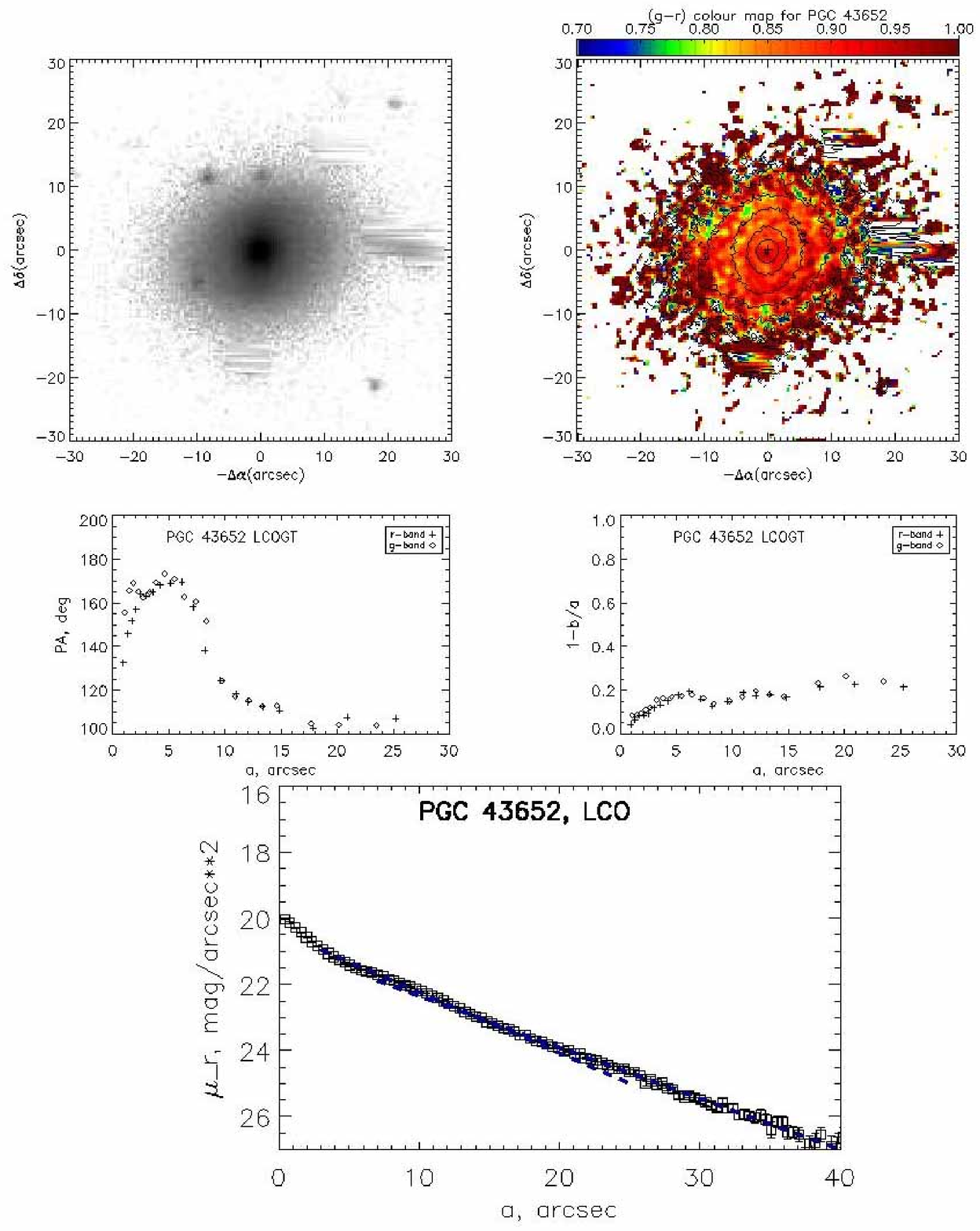}
\end{figure*}

\clearpage

\bigskip

\begin{figure*}
\centering
\includegraphics[width=\textwidth]{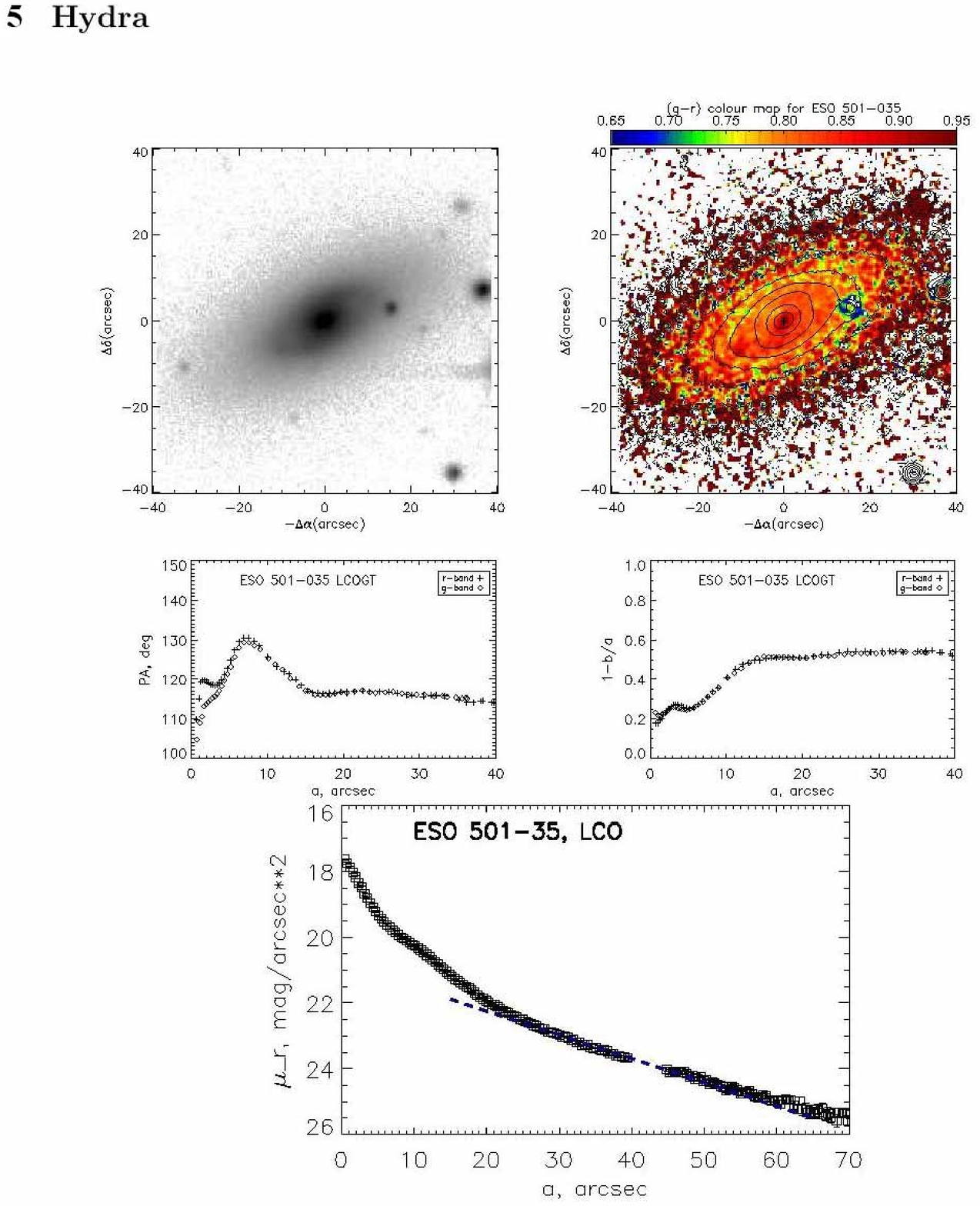}
\end{figure*}

\clearpage

\bigskip
\bigskip

\begin{figure*}
\centering
\includegraphics[width=\textwidth]{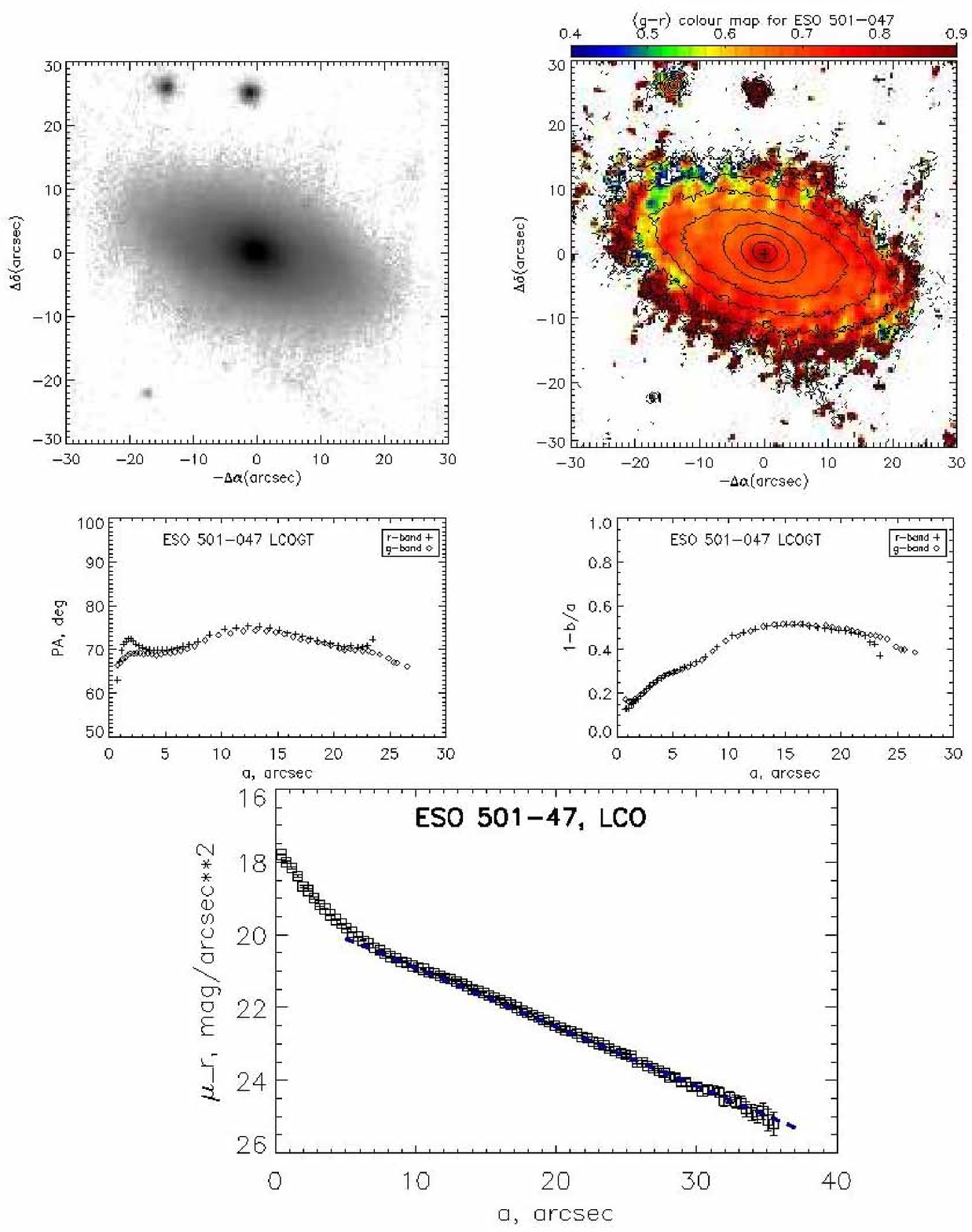}
\end{figure*}

\clearpage

\bigskip
\bigskip

\begin{figure*}
\centering
\includegraphics[width=\textwidth]{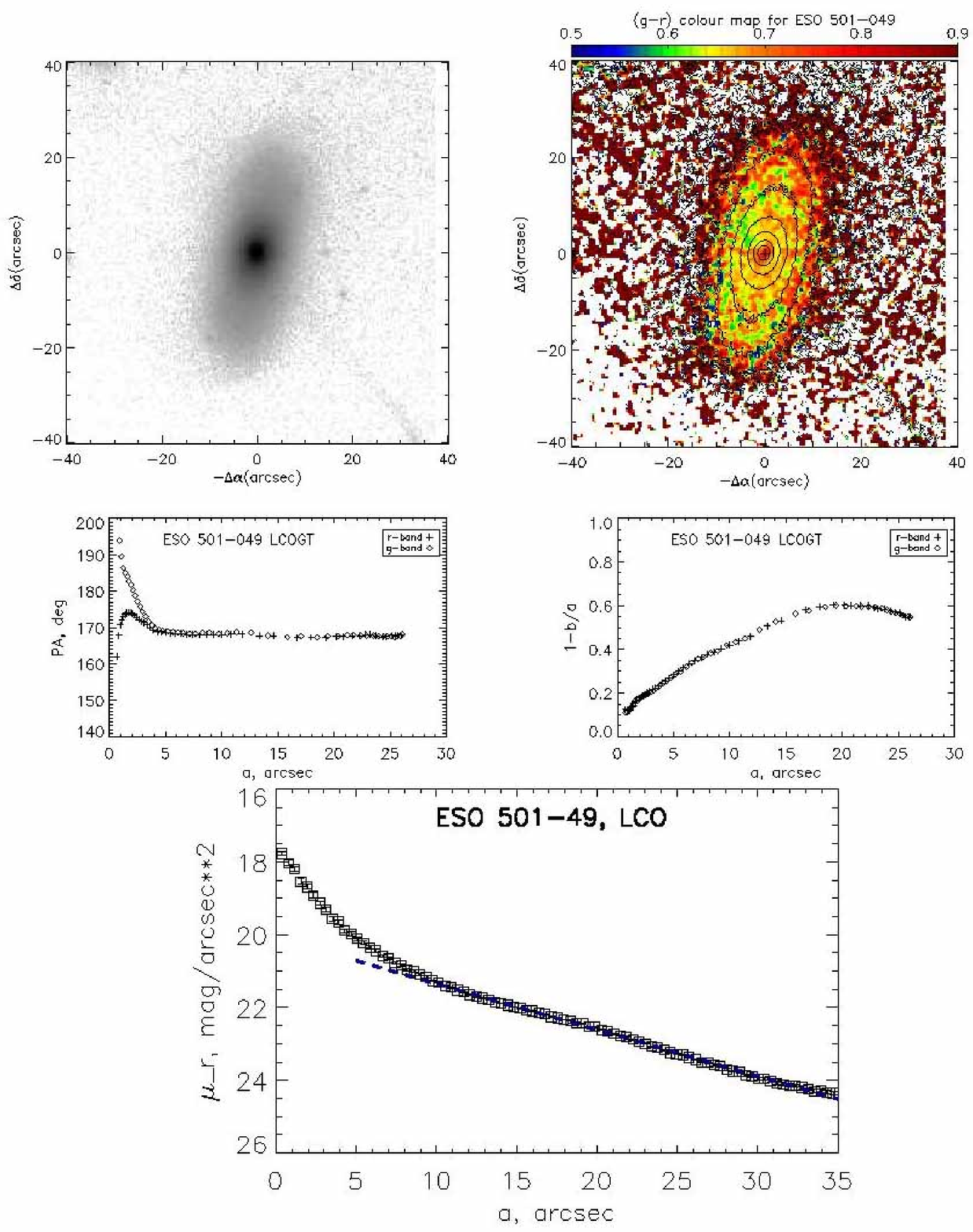}
\end{figure*}

\clearpage

\bigskip
\bigskip

\begin{figure*}
\centering
\includegraphics[width=\textwidth]{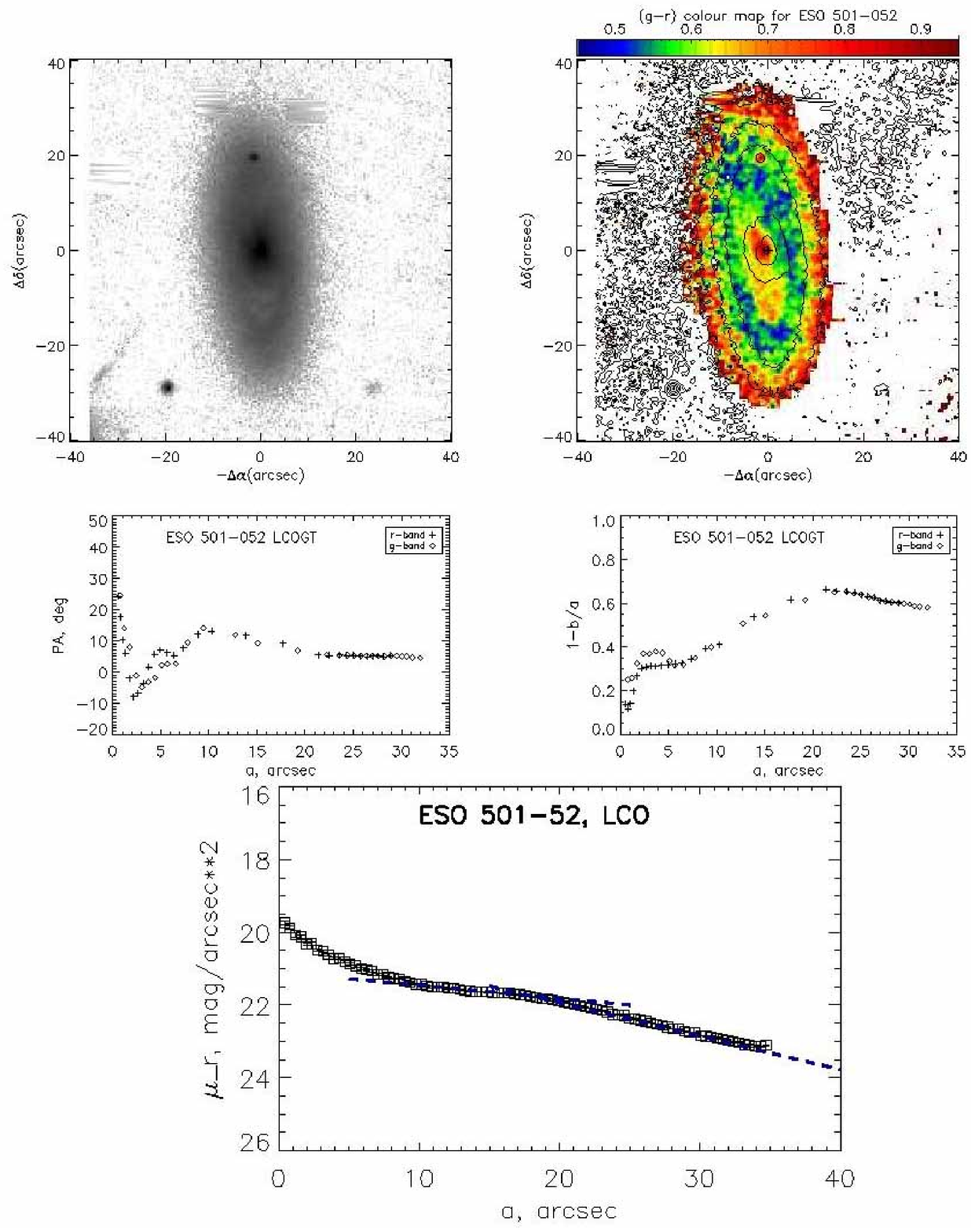}
\end{figure*}

\clearpage

\bigskip
\bigskip

\begin{figure*}
\centering
\includegraphics[width=\textwidth]{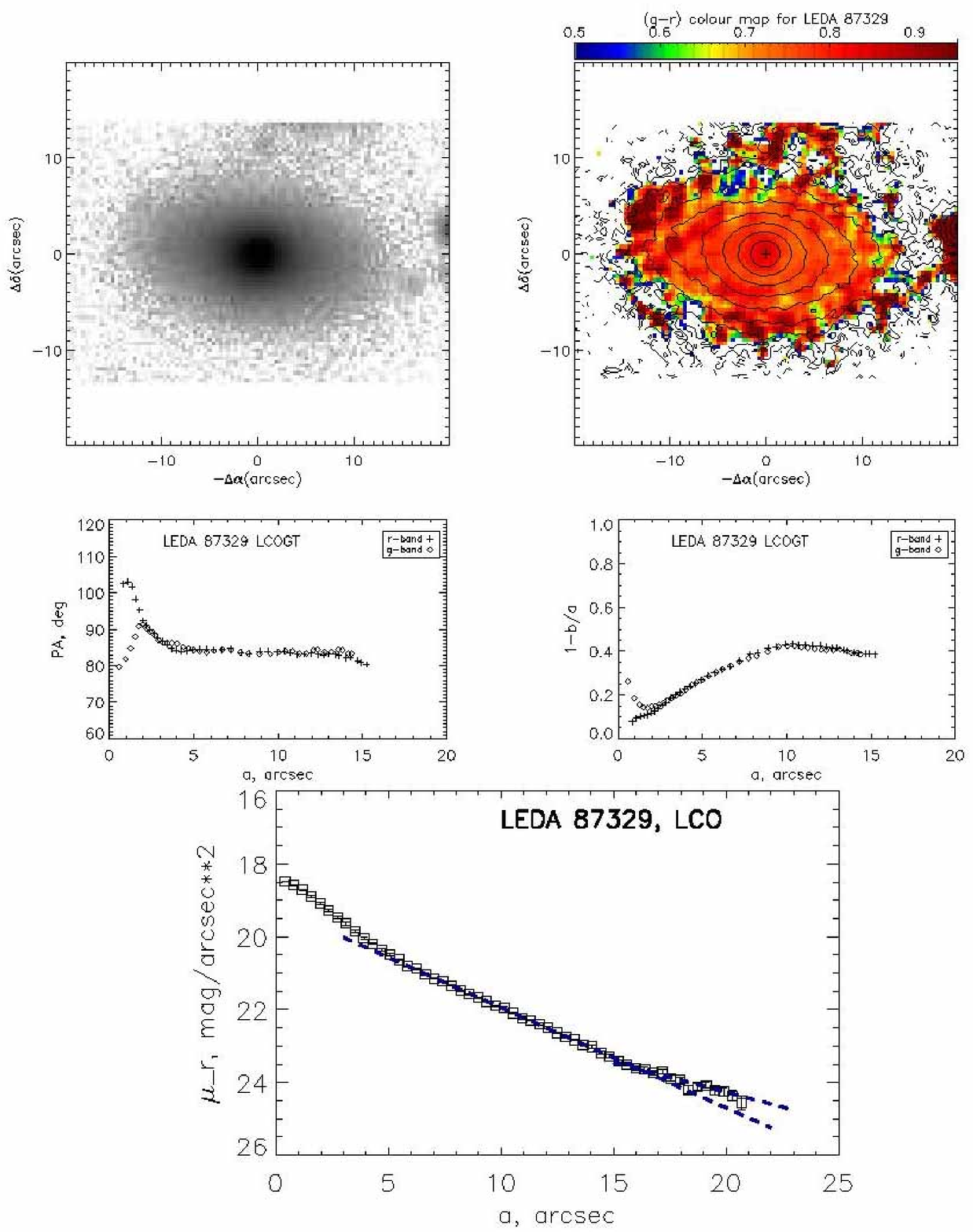}
\end{figure*}

\clearpage

\bigskip
\bigskip

\begin{figure*}
\centering
\includegraphics[width=\textwidth]{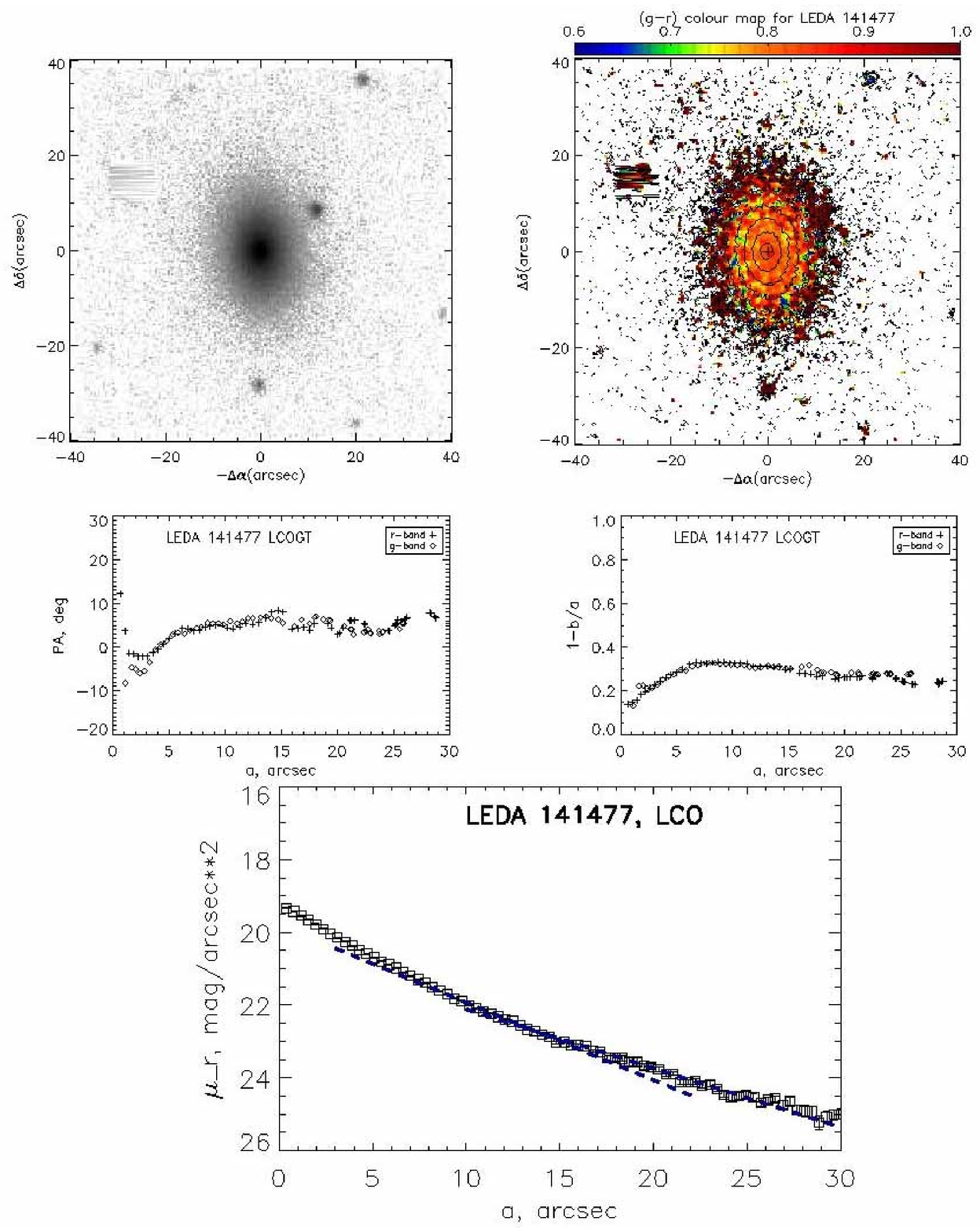}
\end{figure*}

\clearpage

\bigskip
\bigskip

\begin{figure*}
\centering
\includegraphics[width=\textwidth]{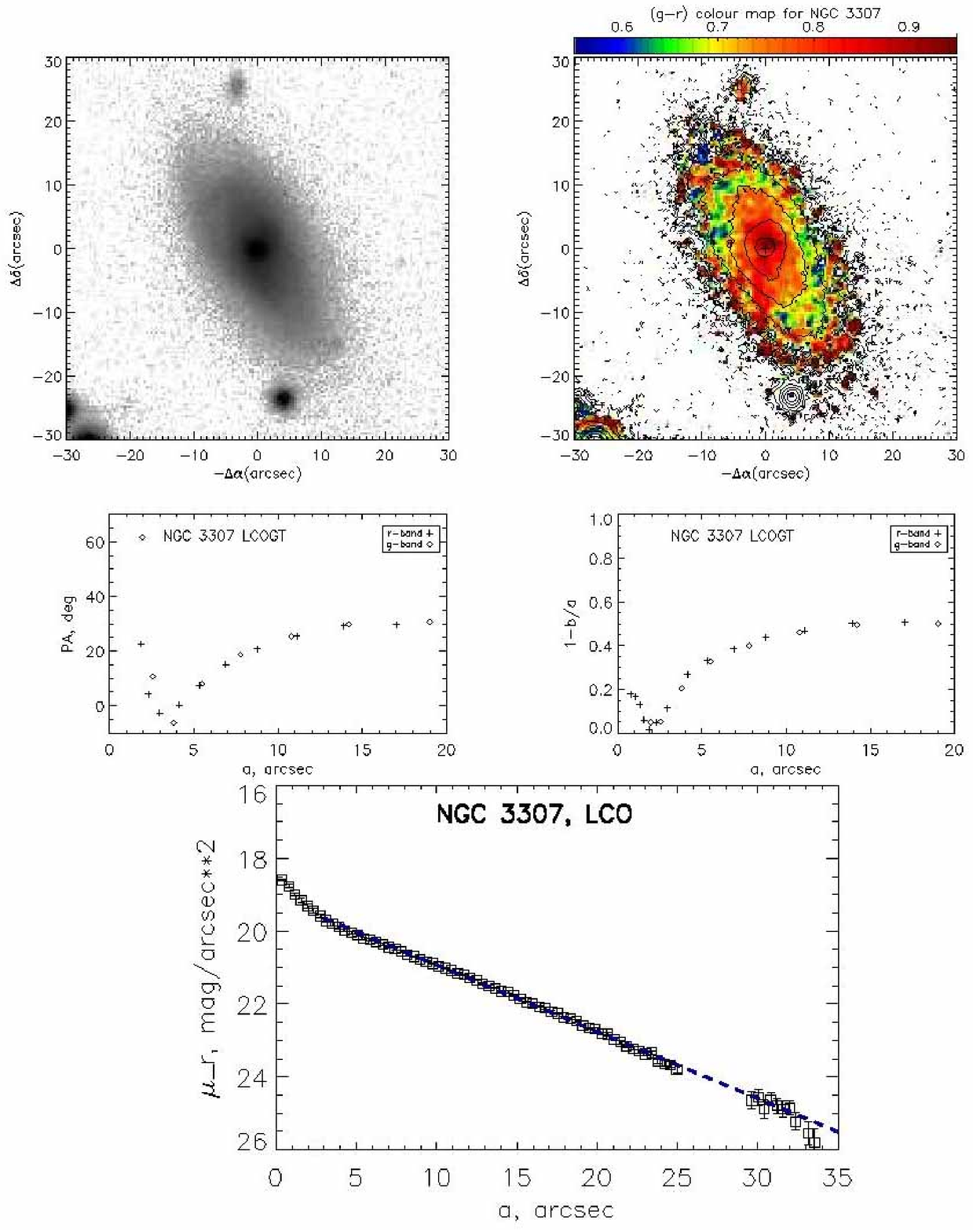}
\end{figure*}

\clearpage

\bigskip
\bigskip

\begin{figure*}
\centering
\includegraphics[width=\textwidth]{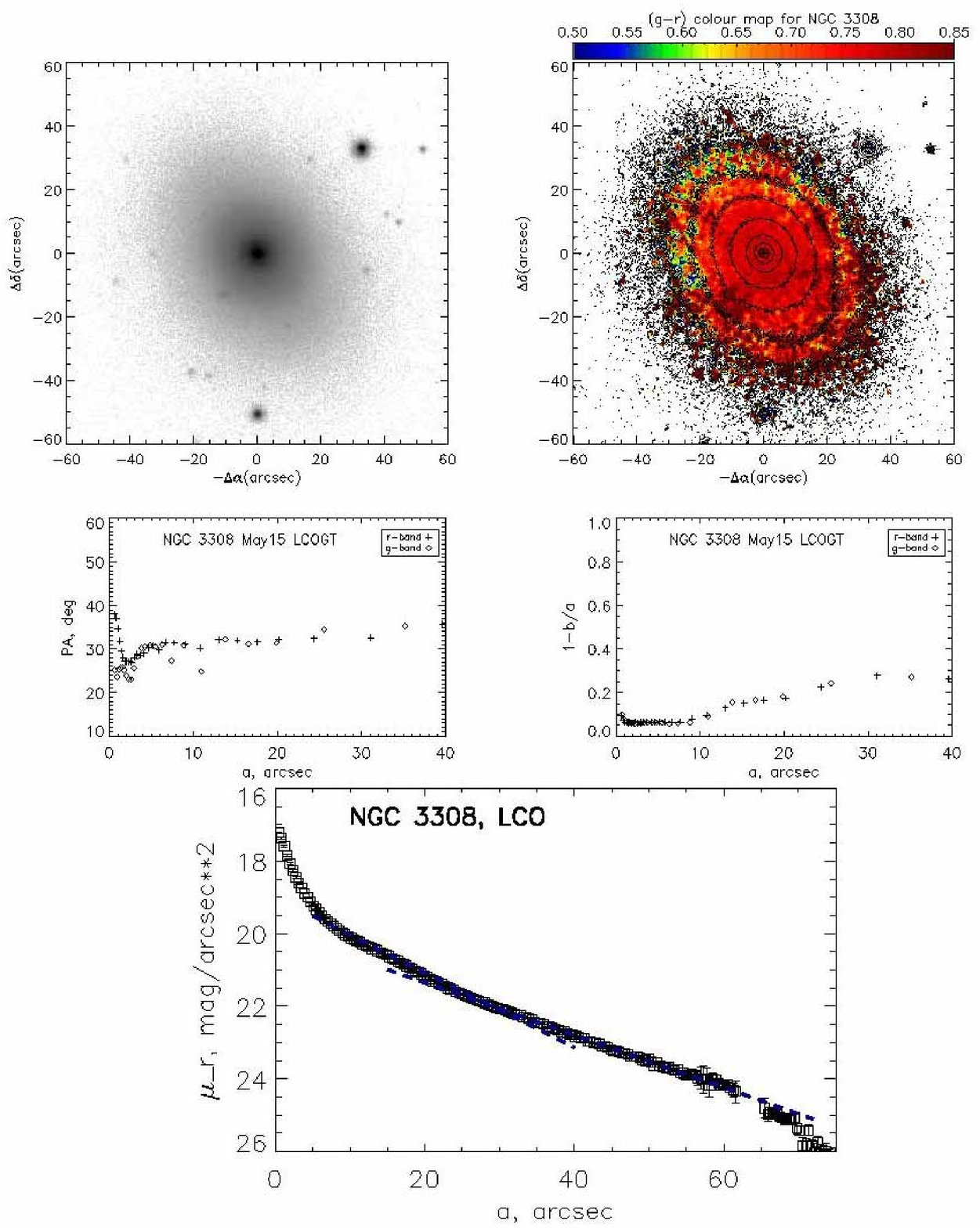}
\end{figure*}

\clearpage

\bigskip
\bigskip

\begin{figure*}
\centering
\includegraphics[width=\textwidth]{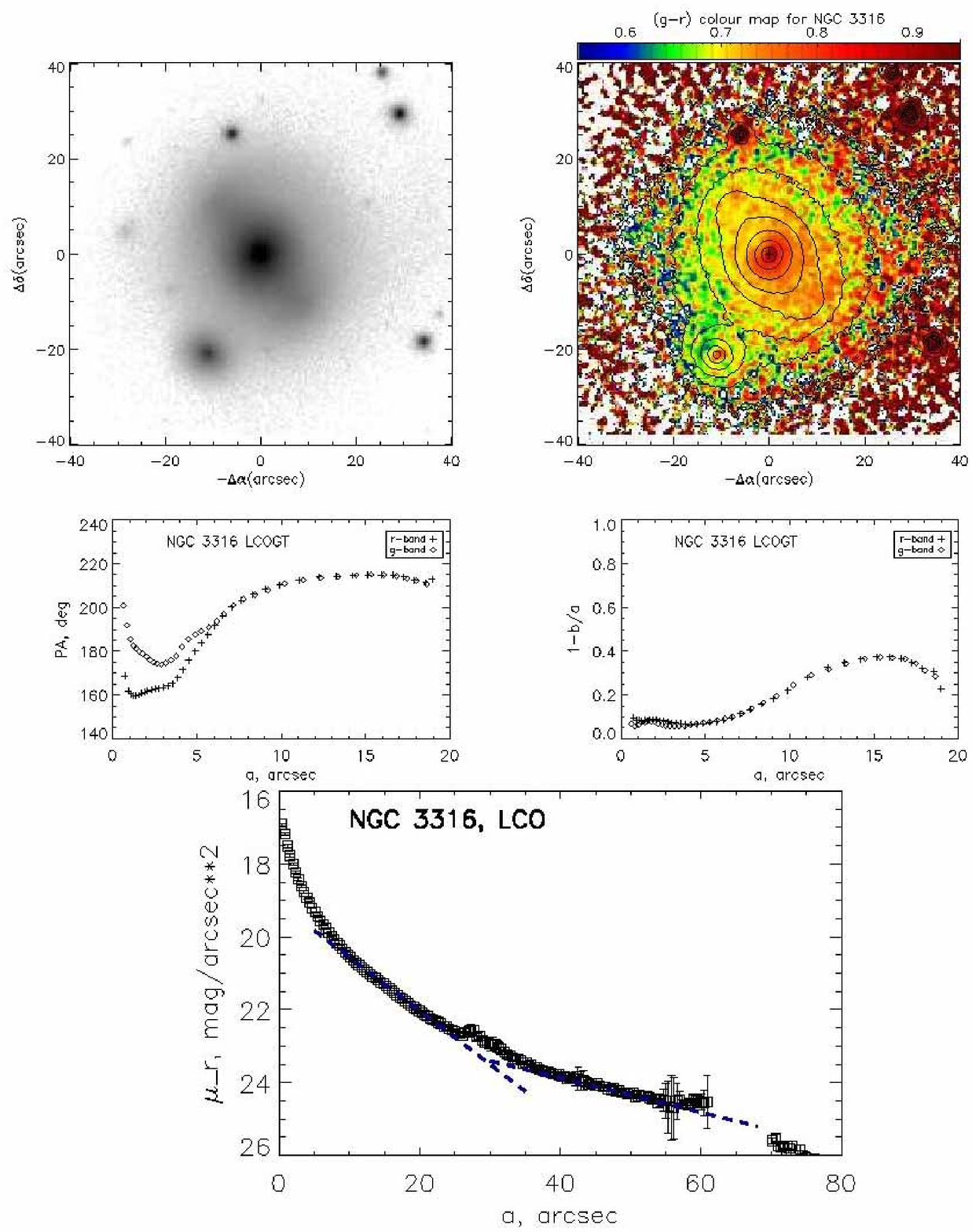}
\end{figure*}

\clearpage

\bigskip
\bigskip

\begin{figure*}
\centering
\includegraphics[width=\textwidth]{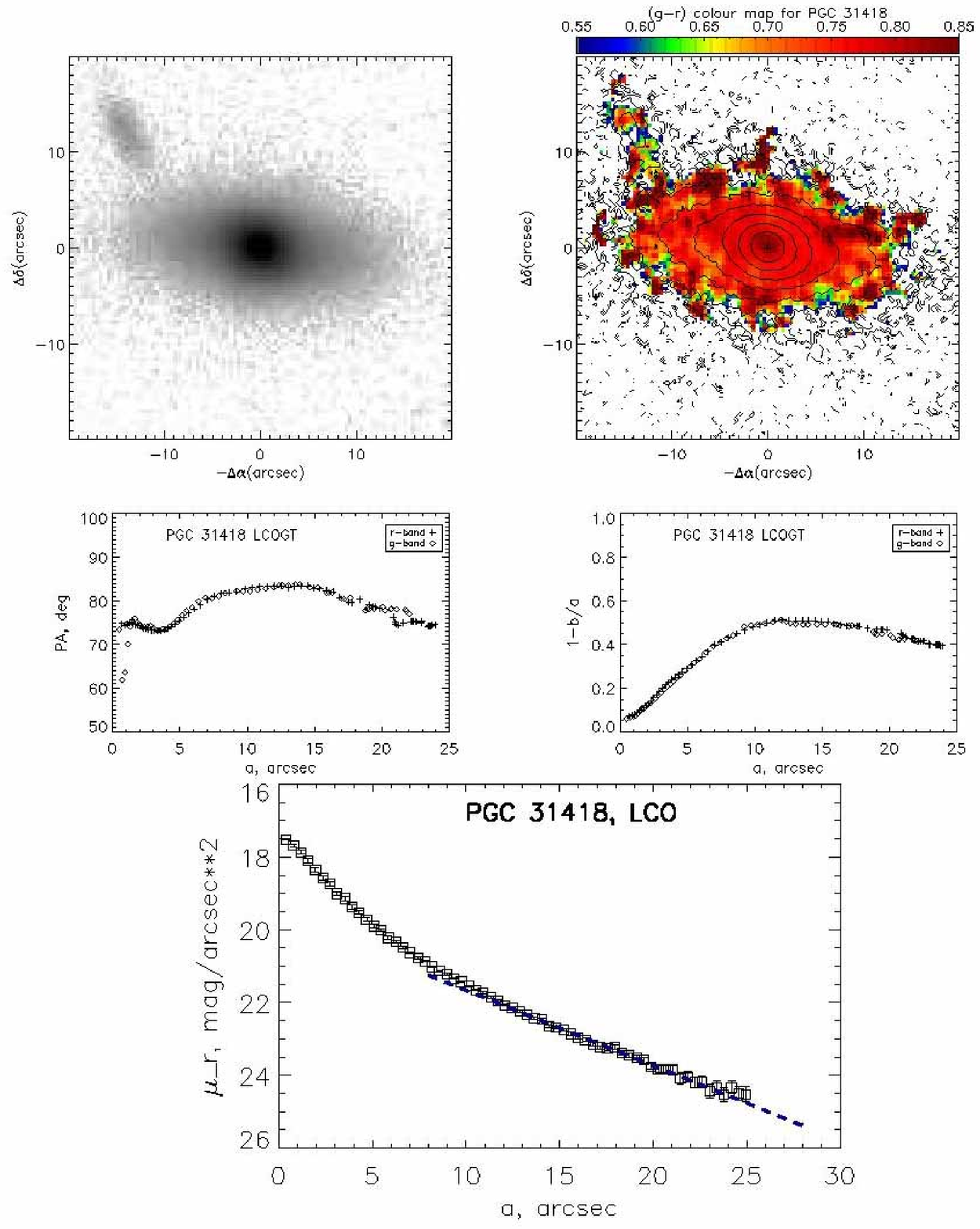}
\end{figure*}

\clearpage

\bigskip
\bigskip

\begin{figure*}
\centering
\includegraphics[width=\textwidth]{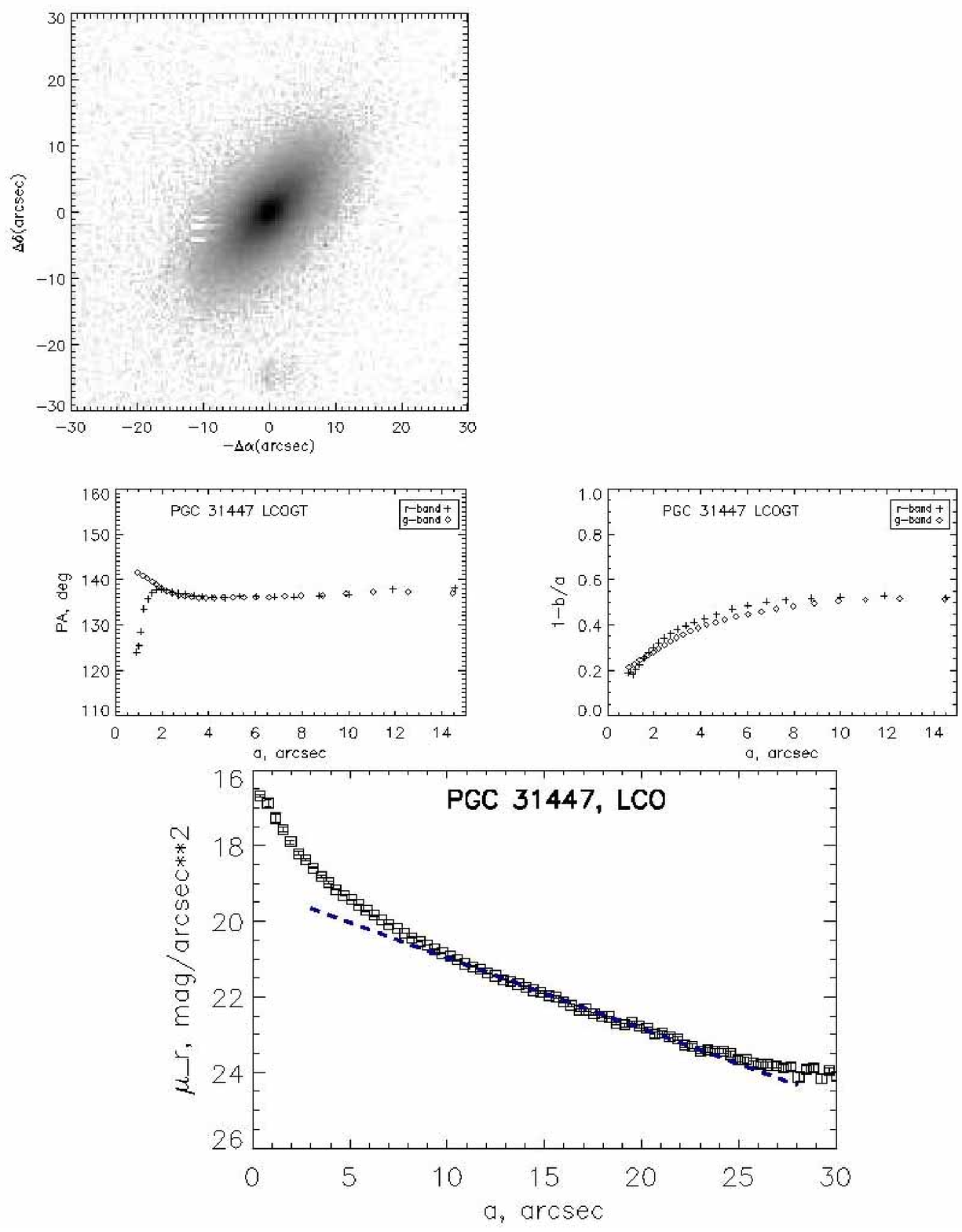}
\end{figure*}

\clearpage

\bigskip
\bigskip

\begin{figure*}
\centering
\includegraphics[width=\textwidth]{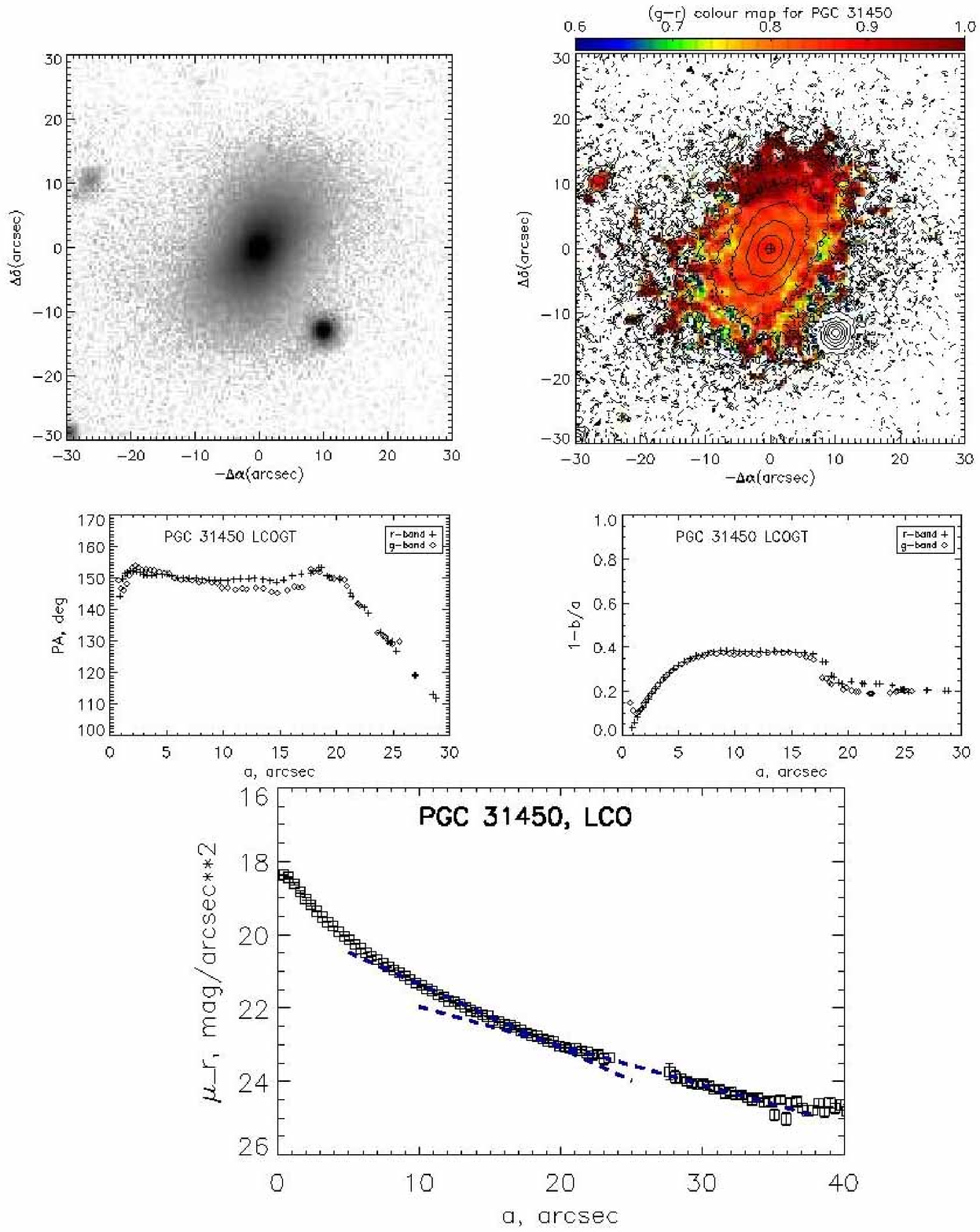}
\end{figure*}

\clearpage

\bigskip
\bigskip

\begin{figure*}
\centering
\includegraphics[width=\textwidth]{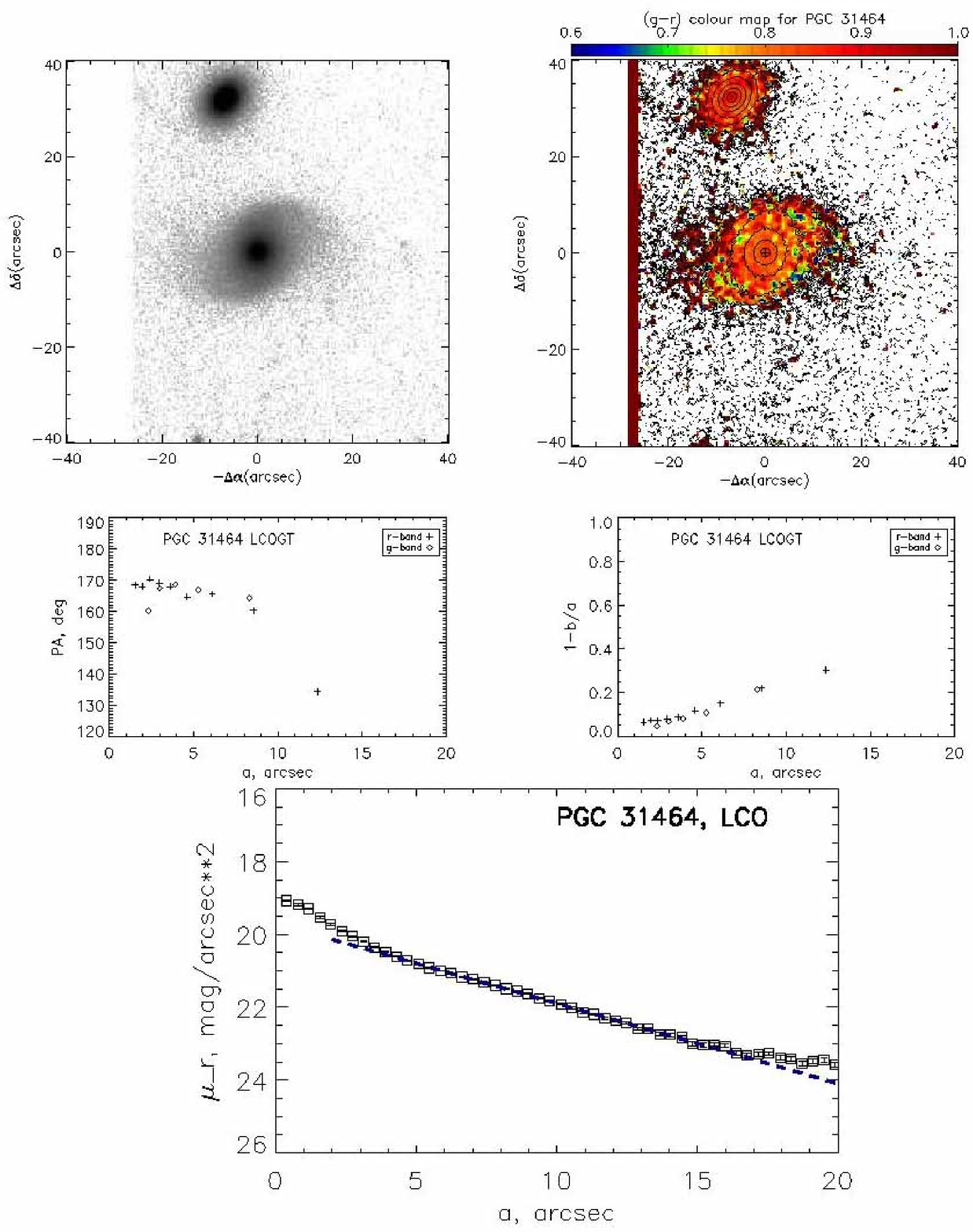}
\end{figure*}

\clearpage

\bigskip

\begin{figure*}
\centering
\includegraphics[width=\textwidth]{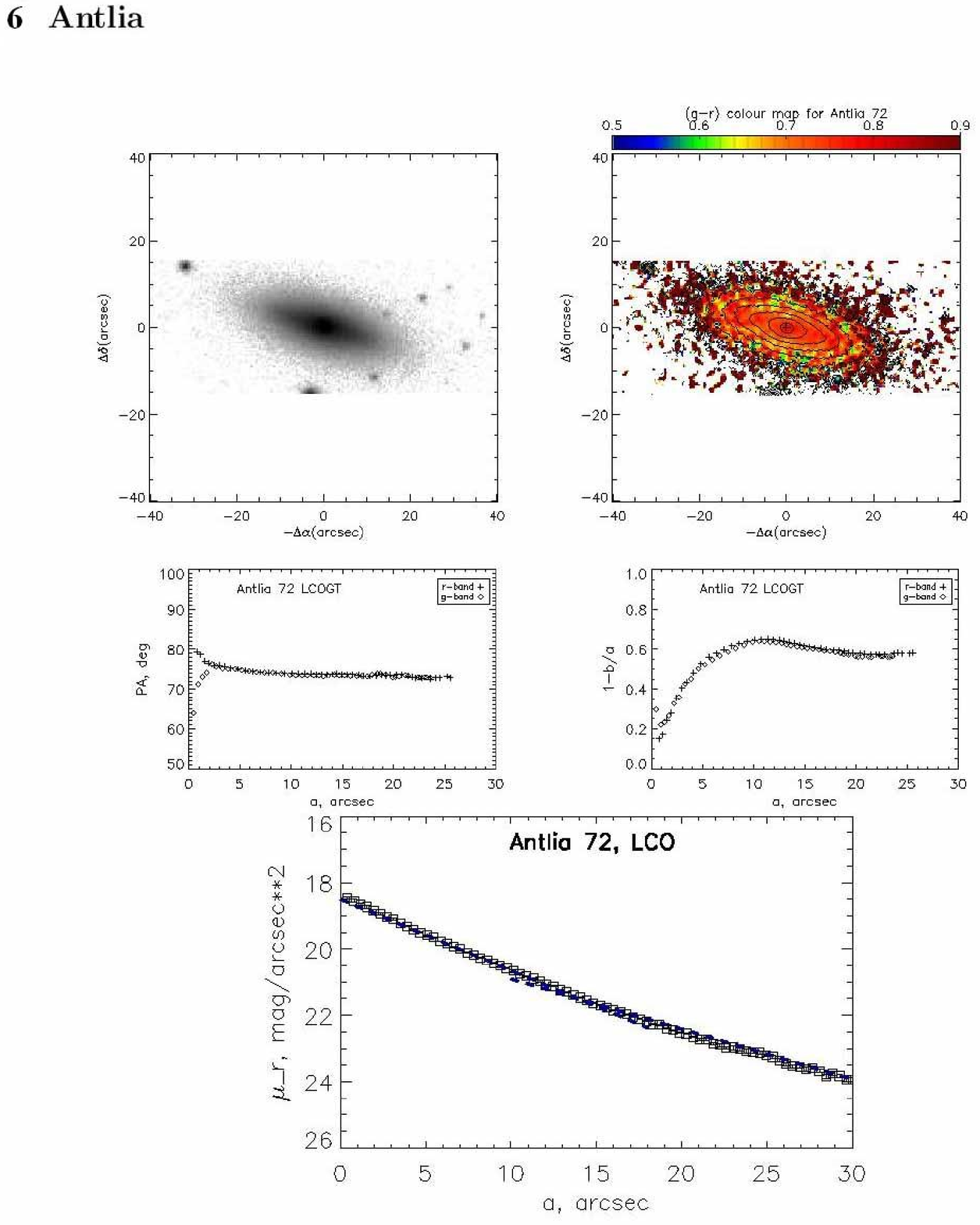}
\end{figure*}

\clearpage

\bigskip
\bigskip

\begin{figure*}
\centering
\includegraphics[width=\textwidth]{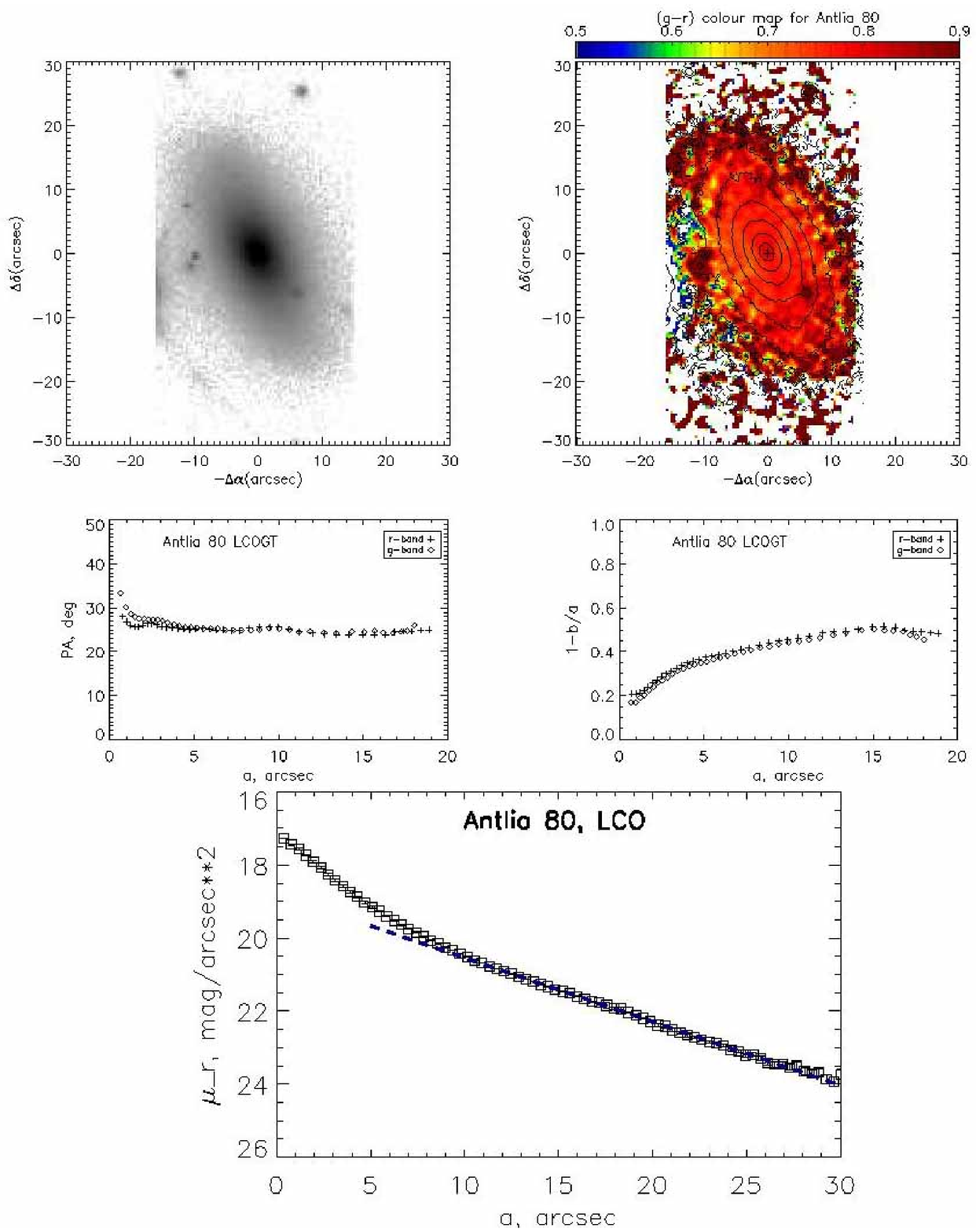}
\end{figure*}

\clearpage

\bigskip
\bigskip

\begin{figure*}
\centering
\includegraphics[width=\textwidth]{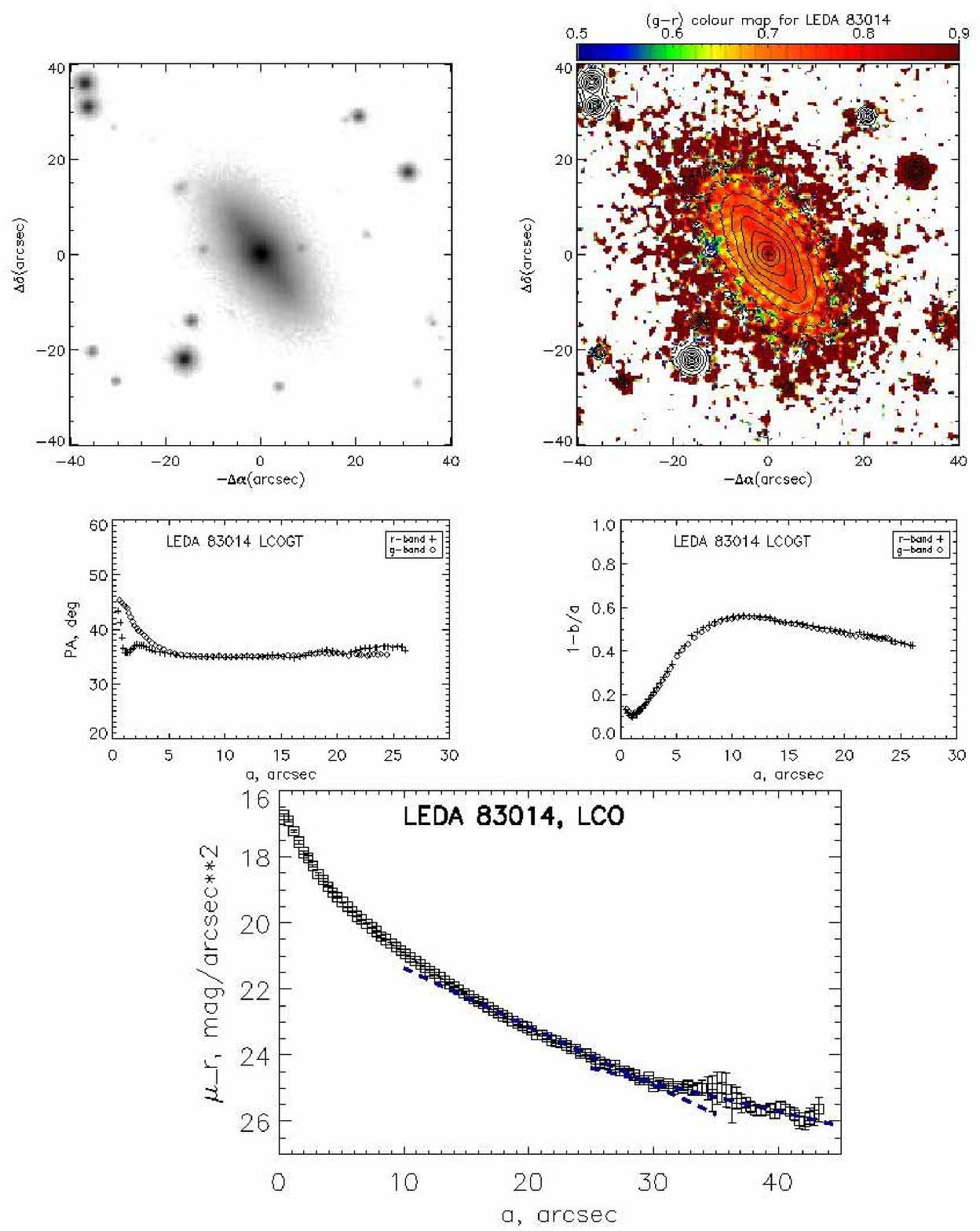}
\end{figure*}

\clearpage

\bigskip
\bigskip

\begin{figure*}
\centering
\includegraphics[width=\textwidth]{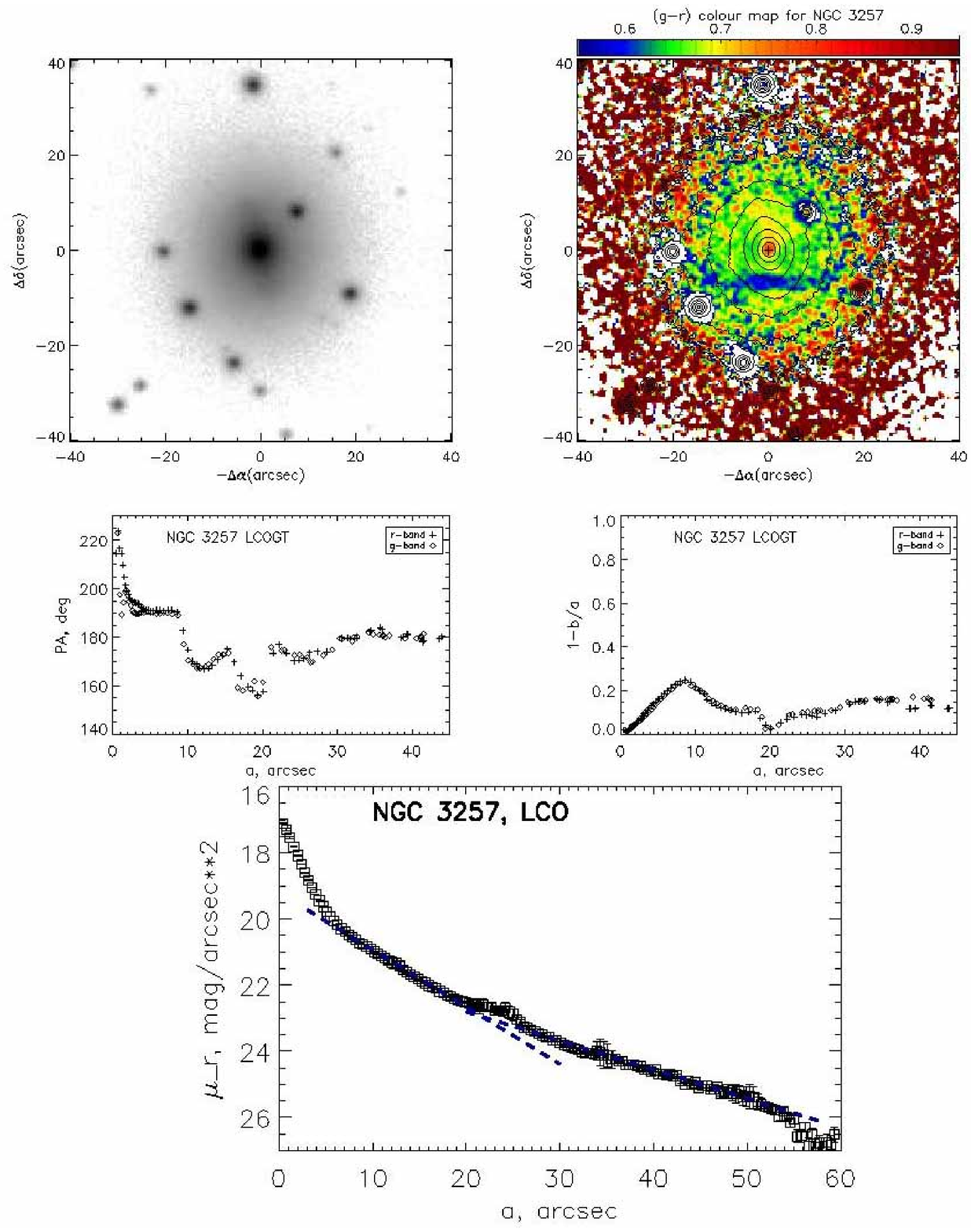}
\end{figure*}

\clearpage

\bigskip
\bigskip

\begin{figure*}
\centering
\includegraphics[width=\textwidth]{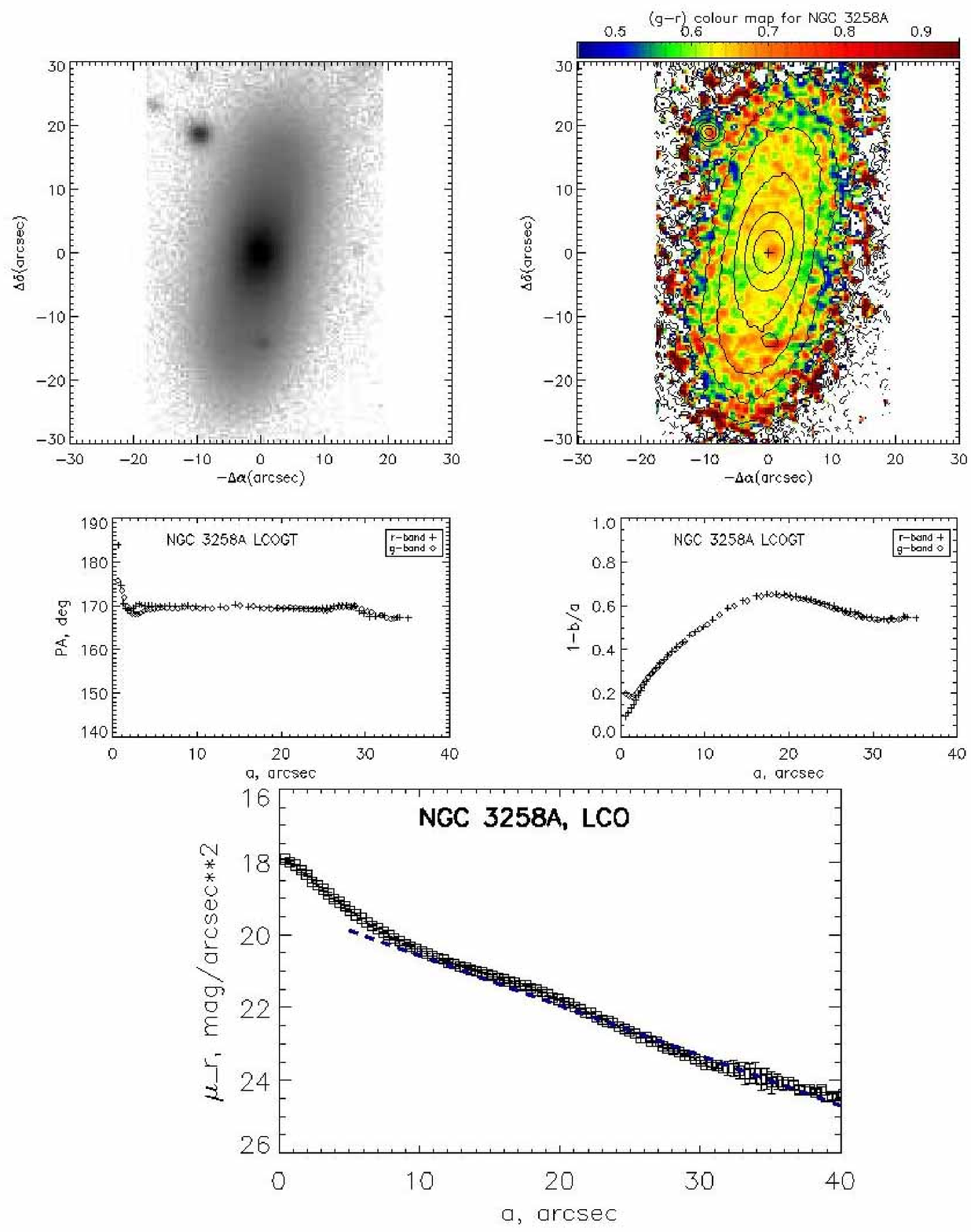}
\end{figure*}

\clearpage

\bigskip
\bigskip

\begin{figure*}
\centering
\includegraphics[width=\textwidth]{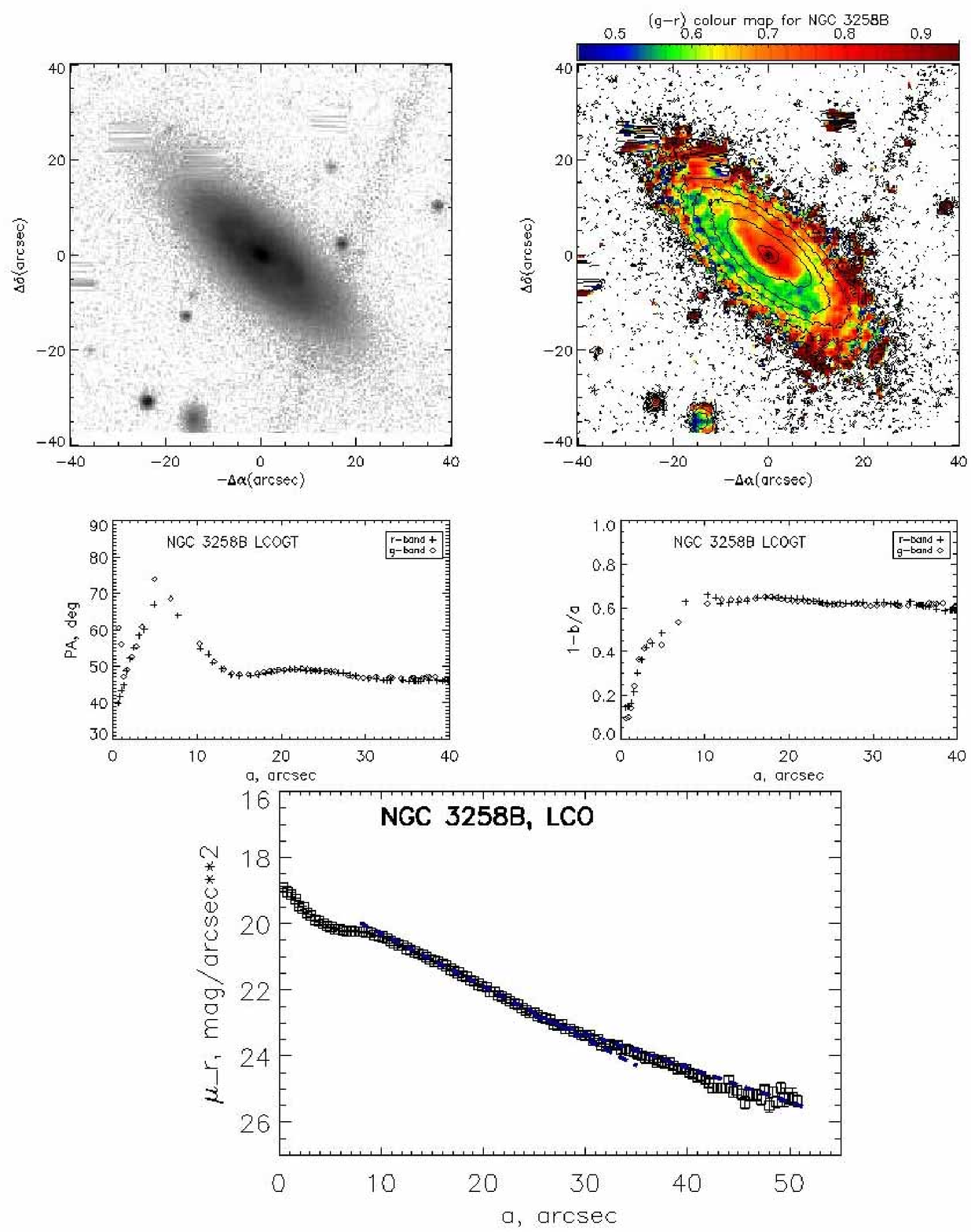}
\end{figure*}

\clearpage

\bigskip
\bigskip

\begin{figure*}
\centering
\includegraphics[width=\textwidth]{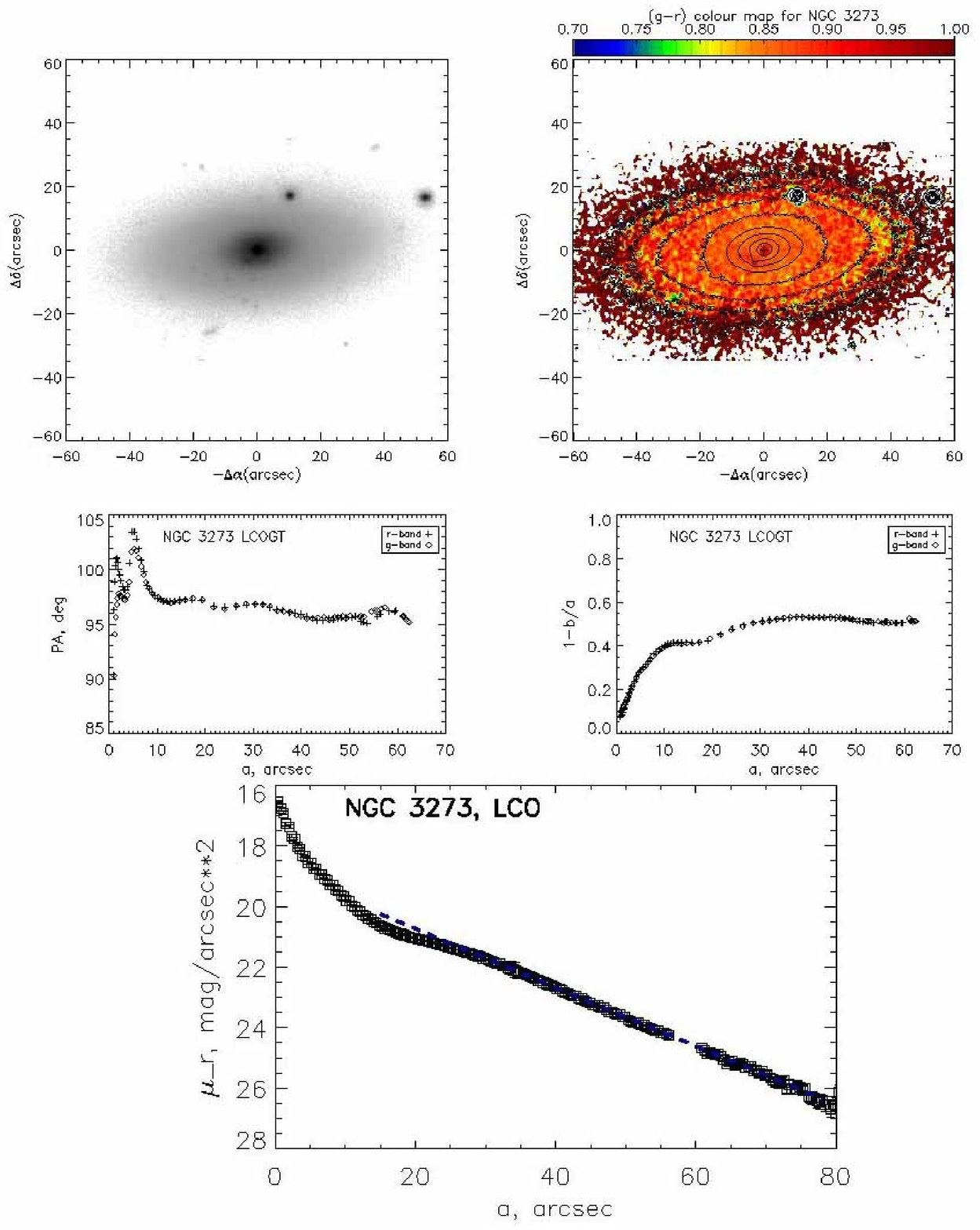}
\end{figure*}

\clearpage

\bigskip
\bigskip

\begin{figure*}
\centering
\includegraphics[width=\textwidth]{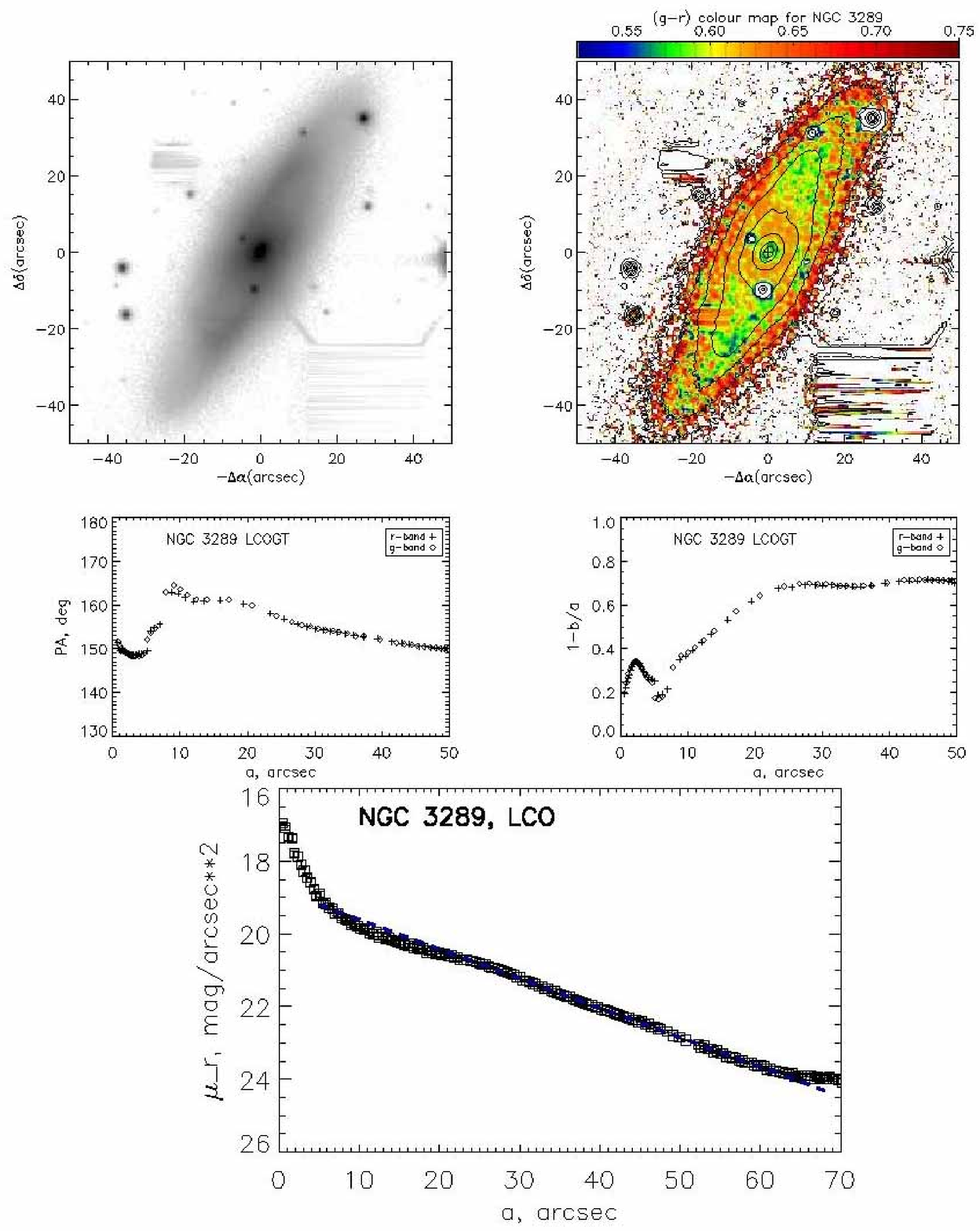}
\end{figure*}

\clearpage

\bigskip

\begin{figure*}
\centering
\includegraphics[width=\textwidth]{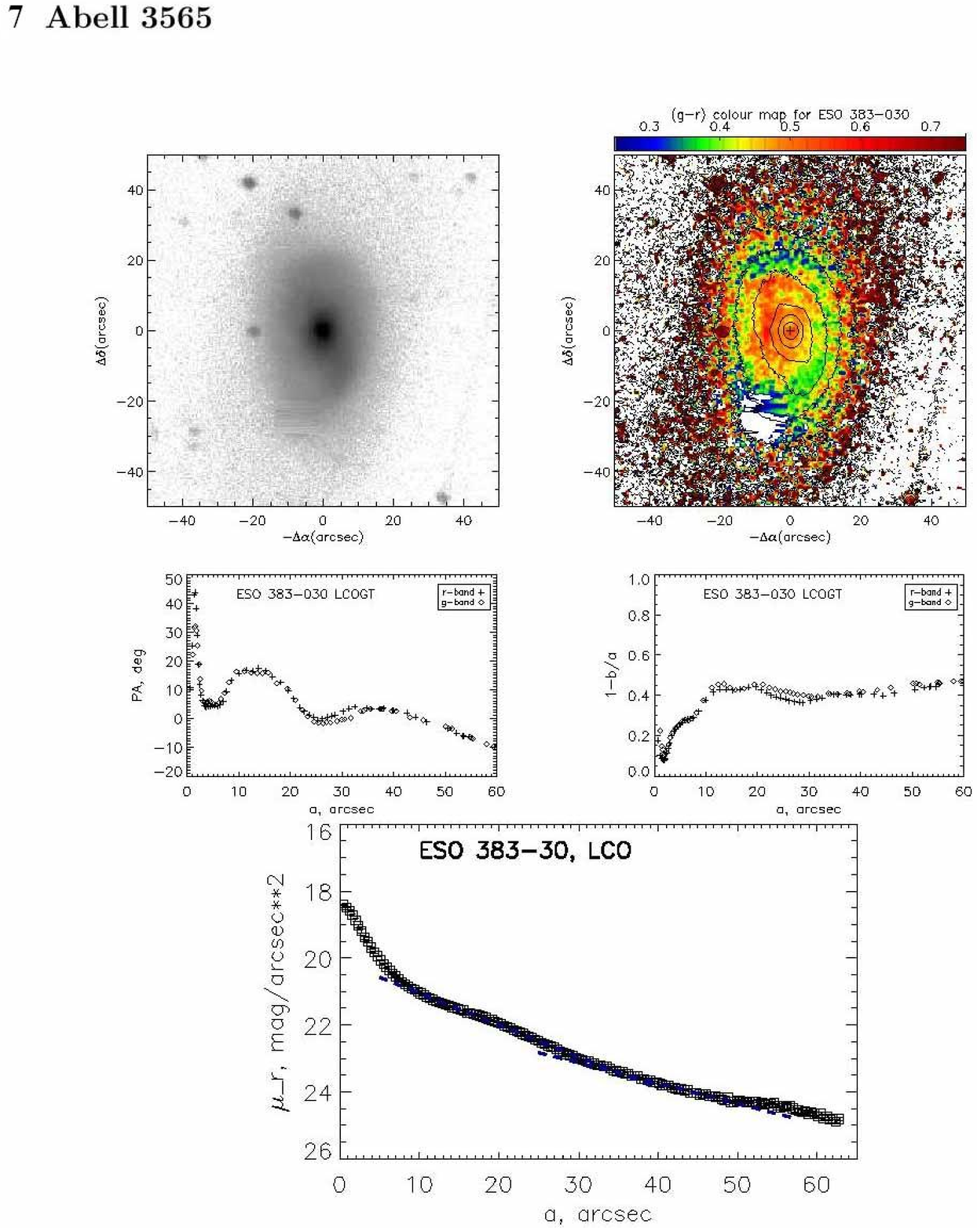}
\end{figure*}

\clearpage

\bigskip
\bigskip

\begin{figure*}
\centering
\includegraphics[width=\textwidth]{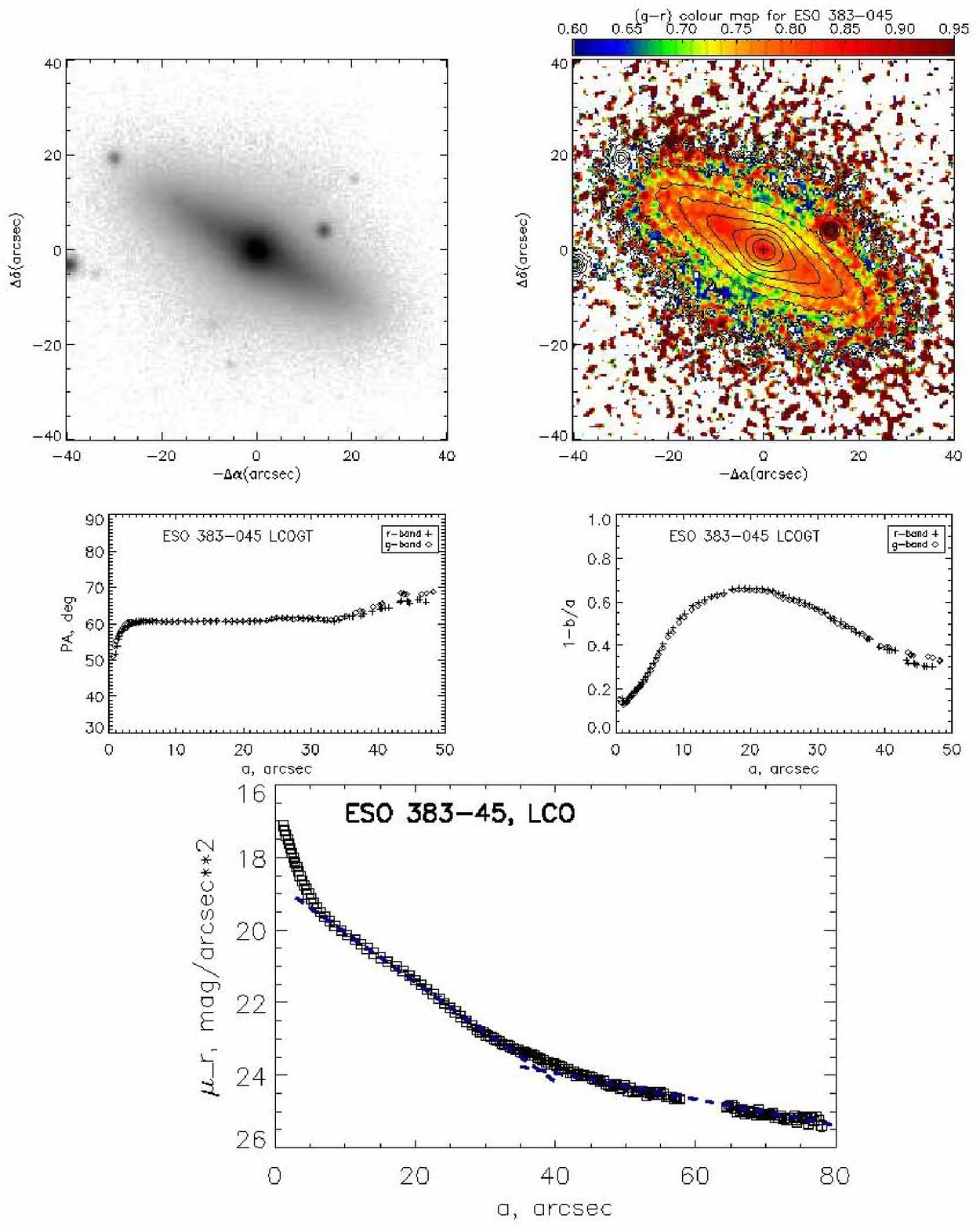}
\end{figure*}

\clearpage

\bigskip
\bigskip

\begin{figure*}
\centering
\includegraphics[width=\textwidth]{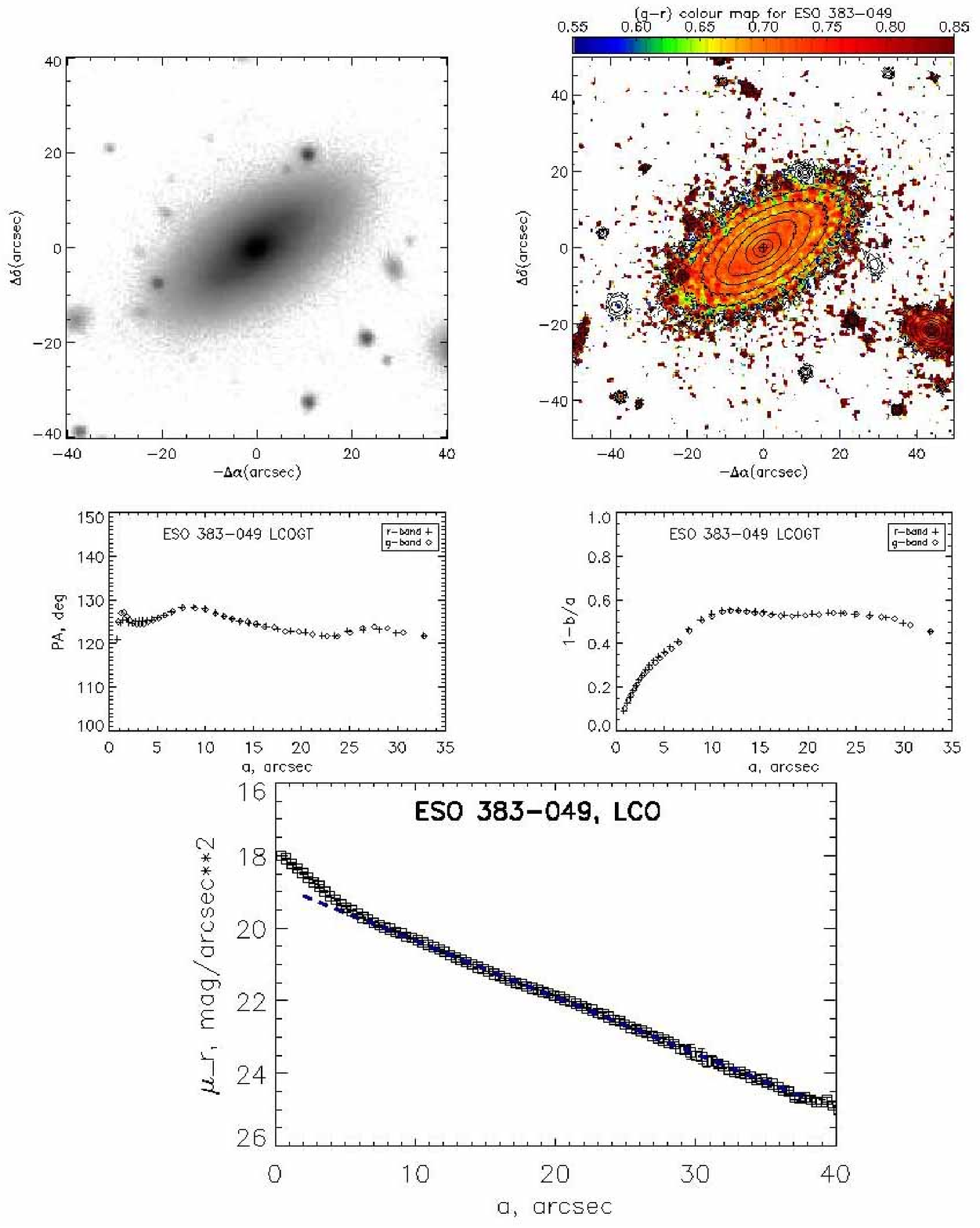}
\end{figure*}

\clearpage

\bigskip
\bigskip

\begin{figure*}
\centering
\includegraphics[width=\textwidth]{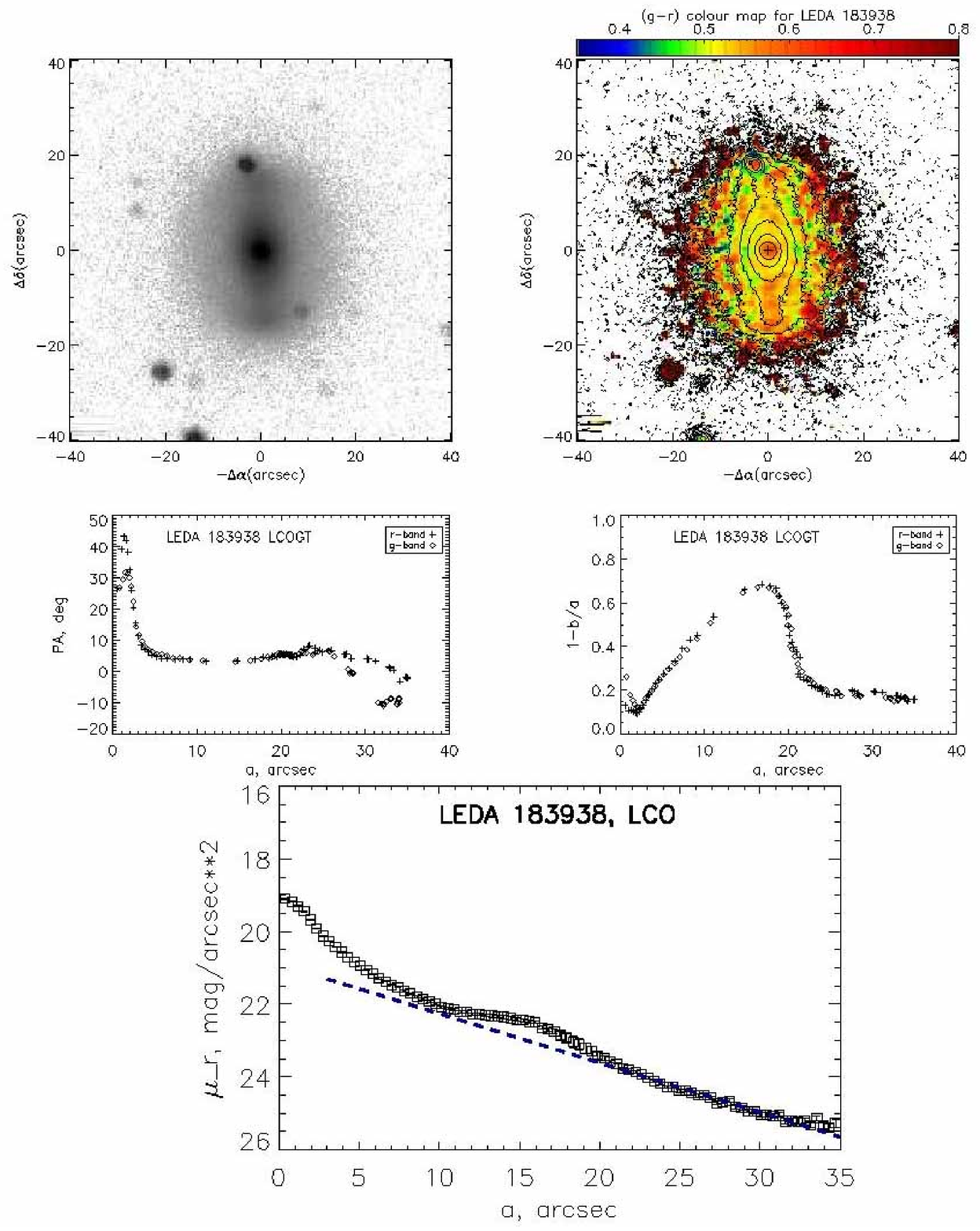}
\end{figure*}

\clearpage

\bigskip

\begin{figure*}
\centering
\includegraphics[width=\textwidth]{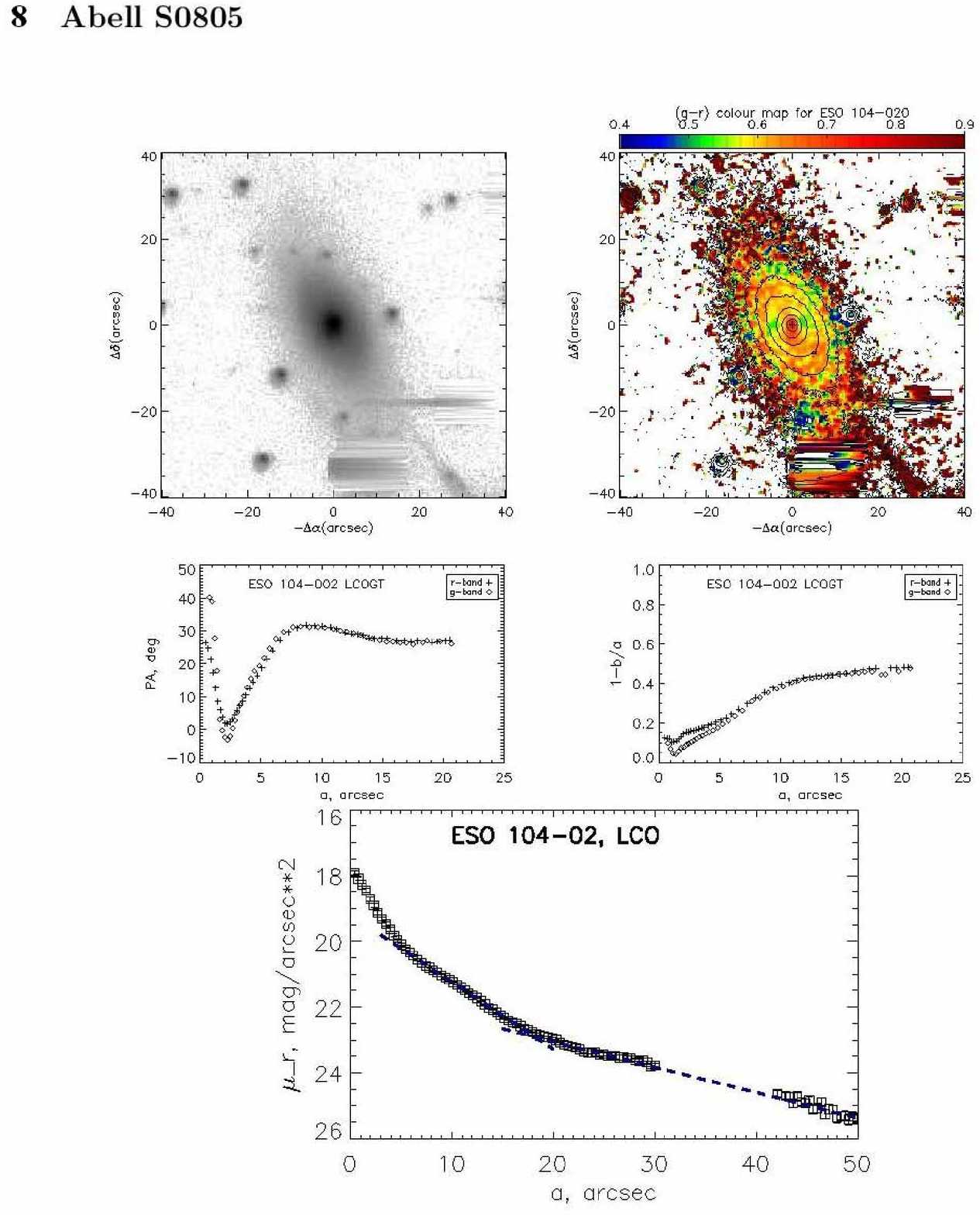}
\end{figure*}

\clearpage

\bigskip
\bigskip

\begin{figure*}
\centering
\includegraphics[width=\textwidth]{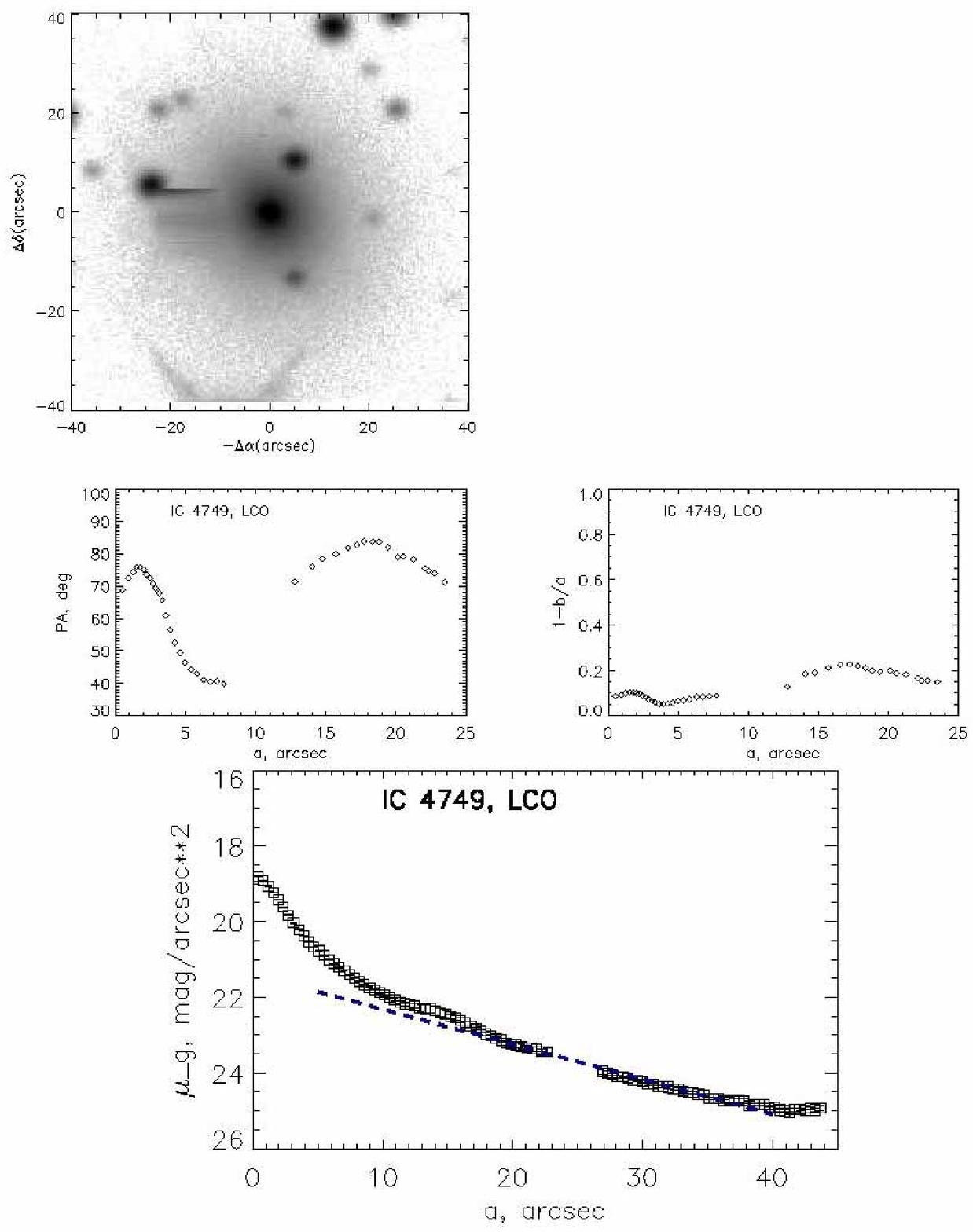}
\end{figure*}

\clearpage

\bigskip
\bigskip

\begin{figure*}
\centering
\includegraphics[width=\textwidth]{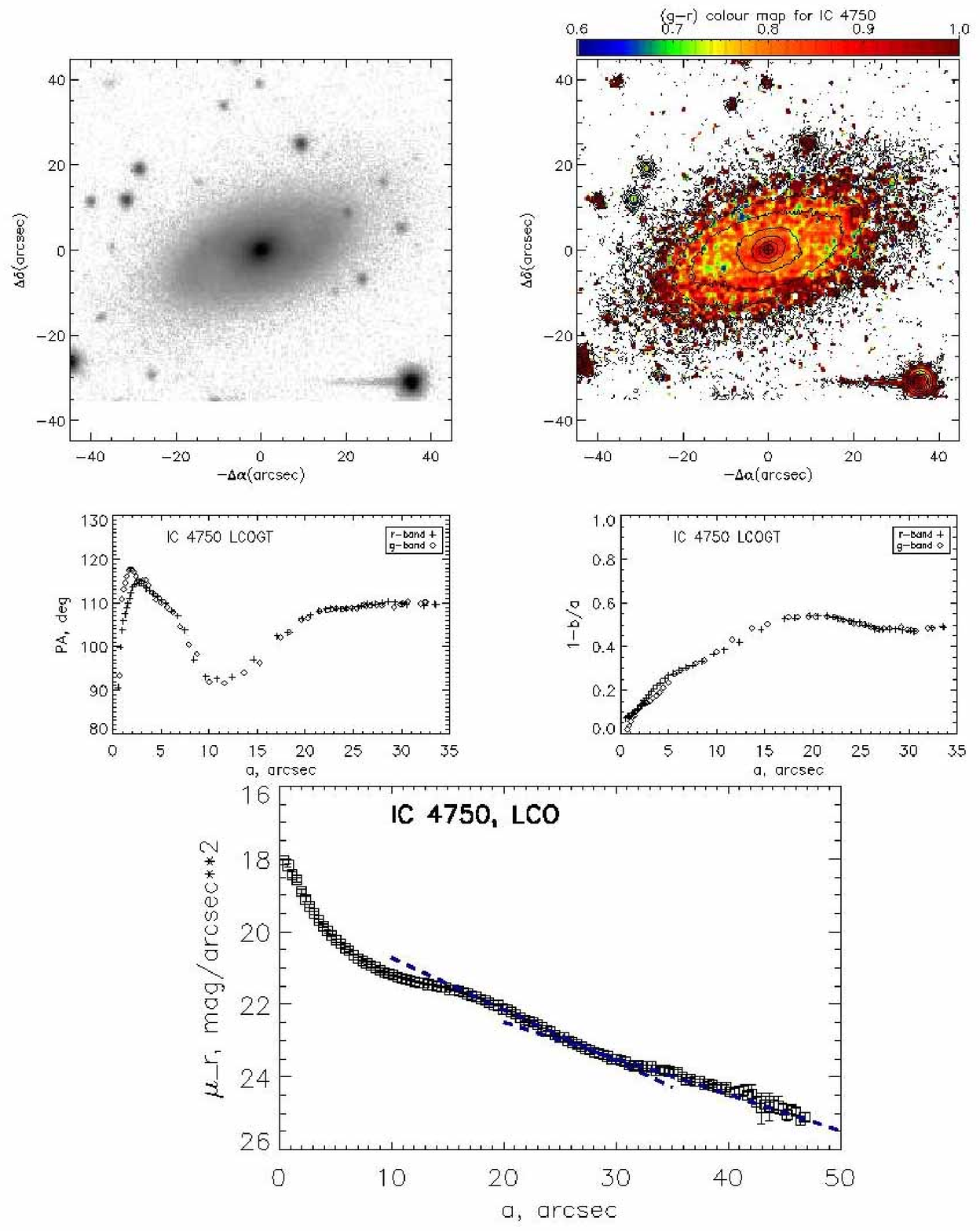}
\end{figure*}

\clearpage

\bigskip
\bigskip

\begin{figure*}
\centering
\includegraphics[width=\textwidth]{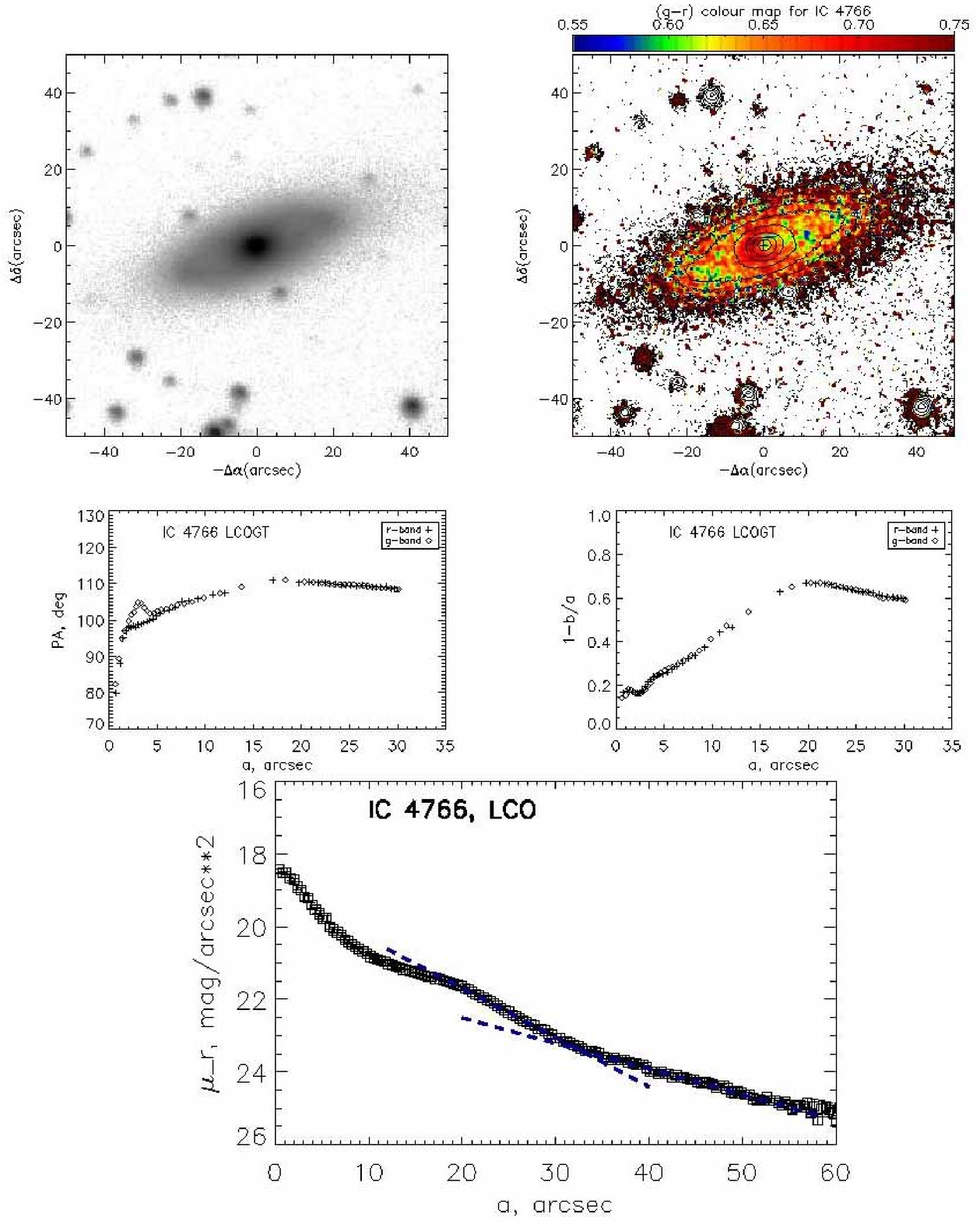}
\end{figure*}

\clearpage

\bigskip
\bigskip

\begin{figure*}
\centering
\includegraphics[width=\textwidth]{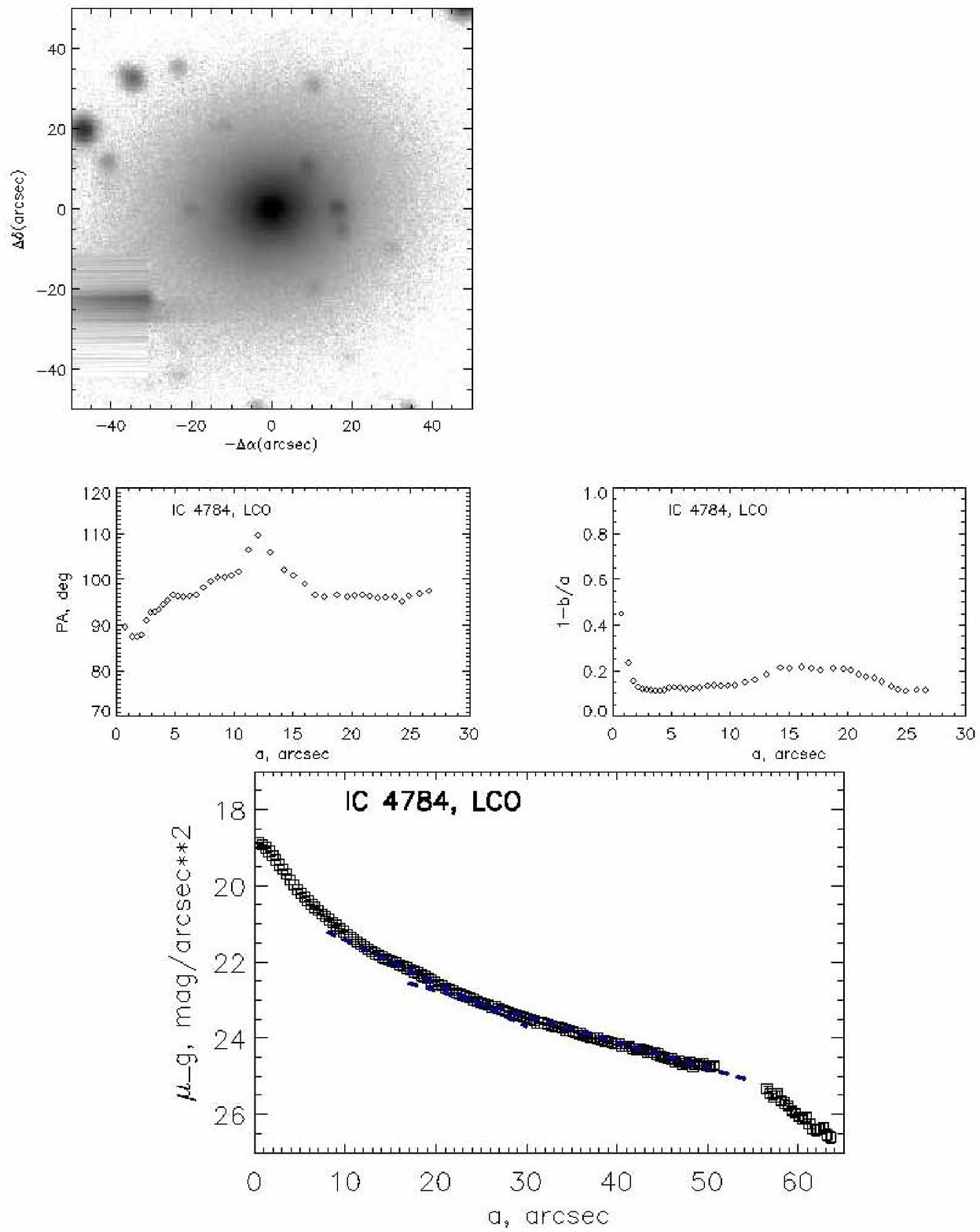}
\end{figure*}

\clearpage

\bigskip
\bigskip

\begin{figure*}
\centering
\includegraphics[width=\textwidth]{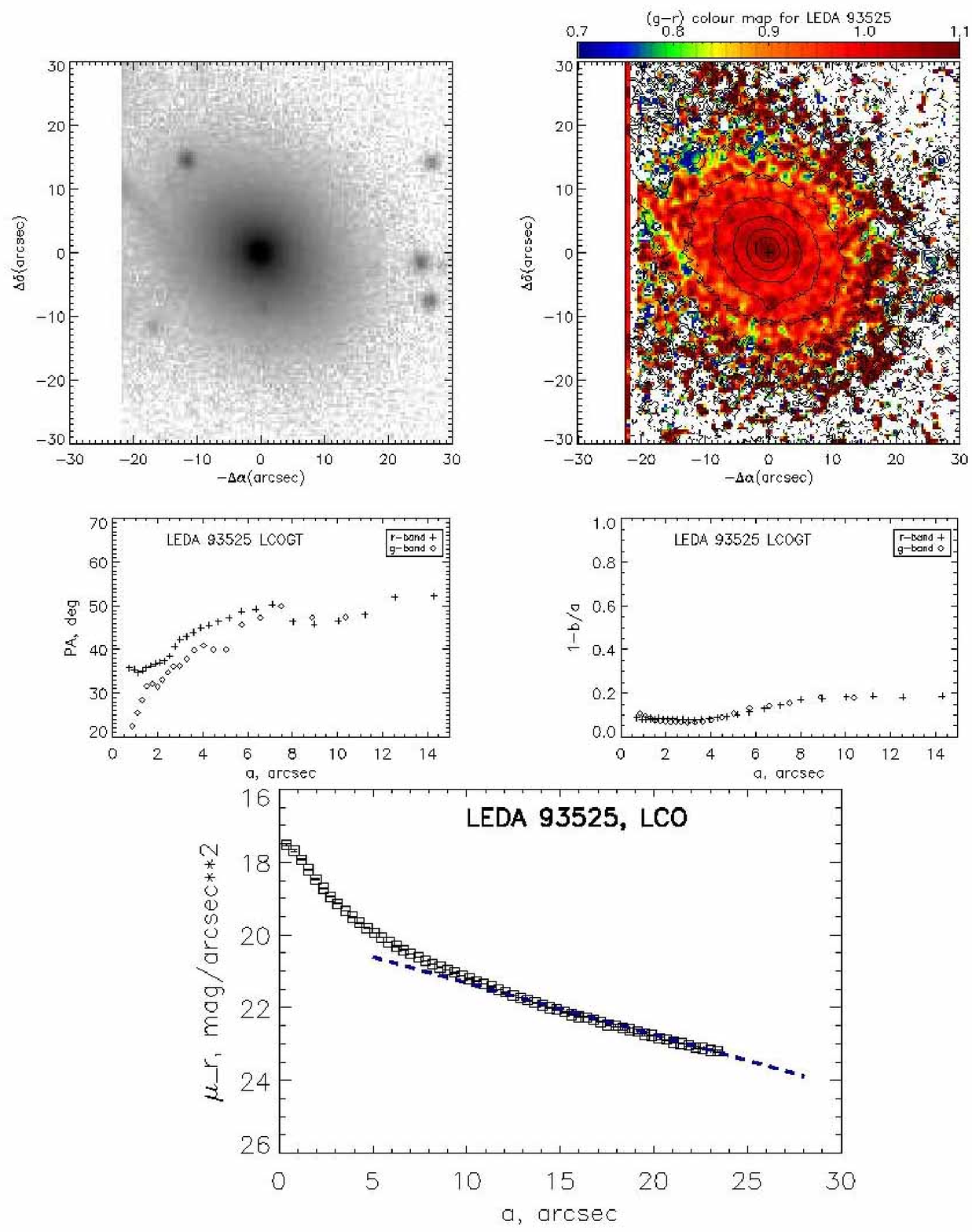}
\end{figure*}

\clearpage

\bigskip
\bigskip

\begin{figure*}
\centering
\includegraphics[width=\textwidth]{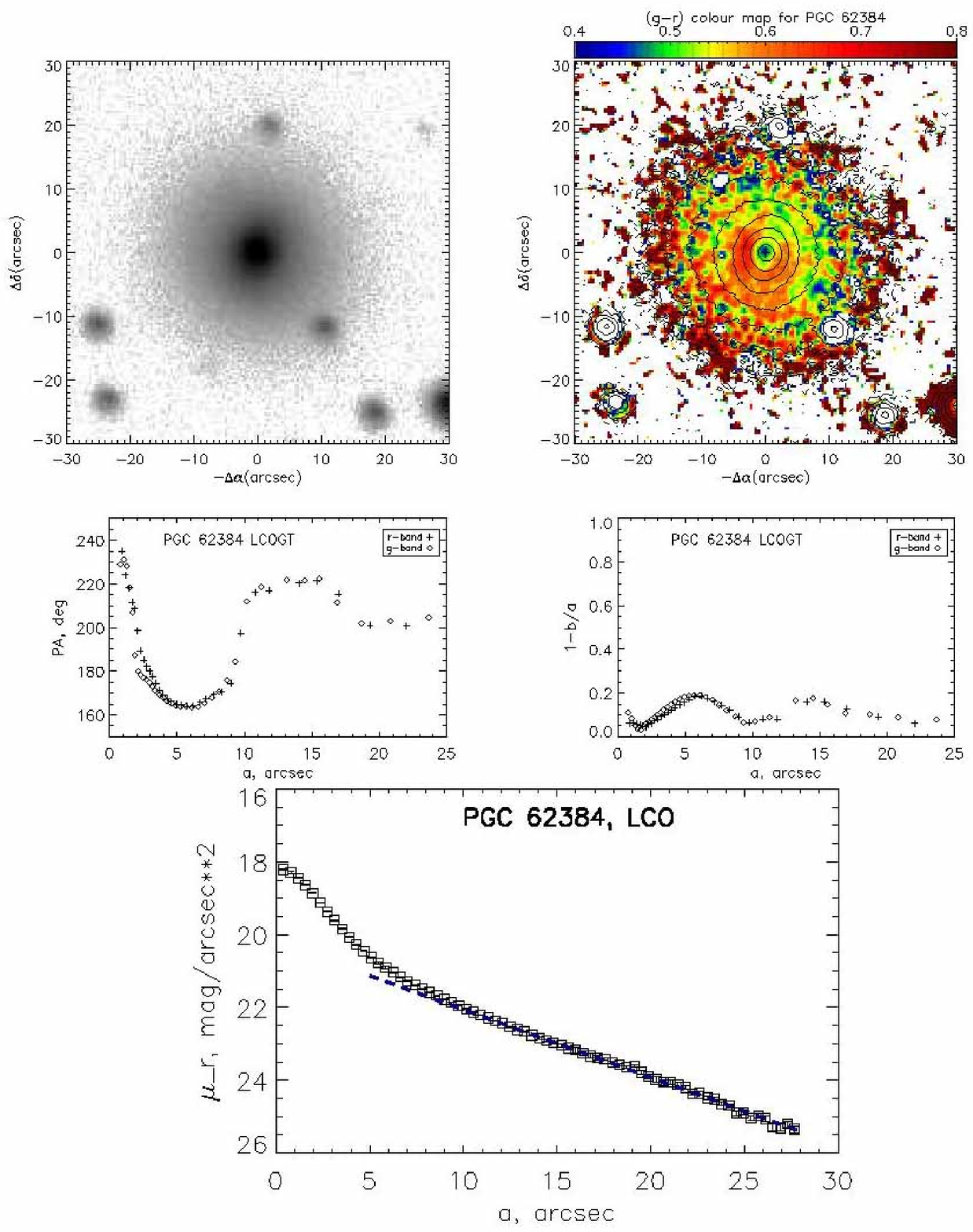}
\end{figure*}

\clearpage

\bigskip
\bigskip

\begin{figure*}
\centering
\includegraphics[width=\textwidth]{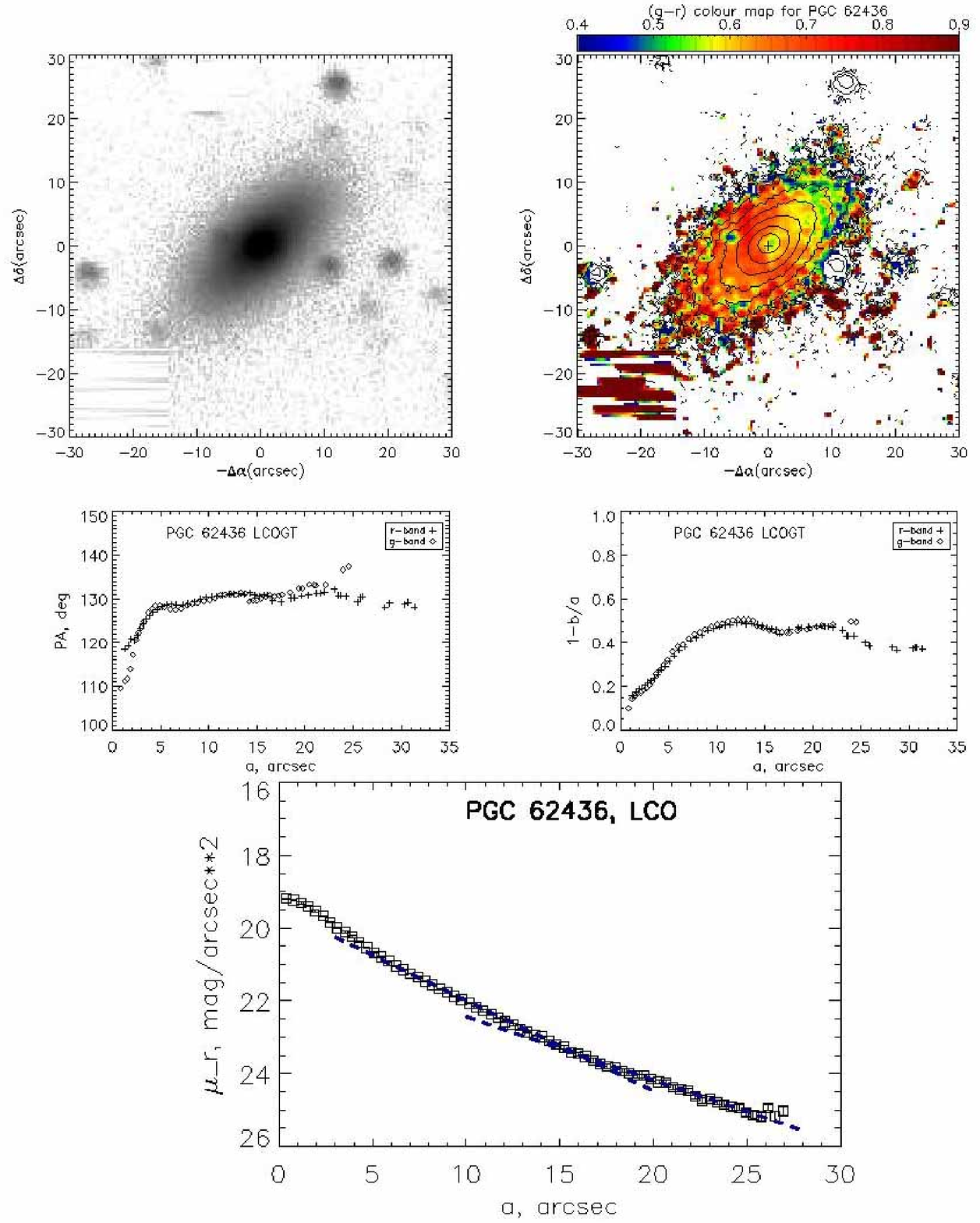}
\end{figure*}

\clearpage

\bigskip
\bigskip

\begin{figure*}
\centering
\includegraphics[width=\textwidth]{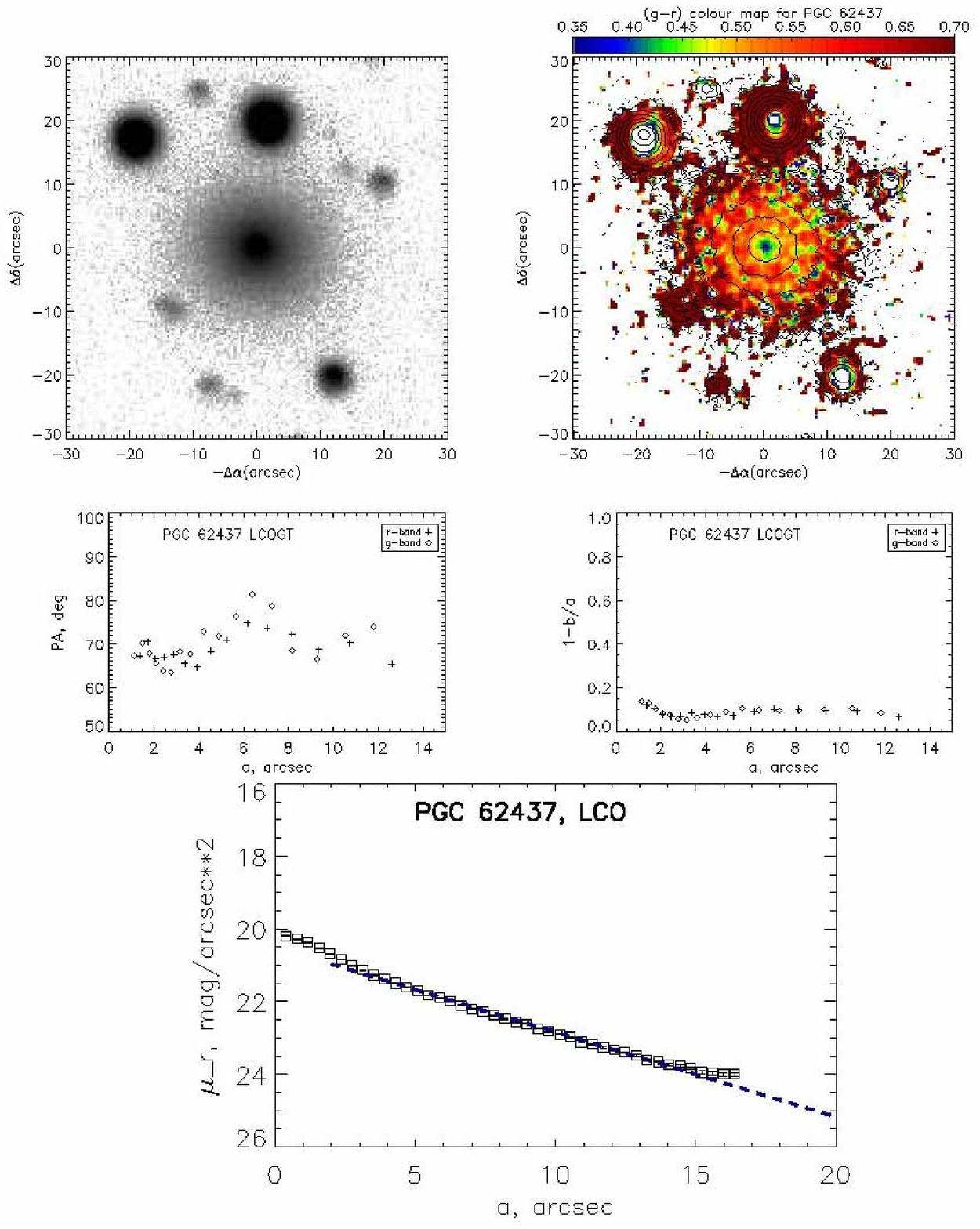}
\end{figure*}

\end{document}